\documentclass[aps,prl,superscriptaddress,longbibliography,twocolumn,reprint,bookmarks=true,colorlinks,linkcolor=blue,citecolor=blue,urlcolor=blue]{revtex4-2}
\usepackage{silence}
\usepackage[english]{babel}
\usepackage{amsmath,amssymb,bbm,graphicx,color,comment}

\usepackage{dsfont}
\usepackage{soul}
\usepackage{subfigure}
\usepackage{extarrows}
\usepackage{float}

\usepackage{xr}

\usepackage{txfonts}
\usepackage[Symbol]{upgreek}
\usepackage{mathtools}
\usepackage[dvipsnames]{xcolor} 
\usepackage[export]{adjustbox}
\usepackage{adjustbox}

\usepackage{dcolumn}   
\usepackage{bm}        
\usepackage{amsfonts}  
\usepackage{lineno}

\usepackage{nicematrix}

\usepackage{orcidlink}

\usepackage{flushend}
\usepackage{siunitx}

\usepackage{pifont}

\DeclareMathOperator{\diag}{diag}

\renewcommand{\vec}[1]{\bm{#1}}

\usepackage{times}

\AtBeginEnvironment{pmatrix}{\setlength{\arraycolsep}{4pt}}

\begin{document}

\title{Altermagnetic Magnons in Dipolar Nanomagnet Arrays}

\author{Rhea Hoyer\,\orcidlink{0000-0003-2285-435X}}
\affiliation{Institut für Festkörpertheorie, Universität Münster, Wilhelm-Klemm-Straße 10, 48149 Münster, Germany}

\author{Ephraim Spindler\,\orcidlink{0009-0001-5656-525X}}
\affiliation{Fachbereich Physik and Landesforschungszentrum OPTIMAS, Rheinland-Pfälzische Technische Universität Kaiserslautern-Landau, 67663 Kaiserslautern, Germany}

\author{Lukas Körber\,\orcidlink{0000-0001-8332-9669}}
\affiliation{Institute for Molecules and Materials, Radboud University, Heyendaalseweg 135, 6525 AJ Nijmegen, The Netherlands}

\author{Tobias Wagner\,\orcidlink{0009-0006-6764-0979}}
\affiliation{Department of Physics, Johannes Gutenberg University Mainz, Staudingerweg 7, 55128 Mainz, Germany}

\author{Mathias Weiler\,\orcidlink{0000-0003-0537-9251}}
\affiliation{Fachbereich Physik and Landesforschungszentrum OPTIMAS, Rheinland-Pfälzische Technische Universität Kaiserslautern-Landau, 67663 Kaiserslautern, Germany}

\author{Alexander Mook\,\orcidlink{0000-0002-8599-9209}}
\affiliation{Institut für Festkörpertheorie, Universität Münster, Wilhelm-Klemm-Straße 10, 48149 Münster, Germany}

\begin{abstract}
    Altermagnetism is conventionally understood in a spin-conserving framework, where symmetry-enforced momentum-dependent spin splitting emerges in collinear magnets with vanishing net magnetization. Here, we show that its defining signatures persist in nanomagnet arrays coupled exclusively by dipolar interactions, despite the intrinsic breaking of spin conservation. Using a macrospin theory of dipolar-coupled ferromagnetic nanoislands, corroborated by micromagnetic simulations, we demonstrate that arrays engineered with altermagnetic symmetries exhibit spin-split magnon bands whose eigenstates can partially remain strongly spin polarized. The resulting spin expectation value displays the characteristic $d$-wave pattern throughout the Brillouin zone, establishing a mesoscopic realization of altermagnetic magnons beyond the conventional spin-conserving paradigm. As a consequence, spin-wave propagation becomes strongly direction dependent, providing a highly tunable platform for anisotropic magnon transport and synthetic altermagnetic functionality.
\end{abstract}

\date{\today}
\maketitle
\textit{Introduction}---Altermagnetism combines the time-reversal symmetry breaking and spin splitting of ferromagnets with the vanishing net magnetization and absence of stray fields of antiferromagnets into a single platform \cite{Smejkal2022Beyond,Smejkal2022Emerging}. These properties have established altermagnets as promising materials for spintronic and magnonic applications \cite{Bai2024,Jungwirth2025,Song2025}, triggering extensive efforts to identify suitable compounds. To date, roughly 180 candidate materials have been predicted \cite{Chen2024,Jiang2024,Liu2024,Xiao2024,Sufyan2026}, and an increasing number of them have been experimentally confirmed through electron spin splitting measured by angle-resolved photoemission spectroscopy (ARPES) \cite{Krempasky2024,Lee2024,Reimers2024,Osumi2024,Hajlaoui2024,Zeng2024,Ding2024,Santhosh2025,Yang2025,Jiang2025,Zhang2025} and magnon spin splitting observed by inelastic neutron scattering (INS) \cite{Liu2024,Sun2025,Faure2025,Singh2026,Sears2026,Asai2026}. Beyond the search for crystalline compounds, an equally appealing direction is the realization of \emph{synthetic} altermagnetic systems \cite{Asgharpour2025, Gallardo2026}, where the underlying symmetry, interactions, and band structure can be engineered at will. Such synthetic systems can provide highly desirable properties for applications, as widely demonstrated by synthetic antiferromagnets~\cite{Duine:Synthetic:2018}. While synthetic antferromagnets provide magnon bands in the GHz range, compatible with magnonic applications~\cite{Ishibashi:Switchable:2020}, a corresponding synthetic altermagnetic platform for magnonics is missing. Here, we demonstrate that nanomagnet arrays can provide such a GHz-magnonics thin-film synthetic altermagnetic platform.

Altermagnetism is conventionally formulated in the spin-conserving limit, where spin remains a good quantum number and the spin splitting is protected by spin-space symmetries \cite{Smejkal2022Beyond,Smejkal2022Emerging,Litvin1974}. It therefore remains an open question whether the characteristic signatures of altermagnetism survive in systems with intrinsically spin-nonconserving interactions \cite{Campos2026} on the mesoscopic scale. Here, we answer this question affirmatively by considering perhaps the most extreme case: a system in which altermagnetic magnon bands emerge \emph{solely} from magnetic dipolar interactions. Since dipolar interactions intrinsically couple spin and real space \cite{Politi2002}, spin conservation is broken and spin symmetry arguments no longer apply. Remarkably, we show that the magnon bands can nevertheless remain strongly spin polarized and retain the characteristic $d$-wave pattern of spin expectation values throughout the Brillouin zone (BZ).

To demonstrate a synthetic altermagnet realized solely by spin-nonconserving interactions, we develop a macrospin theory of dipolar-coupled ferromagnetic nanoisland arrays inspired by magnonic crystals \cite{Puszkarski2003,Krawczyk2014,Chumak2017} and artificial spin ices \cite{Skjaervo2020,Lendinez2020,Kaffash2021}. Starting from the circular-island antiferromagnetic dot lattice introduced in Ref.~\cite{Verba2012}, whose magnon bands are spin unpolarized, we show that a simple geometric modification---elongating the islands into ellipses with sublattice-dependent orientations---realizes altermagnetic symmetries and generates momentum-dependent magnon spin polarization for two distinct lattice geometries. We benchmark the resulting band structures against micromagnetic simulations and discuss experimentally realistic parameters. Our work establishes dipolar nanomagnet arrays as a programmable platform for synthetic altermagnetic magnonics, demonstrating that the defining symmetry principles of altermagnetism can be engineered on the mesoscopic scale despite the absence of spin conservation.

\textit{Model}---We consider a periodic array of dipolar-coupled flat insulating nanoislands \cite{Verba2012} and assume that the islands are made of a low-damping ferromagnetic material with out-of-plane easy-axis 
anisotropy, such as bismuth-doped yttrium iron garnet (Bi$_x$Y$_{3-x}$Fe$_5$O$_{12}$ or Bi:YIG) \cite{Hansen1984,Soumah2018,Deb2019,Bhatti2026}.
The total magnetic energy is approximately given by
\begin{align}
    H & = \frac{\mu_0 M_\mathrm{s}^2  V}{2} \sum_{i, j} \sum_{\alpha, \beta} \bm{m}_{i}^{\alpha} \cdot \hat{\bm{N}}^{\alpha \beta}(\bm{r}_i - \bm{r}_j) \cdot  \bm{m}_{j}^{\beta}  \nonumber \\
     & \quad -  V \sum_i \sum_\alpha K \left( \bm{e}_z \cdot \bm{m}_{i}^{\alpha} \right)^2,
     \label{eq:HAM}
\end{align}
where $\alpha, \beta = \text{A},\text{B}$ denote the two sublattices, $\mu_0$ is the vacuum permeability, $M_\mathrm{s}$ the saturation magnetization, $V$ the volume of an island, $\bm{e}_i$ is a unit vector in direction $i$, $K > 0$ is an easy-axis anisotropy, and $\bm{m}_i^\alpha$ is a unit-vector macrospin on sublattice $\alpha$ with $|\bm{m}_i^\alpha| = 1$. Furthermore, $\boldsymbol{N}^{\alpha\beta}$ is the demagnetization tensor leading to onsite shape anisotropy and mediating the dipolar interactions between different nanoislands (see End Matter).
We assume the nanoislands to be small enough such that each can be treated as a single macrospin, neglecting the intra-island exchange interaction between individual magnetic moments, as well as inhomogeneous modes appearing at higher frequencies.

Depending on the saturation magnetization, crystalline anisotropy, and geometric parameters of the nanoislands, various ground-state configurations can be stabilized. Here, we choose parameters such that a staggered collinear out-of-plane magnetization is the ground state, confirmed using the Luttiger-Tisza method \cite{Luttinger1946,Luttinger1951} (see End Matter).

We apply the Holstein-Primakoff transformation \cite{Holstein1940} up to first order in bosons as $\hat{m}_{i,x}^\alpha \approx -\frac{1}{\sqrt{2 S}}\left( \hat{c}_i +  \hat{c}_i^\dagger\right)$, $\hat{m}_{i,y}^\alpha \approx (- 1)^n  \frac{\mathrm{i}}{\sqrt{2 S}}\left( \hat{c}_i - \hat{c}_i^\dagger \right)$, $\hat{m}_{i,z}^\alpha = (- 1)^n \left(-1 + \frac{1}{S}  \hat{c}_i^\dagger \hat{c}_i \right)$, with $S = \frac{M_\text{s} V}{|\gamma| \hbar}$ and the gyromagnetic ratio $\gamma$ (see End Matter).
For sublattice A (B), we set $n$ even (odd) and $\hat{c}_i^\dagger=\hat{a}_i^\dagger$ ($\hat{b}_i^\dagger)$ are the boson creation operators. 
After a Fourier transformation, $\hat{c}_i^\dagger = \frac{1}{\sqrt{N}} \sum_{\vec{k}} \mathrm{e}^{-\mathrm{i} \bm{k} \cdot \bm{r}_i} \hat{c}_{\bm{k}}^\dagger$, and $\hat{c}_i= \frac{1}{\sqrt{N}}\sum_{\vec{k}}\mathrm{e}^{\mathrm{i} \bm{k} \cdot \bm{r}_i} \hat{c}_{\bm{k}}$, with $N$ the total number of unit cells, we find the harmonic Hamiltonian $H_2 = \frac{1}{2} \sum_{\bm{k}} \bm{\Psi}^\dagger_{\bm{k}} \mathcal{H}_{\bm{k}} \bm{\Psi}_{\bm{k}}$, written in the Nambu basis $ \bm{\Psi}_{\bm{k}}^{\dagger} = \left(\hat{a}_{\bm{k}}^{\dagger}, \hat{b}_{\bm{k}}^\dagger, \hat{a}_{-\bm{k}}, \hat{b}_{-\bm{k}}\right)$. Magnon-magnon interactions can be neglected at low temperatures. 
The Hamilton kernel reads
\begin{align}
\label{eq:ham-kernel-dipolar}
	\mathcal{H}_{\bm{k}} 
    & = \begin{pmatrix} 
            A(\bm{k}) & B(\bm{k})\\
            B(\bm{k})^\dagger & A(-\bm{k})^\mathrm{T}
        \end{pmatrix},
\end{align}
where $A(-\bm{k})^\mathrm{T} = A(\bm{k})$ and
\begin{align}
    A(\bm{k}) = \begin{pmatrix}
        V_{\bm{k}}^\text{A}  & W_{\bm{k}} \\
        W_{\bm{k}}^*  & V_{\bm{k}}^\text{B}
    \end{pmatrix}, \quad 
    B(\bm{k}) = \begin{pmatrix}
        X_{\bm{k}}^\text{A}& Y_{\bm{k}} \\
        Y_{\bm{k}}^*& X_{\bm{k}}^\text{B}
    \end{pmatrix},
\end{align}
with the corresponding matrix elements
\begin{subequations}
\label{eq:matrix-elements}
\begin{align}
    V_{\bm{k}}^\alpha &= \frac{\hbar \omega_{\text{M}}}{2} \left( \hat{F}_{\bm{k},xx}^{\alpha\alpha} + \hat{F}_{\bm{k} ,yy}^{\alpha\alpha}  - 2 \hat{F}_{\bm{0} ,zz}^{\alpha\alpha} + 2 \hat{F}_{\bm{0} ,zz}^{\text{AB}} \right) + \hbar \omega_{\text{K}}, \\ 
    W_{\bm{k}} & =  \frac{\hbar \omega_{\text{M}}}{2} \left( \hat{F}_{\bm{k} ,xx}^{\text{AB}} - \hat{F}_{\bm{k} ,yy}^{\text{AB}} + 2 \mathrm{i} \hat{F}_{\bm{k} ,xy}^{\text{AB}} \right), \\
    X_{\bm{k}}^\alpha & =\frac{\hbar \omega_{\text{M}}}{4} \left[ \hat{F}_{\bm{k} ,xx}^{\alpha\alpha} - \hat{F}_{\bm{k} ,yy}^{\alpha\alpha}  + (-1)^n 2 \mathrm{i} \hat{F}_{\bm{k} ,xy}^{\alpha\alpha}\right], \\
    Y_{\bm{k}} & =\frac{\hbar \omega_{\text{M}}}{4} \left( \hat{F}_{\bm{k} ,xx}^{\text{AB}} + \hat{F}_{\bm{k} ,yy}^{\text{AB}} \right),
\end{align}
\end{subequations}
where $\omega_{\text{M}} = |\gamma| \mu_0 M_\text{s}$ and $\omega_\text{K} = |\gamma| 2 K / M_\text{s}$.
We define $\hat{\bm{F}}_{\bm{k}}^{\alpha \beta} = \sum_{\bm{r}_{ij}} \hat{\bm{N}}_{ij}^{\alpha \beta} \mathrm{e}^{- \mathrm{i} \bm{k} \cdot \bm{r}_{ij}} \mathrm{e}^{- \mathrm{i} \bm{k} \cdot \bm{r}_{\alpha \beta}}$, where $\hat{\bm{N}}_{ij}^{\alpha \beta}$ is the mutual demagnetizing tensor of islands at sites $i$ and $j$ [cf.~Eq.~\eqref{eq:demag_2}] and
$\bm{r}_{\text{AB}} = \bm{r}_{\text{A}} - \bm{r}_{\text{B}}=- \frac{1}{2}\left(\bm{a}_1 + \bm{a}_2 \right)$ with appropriate lattice vectors $\bm{a}_1$ and $\bm{a}_2$. 
Using $\sum_{\bm{r}_{ij}} \mathrm{e}^{- \mathrm{i} \bm{k} \cdot \bm{r}_{ij}} = \frac{\left(2 \uppi\right)^2}{A_\text{UC}} \sum_{\bm{G}} \delta (\bm{k} + \bm{G})$ \cite{Ashcroft1978}, where $A_{\text{UC}}$ is the area of the unit cell, and $\bm{G}$ is a reciprocal lattice vector, we find 
\begin{align}
    \hat{\bm{F}}_{\bm{k} }^{\alpha \beta} &= \frac{1}{A_\text{UC}} \sum_{\bm{G}} \hat{\bm{N}}_{\bm{k} + \bm{G}}^{\alpha \beta} \,\mathrm{e}^{-\mathrm{i} \left(\bm{k} + \bm{G}\right) \cdot \bm{r}_{\alpha \beta}}.
    \label{eq:F-tensor}
\end{align}
The Fourier image of the demagnetization tensor, $\hat{\bm{N}}_{\bm{k}}^{\alpha \beta}$, contains the shape factors $\sigma_{\bm{k}}^\alpha$ of the two interacting islands in Eq.~\eqref{eq:demag_tensor_k}, as derived in the End Matter.
We numerically diagonalize the Hamilton kernel \eqref{eq:ham-kernel-dipolar} using a paraunitary Bogoliubov transformation \cite{Colpa1978} to find the magnon spectrum $ H_2 = \sum_{\boldsymbol{k}} \left[ \varepsilon_{\boldsymbol{k},1} \left( \hat{\alpha}_{\boldsymbol{k}}^{\dagger} \hat{\alpha}_{\boldsymbol{k}} + \frac{1}{2} \right) + \varepsilon_{\boldsymbol{k},2} \left( \hat{\beta}_{\boldsymbol{k}}^{\dagger} \hat{\beta}_{\boldsymbol{k}} + \frac{1}{2} \right) \right]
$ in form of harmonic oscillators, where $\varepsilon_{\boldsymbol{k},1}$ $\left(\varepsilon_{\boldsymbol{k},2}\right)$ denotes the energy of the magnon normal mode $\hat{\alpha}_{\boldsymbol{k}}$ $\left(\hat{\beta}_{\boldsymbol{k}}\right)$.

Due to the dipolar interaction, spin is not conserved. In terms of the free theory of magnons, that means that $\mathcal{H}_{\vec{k}}$ does not $G$-commute with the spin operator $\Sigma = \text{diag} (1,-1,1,-1)$, where $G = \text{diag} (1,1,-1,-1)$ is the  bosonic metric:
\begin{align}
    [\mathcal{H}_{\vec{k}}, \Sigma]_G = \mathcal{H}_{\vec{k}} G \Sigma - \Sigma G \mathcal{H}_{\vec{k}} \ne 0.
\end{align}
Note that $[\mathcal{H}_{\vec{k}}, \Sigma]_G = 0$ only if $W_{\vec{k}} = X^\text{A}_{\vec{k}} = X^\text{B}_{\vec{k}} = 0$, which is, however, not the case. The limit in which altermagnetism is cleanly defined is therefore never assumed. Nonetheless, we do find points in the spectrum, the X and Y points, where $W_{\vec{k}} = Y_{\vec{k}} = 0$ and the Hamilton kernel becomes block diagonal in the sublattices [as in Ref.~\cite{Verba2012} and shown in the Supplemental Material (SM) \cite{SM}].
The magnon spin expectation value at the X point reads \cite{SM}
\begin{align}
        \langle m_z^\alpha \rangle &= \mp \frac{V_{\bm{X}}^\alpha}{\varepsilon_{\bm{X}}^\alpha},
\end{align}
where ``$-$'' (``$+$'') belongs to sublattice $\alpha = \text{A}$ ($\alpha = \text{B}$), and $\varepsilon_{\bm{X}}^\alpha = \sqrt{(V_{\bm{X}}^\alpha )^2- | X^\alpha_{\bm{X}}|^2}$ denotes the energies at X (see SM~\cite{SM}).
Interestingly, it follows that $|\langle m_z^\alpha \rangle| > 1 $ is \textit{larger} than in the spin-conserving case of antiferromagnets or canonical altermagnets where $|\langle m_z^\alpha \rangle| = 1$ \cite{Hoyer2025a} (see SM~\cite{SM}). The sublattice-decoupled case is similar to two oppositely magnetized copies of an anisotropic ferromagnet with elliptical spin waves, for which the magnon spin enhancement beyond unity has been reported previously in Ref.~\cite{Kamra2017}.

\begin{figure}
    \centering
    \includegraphics[width=\linewidth]{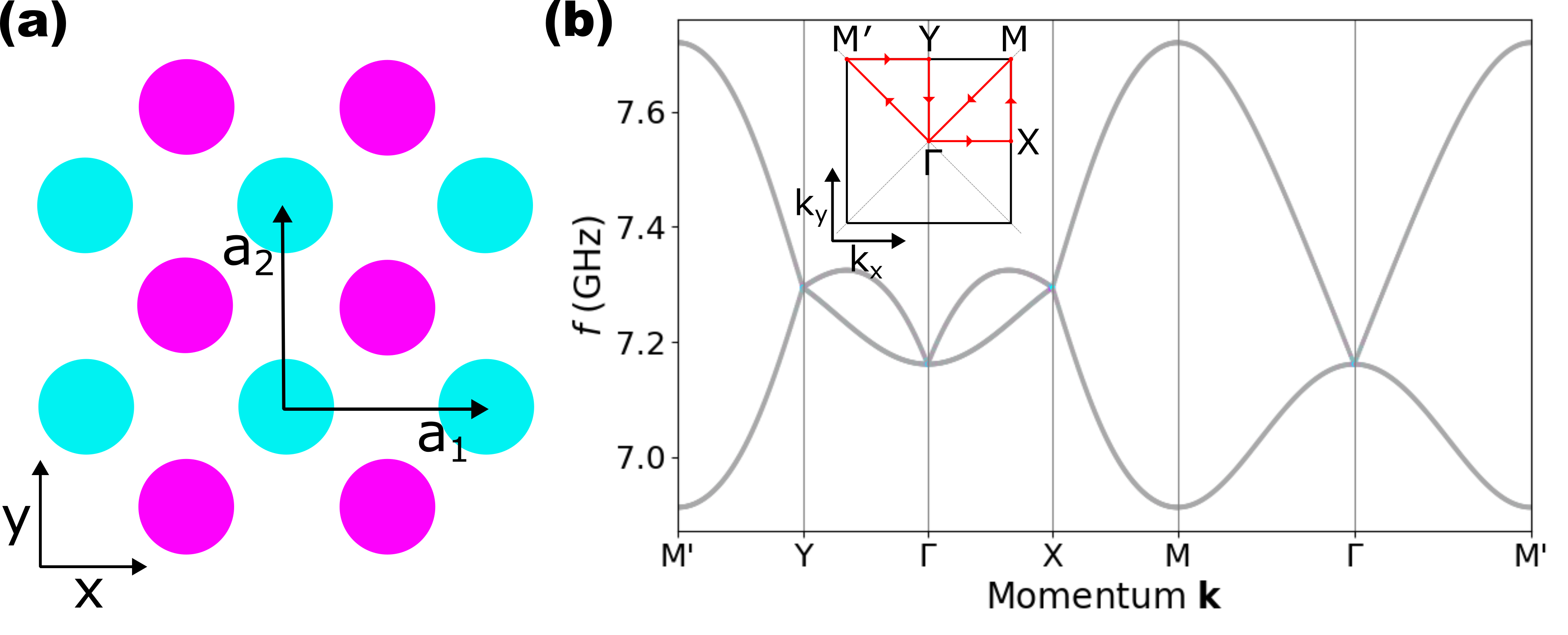}
    \caption{(a) Antiferromagnetic nanomagnet array of circular islands with lattice vectors $\bm{a}_1$ and $\bm{a}_2$. (b) Magnon dispersion along a high-symmetry path; inset shows the first BZ with dashed diagonal lines marking compensating mirror symmetries. Parameters deviating from Table~\ref{tab:parameters-1} (End Matter) are $h_1=h_2$ and $a = 6.0 \times 10^{-8}$~m.}
    \label{fig:model-0}
\end{figure}

\textit{Antiferromagnetic geometry.} We introduce the antiferromagnetic lattice made of circular islands with magnetization in $z$ ($-z$) for cyan (magenta) islands, as shown in Fig.~\ref{fig:model-0}(a). We set the half-axes of the elliptical shape factor $\sigma^{\alpha}$ to be equal: $h_1 = h_2$.
The array has lattice vectors $\bm{a}_1=(a, 0)$ and $\bm{a}_2=(0,a)$ with lattice constant $a = 6.0 \times 10^{-7}$~m. The model belongs to the nonsymmorphic magnetic layer group (MLG) $p_c4/mmm$, whose $\mathcal{P}\mathcal{T}\bm{\tau}$ and $\mathcal{T}\vec{\tau}$ symmetry (with $\mathcal{P}$ inversion, $\mathcal{T}$ time-reversal symmetry, and $\vec{\tau}$ a nontrivial lattice translation) individually enforce magnetic compensation.
In Fig.~\ref{fig:model-0}(b) the magnon dispersion with frequency $f$ along a high-symmetry path in the first BZ is fully split due to dipolar interactions except at the X, Y, and $\Gamma$ points, congruent with Ref.~\cite{Verba2012}. The $\Gamma$ point degeneracy is tied to the $\mathcal{C}_{4z}$ symmetry, whereas the X and Y point degeneracies involve nonsymmorphic symmetries (see SM~\cite{SM}). The magnon spin polarization $\langle m_z^\alpha \rangle$ vanishes entirely (except for band degeneracies).

\begin{figure}
    \centering
    \includegraphics[width=\linewidth]{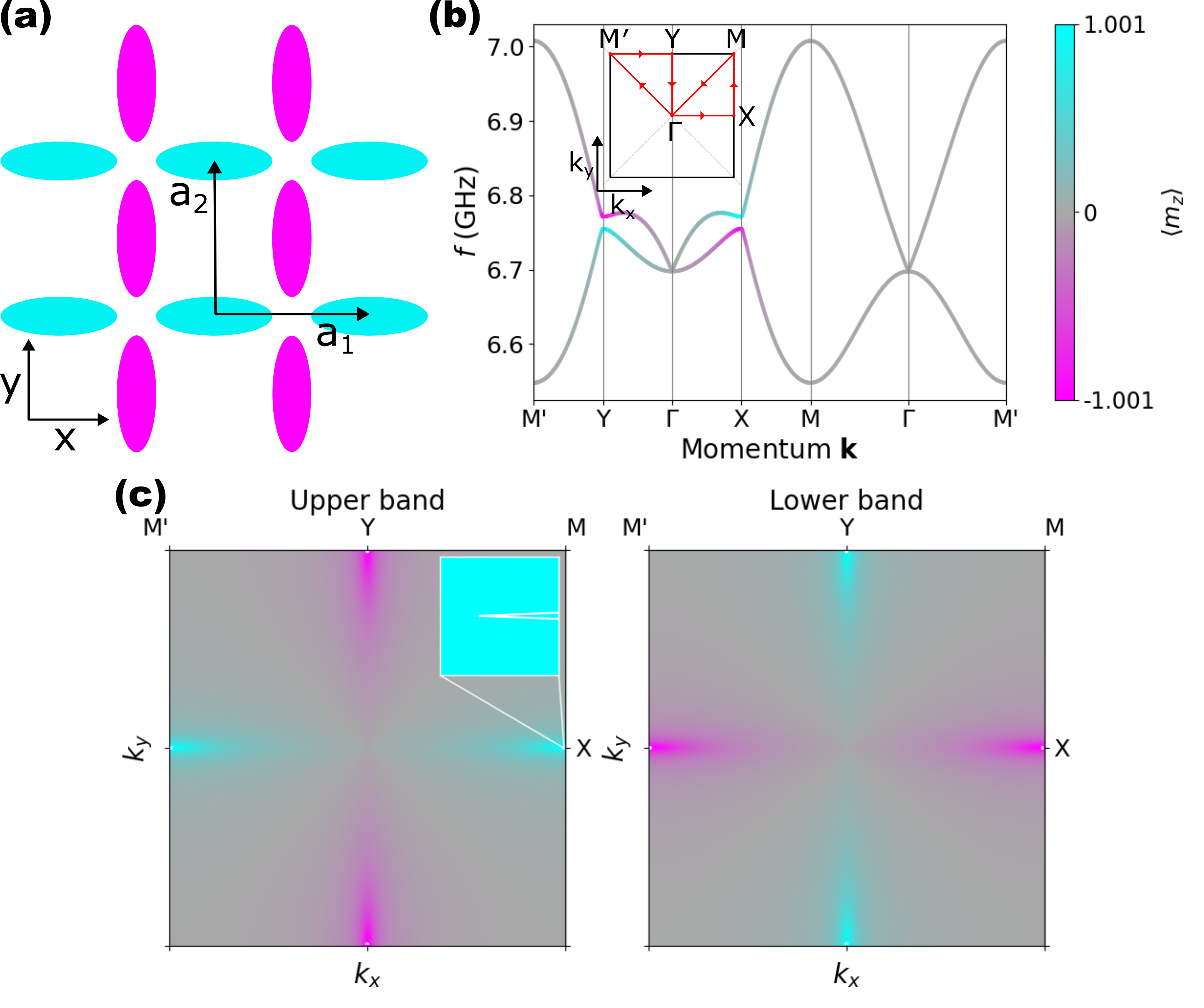}
    \caption{(a) Altermagnetic nanomagnet array of elliptical islands and lattice vectors $\bm{a}_1$ and $\bm{a}_2$. (b) Magnon dispersion along a high-symmetry path; inset shows the first BZ with dashed diagonal lines marking compensating mirror symmetries. Color represents the spin expectation value in $z$ direction. (c) Spin expectation value for upper and lower energy magnon bands projected onto a plane with contour lines of $|\langle m_z \rangle |= 1$ in white; inset shows more detail for the upper band at X. Parameters in Table~\ref{tab:parameters-1} (End Matter).}
    \label{fig:model-1}
\end{figure}

\textit{Altermagnetic geometry 1.} To generate finite $d$-wave magnon spin polarization, we break $\mathcal{P}\mathcal{T} \bm{\tau}$ and $\mathcal{T}\vec{\tau}$ symmetry by elongating the dots into ellipses [cf.~Fig.~\ref{fig:model-1}(a)]. This manipulation creates an anisotropy in real space that breaks time-reversal symmetry of the magnon band structure. The obtained checkerboard lattice belongs to the symmorphic MLG $p4'/mm'm$. The compensating symmetries are two mirrors, $\mathcal{M}_{110}$ and $\mathcal{M}_{1\bar{1}0}$, and a $\mathcal{C}_{4z}'$ symmetry. In a canonical spin-conserving altermagnet, these symmetries lead to nodal lines along the diagonals in the BZ. However, as the dipolar interaction couples spin and real space, these degeneracies are lifted.
 
Figure~\ref{fig:model-1}(b) shows the dispersion relation along a high-symmetry path in the first BZ marked in red in the inset. In addition to the splitting already present in the antiferromagnetic dot lattice (recall Fig.~\ref{fig:model-0}), the lack of nonsymmorphic symmetries (and compensating symmetries parallel to the BZ edges) lifts the X and Y point degeneracies with a splitting of $\Delta f \approx 16$~MHz. However, the sublattices remain decoupled at these points, and the dispersion acquires the characteristic alternating spin polarization along the path Y-$\Gamma$-X, which lies in-between the compensating mirror planes shown as diagonal dashed lines in the inset. Along the path M-$\Gamma$-M$'$, which lies inside these mirror planes, the spin polarization vanishes (see SM~\cite{SM}). Figure~\ref{fig:model-1}(c) shows the spin polarization of the energetically upper and lower magnon bands across the entire BZ. The band-opposite $d$-wave pattern is clearly visible. Contour lines in white at $|\langle m_z \rangle |= \pm 1$ delimit the area with above-unity magnon spin that grows to $| \langle m_z \rangle | = 1.001$ at X and Y.

\begin{figure}
    \centering
    \includegraphics[width=\linewidth]{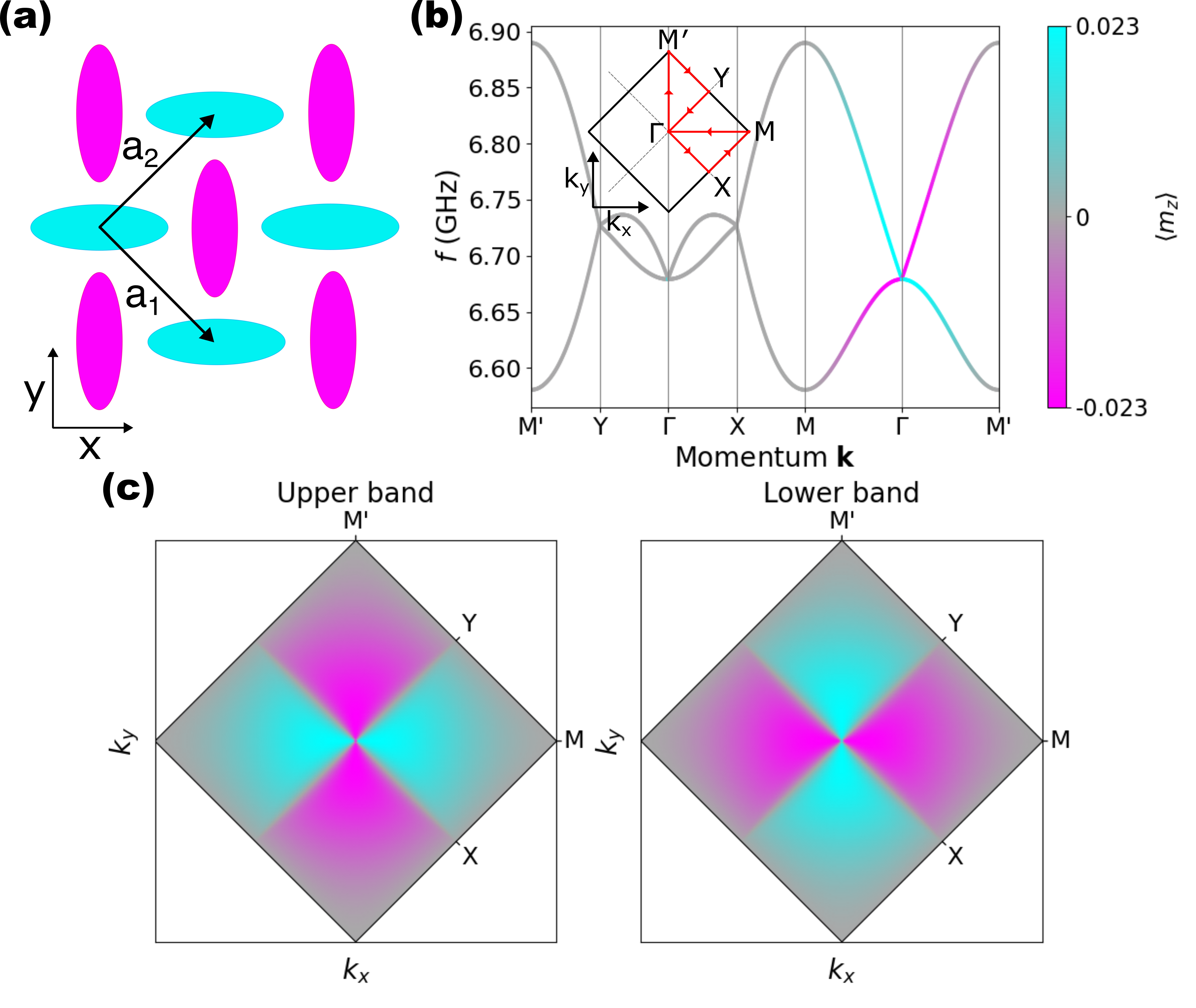}
    \caption{(a) Altermagnetic  nanomagnet array of elliptical islands with lattice vectors $\bm{a}_1=a\left(1, -1\right)$ and $\bm{a}_2=a\left(1, 1\right)$. (b) Magnon dispersion along a high-symmetry path; inset shows the first BZ with dashed diagonal lines marking compensating mirror symmetries. Color represents the spin expectation value in $z$ direction. (c) Spin expectation value for upper and lower energy magnon bands projected onto a plane. The from Table~\ref{tab:parameters-1} differing parameter reads $a=8.5 \times 10^{-8}$~m.}
    \label{fig:model-2}
\end{figure}

\textit{Altermagnetic geometry 2.} To highlight the influence of geometry and symmetry, we investigate another island configuration [cf.~Fig.~\ref{fig:model-2}(a)]. Here, the nonsymmorphic MLG is $p4'/mbm'$ and the compensating symmetries are glide-mirrors $\mathcal{M}_{100,\bm{\tau}}$ and $\mathcal{M}_{010,\bm{\tau}}$, as well as a $\mathcal{C}_{4z, \bm{\tau}}'$ symmetry with $\bm{\tau} = \left(1/2,1/2\right)$ in fractional coordinates. We choose the lattice constant $a=8.5 \times 10^{-7}$~m and define $\bm{a}_1=a\left(1, -1\right)$ and $\bm{a}_2=a\left(1, 1\right)$, keeping the rest of the parameters in Table~\ref{tab:parameters-1}. The spin polarization of the magnon bands is along the high-symmetry path M-$\Gamma$-M$'$ [see Fig.~\ref{fig:model-2}(b)], while it vanishes for the path Y-$\Gamma$-X, as shown in the SM~\cite{SM}. Although the spin polarization reaches a maximum absolute value of only $|\langle m_z \rangle |= 0.023$ it also exhibits $d$-wave character [see Fig.~\ref{fig:model-2}(c)]. Note that the spectrum is degenerate at the X and Y points as in the antiferromagnetic configuration, as discussed in the SM~\cite{SM}.

\begin{figure*}
    \centering
    \includegraphics[width=1\textwidth]{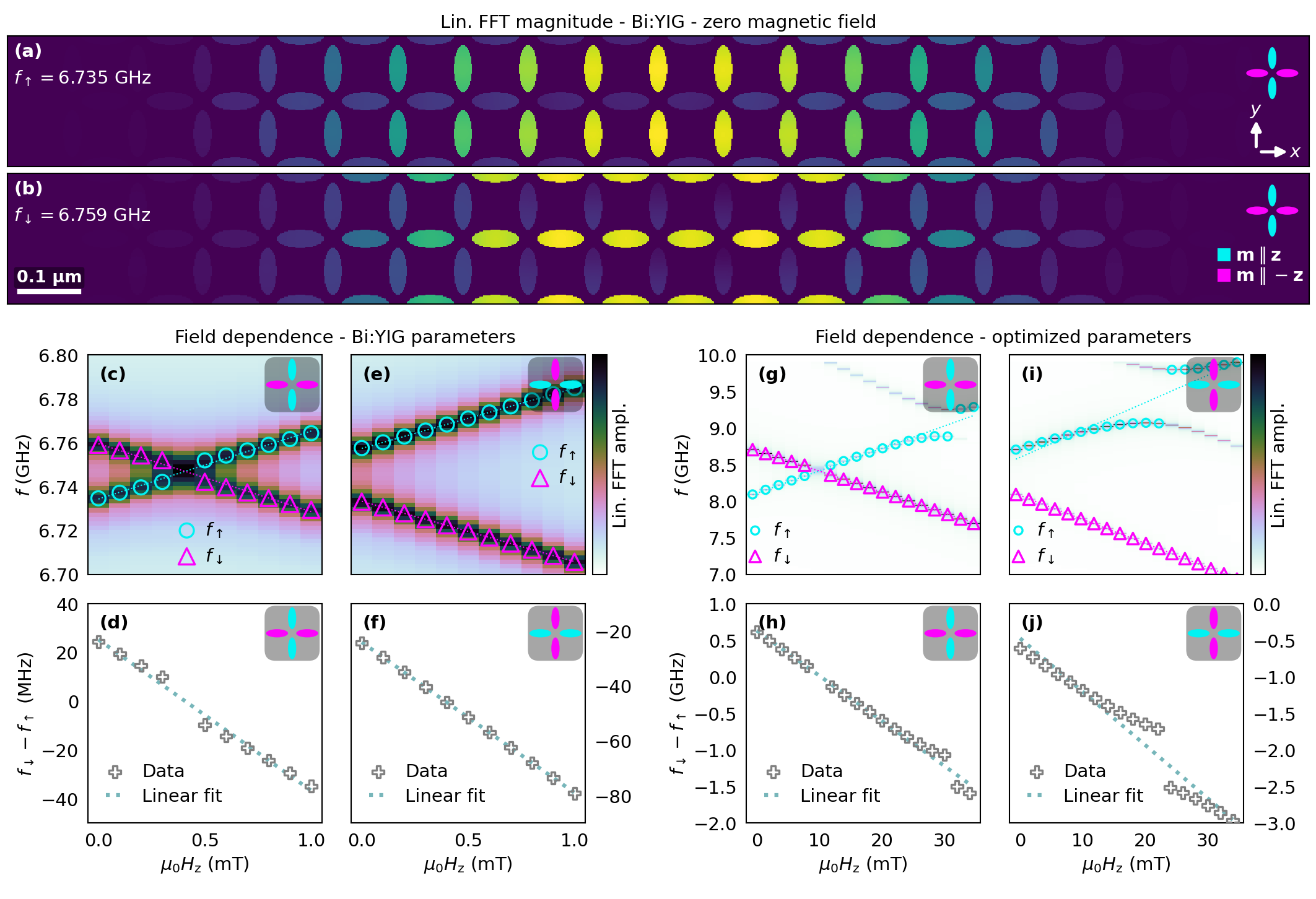}
    \caption{Micromagnetic simulations of the synthetic altermagnetic magnon spin splitting at the BZ boundary. Spatial maps of the dynamic magnetization amplitude $|m_x|$ in zero magnetic field for (a) the $f_{\uparrow} = \qty{6.735}{\giga\hertz}$ mode localized on the vertical islands and (b) the $f_{\downarrow} = \qty{6.759}{\giga\hertz}$ mode localized on the horizontal islands. (c--f) Field dependence for the scaled Bi:YIG lattice: (c, e) color-coded FFT amplitude spectra as a function of the out-of-plane field $\mu_0 H_z$ for two conjugate antiferromagnetic configurations (see insets) and (d, f) the corresponding linear evolution of the frequency splitting $f_{\downarrow} - f_{\uparrow}$ crossing zero near \qty{0.4}{\milli\tesla}. (g--j) Equivalent results for the optimized parameter set ($10\times M_\text{s}$, $20\times K$), showing a zero-field splitting of $\Delta f \approx \qty{0.6}{\giga\hertz}$. At higher fields, the fundamental modes undergo strong hybridization and distinct anti-crossings with higher-order intra-island spin-wave modes.}
    \label{fig:micromagnetics}
\end{figure*}

\textit{Numerical simulations}---To validate our analytical framework beyond the single-macrospin approximation, we perform micromagnetic simulations on altermagnetic geometry 1 (see End Matter). This numerical approach is vital to verify if the predicted altermagnetic magnon spin splitting survives in the presence of inhomogeneous demagnetizing fields, spatial profile variations within the individual nanoislands, and boundary effects.

We first examine the Bi:YIG island lattice with the parameter set in Table~\ref{tab:parameters-1}. Figures~\ref{fig:micromagnetics}(a) and (b) display the spatial maps of the dynamic magnetization amplitude at the two peak frequencies, $f_{\uparrow} = \qty{6.735}{\giga\hertz}$ and $f_{\downarrow} = \qty{6.759}{\giga\hertz}$, in zero external magnetic field. Here, $f_{\uparrow}$ and $f_{\downarrow}$ denote the frequencies of the modes residing on the sublattices with spin-up ($\mathbf{m} \parallel +z$) and spin-down ($\mathbf{m} \parallel -z$) static magnetization, respectively. 

The data presented reflects the system response to a spatial excitation with a wavevector of $k = \uppi/a$, which corresponds to the X point of the BZ where both the magnon spin polarization and the altermagnetic splitting are maximized (compare Fig.~\ref{fig:model-1}). The profiles reveal a distinct localization that reflects this sublattice mapping: the lower-frequency mode $f_{\uparrow}$ is confined to the vertically oriented islands [Fig.~\ref{fig:micromagnetics}(a)], while the higher-frequency mode $f_{\downarrow}$ is localized on the horizontal islands [Fig.~\ref{fig:micromagnetics}(b)]. This spatial separation of the eigenmodes at the BZ boundary perfectly mirrors the absolute magnon spin polarization predicted by our analytical framework and yields a finite, purely dipolar-driven zero-field splitting of $\Delta f = f_{\downarrow} - f_{\uparrow} \approx \qty{24}{\mega\hertz}$. This splitting slightly exceeds the macrospin model’s predicted value of $ \Delta f \approx 16$~MHz.

Figures~\ref{fig:micromagnetics}(c)--(f) show the evolution of these modes under an out-of-plane magnetic field $\mu_0 H_z$ for two conjugate antiferromagnetic ground states. As seen in Fig.~\ref{fig:micromagnetics}(c), the splitting decreases linearly with the field and crosses zero near $\mu_0 H_z \approx \qty{0.4}{\milli\tesla}$ [Fig.~\ref{fig:micromagnetics}(d)]. As the coupling between these pure, sublattice-polarized fundamental altermagnetic modes with wavevector $k = \uppi/a$ is weak, their mutual mode repulsion is suppressed. Consequently, the field-induced splitting follows an almost perfectly linear trajectory, with a small deviation from linearity directly at the crossing point due to residual dipolar cross-talk between the orthogonal sublattices. We observe a frequency shift of $|\Delta f| \approx 31\,\text{MHz}$ in a field of $\Delta \mu_0 H_z = 1\,\text{mT}$ for both modes, corresponding to a magnon spin polarization of $|\langle m_z \rangle| = h(\Delta f/ \Delta \mu_0 H_z)/(2 \mu_\text{B}) \approx 1.11$.

Reversing the sublattice magnetization flips the sign of the initial splitting, which then decreases monotonically with the applied field [Fig.~\ref{fig:micromagnetics}(e),(f)]. In this time-reversed configuration the spin polarization is reduced to $|\langle m_z \rangle| \approx 0.98$ for both modes. 

Correcting for finite-size effects by using a square simulation domain (see SM~\cite{SM}), 
we find excellent agreement with the theoretical model, as the absolute spin polarization approaches $|\langle m_z \rangle| \approx 1.00$ in all four cases and the weak anti-crossing from Fig.~\ref{fig:micromagnetics}(d) vanishes.

To demonstrate that the altermagnetic splitting can be scaled significantly, we investigate an optimized parameter set, increasing $M_\text{s}$ by a factor of 10 and $K$ by a factor of 20. This increased $M_\text{s}$ value ($10 \times M_\text{s,~Bi:YIG} = \qty{1000}{\kilo\ampere\per\meter}$) is about half as large as values reported for low-damping metallic alloys such as Co$_{25}$Fe$_{75}$ ($M_\text{s,~CoFe} \approx \qty{1958}{\kilo\ampere\per\meter}$) \cite{lee_metallic_2017}. While $M_\text{s}$ directly scales the strength of the dipolar interaction, the simultaneous increase in $K$ is required to preserve the perpendicular easy axis against the destabilizing in-plane shape anisotropy of the elongated islands. In general, we find that a strong uniaxial anisotropy tends to suppress the altermagnetic splitting, which is maximized when the anisotropy is tuned just above the threshold necessary to guarantee a stable out-of-plane configuration (see SM~\cite{SM}, Fig.~S1). We note that our chosen factor of 20 exceeds this minimum requirement; however, it spectrally isolates the fundamental altermagnetic modes from higher-order intra-island spin-wave modes (see SM~\cite{SM}, Fig.~S1). As shown in Figs.~\ref{fig:micromagnetics}(g)--(j), this configuration enhances the zero-field splitting to $\Delta f \approx \qty{0.6}{\giga\hertz}$ [see Fig.~\ref{fig:micromagnetics}(h),(j)]. Notably, at higher fields ($\mu_0 H_z > \qty{20}{\milli\tesla}$), we observe a pronounced avoided crossing behavior [see Fig.~\ref{fig:micromagnetics}(g),(i)]. While the fundamental $k = \uppi/a$ modes do not couple with each other, they experience strong hybridization with higher-order intra-island spin-wave modes possessing one or more nodes within a single nanomagnet, which yields these prominent anti-crossings. We note that the magnon spin polarization of the fundamental modes is extracted to $|\langle m_z\rangle | \approx 1.10$ as average value from Fig.~\ref{fig:micromagnetics}(g) and to $|\langle m_z\rangle | \approx 1.30$ as average value from Fig.~\ref{fig:micromagnetics}(i).

\textit{Discussion and Conclusion}---We have shown that engineering a checkerboard array of elliptical ferromagnetic nanoislands with staggered out-of-plane magnetization enables the realization of an altermagnetic magnonic crystal, in which dipolar interactions generate an alternating $d$-wave spin polarization of the magnon bands. The altermagnetic magnon spin polarization predicted within the single-macrospin approximation is confirmed by micromagnetic simulations, demonstrating the robustness of this phenomenon beyond the idealized analytical model. Correcting for finite-size aspect-ratio effects by employing square simulation domains yields quantitative agreement with the theoretical model ($|\langle m_z \rangle| \approx 1.00$, see SM~\cite{SM}, Fig.~S2), verifying that minor polarization asymmetries under time inversion stem entirely from domain geometry of the the $x$-elongated simulation window rather than intrinsic physical symmetry breaking.

While bringing this synthetic altermagnetic platform to experimental realization demands precise nanofabrication, our micromagnetic simulations directly outline viable engineering strategies. For Bi:YIG, resolving the zero-field splitting of $\sim \qty{24}{\mega\hertz}$ requires high array homogeneity to limit inhomogeneous broadening. Promisingly, our anisotropy-tuning analysis reveals a clear optimization route: reducing the perpendicular magnetic anisotropy $K$ toward the out-of-plane stability threshold enhances dipolar dynamic coupling, increasing the splitting in Bi:YIG to more than \qty{150}{\mega\hertz} (see SM~\cite{SM}, Fig.~S1). Continuous, post-growth tuning of perpendicular magnetic anisotropy in Bi:YIG thin films without affecting the saturation magnetization has already been demonstrated experimentally using $\text{He}^+$ ion irradiation~\cite{PhysRevMaterials.8.114419}. Furthermore, while out-of-plane spin-wave dispersion in real-world samples introduces low-frequency surface and edge modes, transitioning to 3D architectures with high vertical aspect ratios ($h \gg h_1$) provides a compelling design strategy: increasing the nanomagnet height directly tunes the effective shape anisotropy to stabilize out-of-plane magnetization without relying solely on intrinsic material anisotropy, while simultaneously increasing the total magnetic volume to significantly amplify the dynamic dipolar coupling and the resulting altermagnetic splitting.

Low-damping $\text{YIG}$ platforms with laser-assisted anisotropy engineering offer a compelling outlook. Direct-write techniques, such as focused UV laser amorphization to pattern isolated magnetic pillars~\cite{florio_programmable_2026} and continuous-wave UV laser irradiation to induce localized perpendicular magnetic anisotropy (PMA) enhancement in single-crystal $\text{YIG}$~\cite{levati_three-dimensional_2025} provide single-step nanofabrication pathways with nanoscale precision. Harnessing these advanced structuring tools to tailor dynamic stray fields establishes a realistic and versatile foundation for functional synthetic altermagnetic metamaterials based on our predictions.

\begin{acknowledgments}
\textit{Acknowledgments}---R.H.~thanks Rodrigo Jaeschke-Ubiergo for insightful discussions about symmetries.
This work was funded by the German Research Foundation (DFG) through TRR 173-268565370 (Project B13); Project 581372738 (Priority Programme 2558); Project 581342865 (Priority Programme 2558); and Project No.~504261060 (Emmy Noether Programme). L.K.~gratefully acknowledges funding by the Radboud Excellence Initiative. 

\textit{Data availability}---The data that support the findings of this article are openly available. 
\end{acknowledgments}

\bibliography{bib}

\begin{thebibliography}{60}%
\makeatletter
\providecommand \@ifxundefined [1]{%
 \@ifx{#1\undefined}
}%
\providecommand \@ifnum [1]{%
 \ifnum #1\expandafter \@firstoftwo
 \else \expandafter \@secondoftwo
 \fi
}%
\providecommand \@ifx [1]{%
 \ifx #1\expandafter \@firstoftwo
 \else \expandafter \@secondoftwo
 \fi
}%
\providecommand \natexlab [1]{#1}%
\providecommand \enquote  [1]{``#1''}%
\providecommand \bibnamefont  [1]{#1}%
\providecommand \bibfnamefont [1]{#1}%
\providecommand \citenamefont [1]{#1}%
\providecommand \href@noop [0]{\@secondoftwo}%
\providecommand \href [0]{\begingroup \@sanitize@url \@href}%
\providecommand \@href[1]{\@@startlink{#1}\@@href}%
\providecommand \@@href[1]{\endgroup#1\@@endlink}%
\providecommand \@sanitize@url [0]{\catcode `\\12\catcode `\$12\catcode
  `\&12\catcode `\#12\catcode `\^12\catcode `\_12\catcode `\%12\relax}%
\providecommand \@@startlink[1]{}%
\providecommand \@@endlink[0]{}%
\providecommand \url  [0]{\begingroup\@sanitize@url \@url }%
\providecommand \@url [1]{\endgroup\@href {#1}{\urlprefix }}%
\providecommand \urlprefix  [0]{URL }%
\providecommand \Eprint [0]{\href }%
\providecommand \doibase [0]{https://doi.org/}%
\providecommand \selectlanguage [0]{\@gobble}%
\providecommand \bibinfo  [0]{\@secondoftwo}%
\providecommand \bibfield  [0]{\@secondoftwo}%
\providecommand \translation [1]{[#1]}%
\providecommand \BibitemOpen [0]{}%
\providecommand \bibitemStop [0]{}%
\providecommand \bibitemNoStop [0]{.\EOS\space}%
\providecommand \EOS [0]{\spacefactor3000\relax}%
\providecommand \BibitemShut  [1]{\csname bibitem#1\endcsname}%
\let\auto@bib@innerbib\@empty
\bibitem [{\citenamefont {\ifmmode~\check{S}\else \v{S}\fi{}mejkal}\ \emph
  {et~al.}(2022{\natexlab{a}})\citenamefont {\ifmmode~\check{S}\else
  \v{S}\fi{}mejkal}, \citenamefont {Sinova},\ and\ \citenamefont
  {Jungwirth}}]{Smejkal2022Beyond}%
  \BibitemOpen
  \bibfield  {author} {\bibinfo {author} {\bibfnamefont {L.}~\bibnamefont
  {\ifmmode~\check{S}\else \v{S}\fi{}mejkal}}, \bibinfo {author} {\bibfnamefont
  {J.}~\bibnamefont {Sinova}},\ and\ \bibinfo {author} {\bibfnamefont
  {T.}~\bibnamefont {Jungwirth}},\ }\bibfield  {title} {\bibinfo {title}
  {Beyond conventional ferromagnetism and antiferromagnetism: A phase with
  nonrelativistic spin and crystal rotation symmetry},\ }\href
  {https://doi.org/10.1103/PhysRevX.12.031042} {\bibfield  {journal} {\bibinfo
  {journal} {Phys. Rev. X}\ }\textbf {\bibinfo {volume} {12}},\ \bibinfo
  {pages} {031042} (\bibinfo {year} {2022}{\natexlab{a}})}\BibitemShut
  {NoStop}%
\bibitem [{\citenamefont {\ifmmode~\check{S}\else \v{S}\fi{}mejkal}\ \emph
  {et~al.}(2022{\natexlab{b}})\citenamefont {\ifmmode~\check{S}\else
  \v{S}\fi{}mejkal}, \citenamefont {Sinova},\ and\ \citenamefont
  {Jungwirth}}]{Smejkal2022Emerging}%
  \BibitemOpen
  \bibfield  {author} {\bibinfo {author} {\bibfnamefont {L.}~\bibnamefont
  {\ifmmode~\check{S}\else \v{S}\fi{}mejkal}}, \bibinfo {author} {\bibfnamefont
  {J.}~\bibnamefont {Sinova}},\ and\ \bibinfo {author} {\bibfnamefont
  {T.}~\bibnamefont {Jungwirth}},\ }\bibfield  {title} {\bibinfo {title}
  {Emerging research landscape of altermagnetism},\ }\href
  {https://doi.org/10.1103/PhysRevX.12.040501} {\bibfield  {journal} {\bibinfo
  {journal} {Phys. Rev. X}\ }\textbf {\bibinfo {volume} {12}},\ \bibinfo
  {pages} {040501} (\bibinfo {year} {2022}{\natexlab{b}})}\BibitemShut
  {NoStop}%
\bibitem [{\citenamefont {Bai}\ \emph {et~al.}(2024)\citenamefont {Bai},
  \citenamefont {Feng}, \citenamefont {Liu}, \citenamefont {Šmejkal},
  \citenamefont {Mokrousov},\ and\ \citenamefont {Yao}}]{Bai2024}%
  \BibitemOpen
  \bibfield  {author} {\bibinfo {author} {\bibfnamefont {L.}~\bibnamefont
  {Bai}}, \bibinfo {author} {\bibfnamefont {W.}~\bibnamefont {Feng}}, \bibinfo
  {author} {\bibfnamefont {S.}~\bibnamefont {Liu}}, \bibinfo {author}
  {\bibfnamefont {L.}~\bibnamefont {Šmejkal}}, \bibinfo {author}
  {\bibfnamefont {Y.}~\bibnamefont {Mokrousov}},\ and\ \bibinfo {author}
  {\bibfnamefont {Y.}~\bibnamefont {Yao}},\ }\bibfield  {title} {\bibinfo
  {title} {Altermagnetism: Exploring new frontiers in magnetism and
  spintronics},\ }\href
  {https://doi.org/https://doi.org/10.1002/adfm.202409327} {\bibfield
  {journal} {\bibinfo  {journal} {Advanced Functional Materials}\ }\textbf
  {\bibinfo {volume} {34}},\ \bibinfo {pages} {2409327} (\bibinfo {year}
  {2024})}\BibitemShut {NoStop}%
\bibitem [{\citenamefont {Jungwirth}\ \emph {et~al.}(2026)\citenamefont
  {Jungwirth}, \citenamefont {Sinova}, \citenamefont {Wadley}, \citenamefont
  {Kriegner}, \citenamefont {Reichlov{\'a}}, \citenamefont {Krizek},
  \citenamefont {Ohno},\ and\ \citenamefont {{\v S}mejkal}}]{Jungwirth2025}%
  \BibitemOpen
  \bibfield  {author} {\bibinfo {author} {\bibfnamefont {T.}~\bibnamefont
  {Jungwirth}}, \bibinfo {author} {\bibfnamefont {J.}~\bibnamefont {Sinova}},
  \bibinfo {author} {\bibfnamefont {P.}~\bibnamefont {Wadley}}, \bibinfo
  {author} {\bibfnamefont {D.}~\bibnamefont {Kriegner}}, \bibinfo {author}
  {\bibfnamefont {H.}~\bibnamefont {Reichlov{\'a}}}, \bibinfo {author}
  {\bibfnamefont {F.}~\bibnamefont {Krizek}}, \bibinfo {author} {\bibfnamefont
  {H.}~\bibnamefont {Ohno}},\ and\ \bibinfo {author} {\bibfnamefont
  {L.}~\bibnamefont {{\v S}mejkal}},\ }\bibfield  {title} {\bibinfo {title}
  {Altermagnetic spintronics},\ }\href
  {https://doi.org/10.1038/s41567-026-03337-w} {\bibfield  {journal} {\bibinfo
  {journal} {Nature Physics}\ }\textbf {\bibinfo {volume} {22}},\ \bibinfo
  {pages} {1012} (\bibinfo {year} {2026})}\BibitemShut {NoStop}%
\bibitem [{\citenamefont {Song}\ \emph {et~al.}(2025)\citenamefont {Song},
  \citenamefont {Bai}, \citenamefont {Zhou}, \citenamefont {Han}, \citenamefont
  {Reichlova}, \citenamefont {Dil}, \citenamefont {Liu}, \citenamefont {Chen},\
  and\ \citenamefont {Pan}}]{Song2025}%
  \BibitemOpen
  \bibfield  {author} {\bibinfo {author} {\bibfnamefont {C.}~\bibnamefont
  {Song}}, \bibinfo {author} {\bibfnamefont {H.}~\bibnamefont {Bai}}, \bibinfo
  {author} {\bibfnamefont {Z.}~\bibnamefont {Zhou}}, \bibinfo {author}
  {\bibfnamefont {L.}~\bibnamefont {Han}}, \bibinfo {author} {\bibfnamefont
  {H.}~\bibnamefont {Reichlova}}, \bibinfo {author} {\bibfnamefont {J.~H.}\
  \bibnamefont {Dil}}, \bibinfo {author} {\bibfnamefont {J.}~\bibnamefont
  {Liu}}, \bibinfo {author} {\bibfnamefont {X.}~\bibnamefont {Chen}},\ and\
  \bibinfo {author} {\bibfnamefont {F.}~\bibnamefont {Pan}},\ }\bibfield
  {title} {\bibinfo {title} {Altermagnets as a new class of functional
  materials},\ }\href {https://doi.org/10.1038/s41578-025-00779-1} {\bibfield
  {journal} {\bibinfo  {journal} {Nature Reviews Materials}\ }\textbf {\bibinfo
  {volume} {10}},\ \bibinfo {pages} {473} (\bibinfo {year} {2025})}\BibitemShut
  {NoStop}%
\bibitem [{\citenamefont {Chen}\ \emph {et~al.}(2024)\citenamefont {Chen},
  \citenamefont {Ren}, \citenamefont {Zhu}, \citenamefont {Yu}, \citenamefont
  {Zhang}, \citenamefont {Liu}, \citenamefont {Li}, \citenamefont {Liu},
  \citenamefont {Li},\ and\ \citenamefont {Liu}}]{Chen2024}%
  \BibitemOpen
  \bibfield  {author} {\bibinfo {author} {\bibfnamefont {X.}~\bibnamefont
  {Chen}}, \bibinfo {author} {\bibfnamefont {J.}~\bibnamefont {Ren}}, \bibinfo
  {author} {\bibfnamefont {Y.}~\bibnamefont {Zhu}}, \bibinfo {author}
  {\bibfnamefont {Y.}~\bibnamefont {Yu}}, \bibinfo {author} {\bibfnamefont
  {A.}~\bibnamefont {Zhang}}, \bibinfo {author} {\bibfnamefont
  {P.}~\bibnamefont {Liu}}, \bibinfo {author} {\bibfnamefont {J.}~\bibnamefont
  {Li}}, \bibinfo {author} {\bibfnamefont {Y.}~\bibnamefont {Liu}}, \bibinfo
  {author} {\bibfnamefont {C.}~\bibnamefont {Li}},\ and\ \bibinfo {author}
  {\bibfnamefont {Q.}~\bibnamefont {Liu}},\ }\bibfield  {title} {\bibinfo
  {title} {Enumeration and representation theory of spin space groups},\ }\href
  {https://doi.org/10.1103/PhysRevX.14.031038} {\bibfield  {journal} {\bibinfo
  {journal} {Phys. Rev. X}\ }\textbf {\bibinfo {volume} {14}},\ \bibinfo
  {pages} {031038} (\bibinfo {year} {2024})}\BibitemShut {NoStop}%
\bibitem [{\citenamefont {Jiang}\ \emph {et~al.}(2024)\citenamefont {Jiang},
  \citenamefont {Song}, \citenamefont {Zhu}, \citenamefont {Fang},
  \citenamefont {Weng}, \citenamefont {Liu}, \citenamefont {Yang},\ and\
  \citenamefont {Fang}}]{Jiang2024}%
  \BibitemOpen
  \bibfield  {author} {\bibinfo {author} {\bibfnamefont {Y.}~\bibnamefont
  {Jiang}}, \bibinfo {author} {\bibfnamefont {Z.}~\bibnamefont {Song}},
  \bibinfo {author} {\bibfnamefont {T.}~\bibnamefont {Zhu}}, \bibinfo {author}
  {\bibfnamefont {Z.}~\bibnamefont {Fang}}, \bibinfo {author} {\bibfnamefont
  {H.}~\bibnamefont {Weng}}, \bibinfo {author} {\bibfnamefont {Z.-X.}\
  \bibnamefont {Liu}}, \bibinfo {author} {\bibfnamefont {J.}~\bibnamefont
  {Yang}},\ and\ \bibinfo {author} {\bibfnamefont {C.}~\bibnamefont {Fang}},\
  }\bibfield  {title} {\bibinfo {title} {Enumeration of spin-space groups:
  Toward a complete description of symmetries of magnetic orders},\ }\href
  {https://doi.org/10.1103/PhysRevX.14.031039} {\bibfield  {journal} {\bibinfo
  {journal} {Phys. Rev. X}\ }\textbf {\bibinfo {volume} {14}},\ \bibinfo
  {pages} {031039} (\bibinfo {year} {2024})}\BibitemShut {NoStop}%
\bibitem [{\citenamefont {Liu}\ \emph {et~al.}(2024)\citenamefont {Liu},
  \citenamefont {Ozeki}, \citenamefont {Asai}, \citenamefont {Itoh},\ and\
  \citenamefont {Masuda}}]{Liu2024}%
  \BibitemOpen
  \bibfield  {author} {\bibinfo {author} {\bibfnamefont {Z.}~\bibnamefont
  {Liu}}, \bibinfo {author} {\bibfnamefont {M.}~\bibnamefont {Ozeki}}, \bibinfo
  {author} {\bibfnamefont {S.}~\bibnamefont {Asai}}, \bibinfo {author}
  {\bibfnamefont {S.}~\bibnamefont {Itoh}},\ and\ \bibinfo {author}
  {\bibfnamefont {T.}~\bibnamefont {Masuda}},\ }\bibfield  {title} {\bibinfo
  {title} {Chiral split magnon in altermagnetic {M}n{T}e},\ }\href
  {https://doi.org/10.1103/PhysRevLett.133.156702} {\bibfield  {journal}
  {\bibinfo  {journal} {Phys. Rev. Lett.}\ }\textbf {\bibinfo {volume} {133}},\
  \bibinfo {pages} {156702} (\bibinfo {year} {2024})}\BibitemShut {NoStop}%
\bibitem [{\citenamefont {Xiao}\ \emph {et~al.}(2024)\citenamefont {Xiao},
  \citenamefont {Zhao}, \citenamefont {Li}, \citenamefont {Shindou},\ and\
  \citenamefont {Song}}]{Xiao2024}%
  \BibitemOpen
  \bibfield  {author} {\bibinfo {author} {\bibfnamefont {Z.}~\bibnamefont
  {Xiao}}, \bibinfo {author} {\bibfnamefont {J.}~\bibnamefont {Zhao}}, \bibinfo
  {author} {\bibfnamefont {Y.}~\bibnamefont {Li}}, \bibinfo {author}
  {\bibfnamefont {R.}~\bibnamefont {Shindou}},\ and\ \bibinfo {author}
  {\bibfnamefont {Z.-D.}\ \bibnamefont {Song}},\ }\bibfield  {title} {\bibinfo
  {title} {Spin space groups: Full classification and applications},\ }\href
  {https://doi.org/10.1103/PhysRevX.14.031037} {\bibfield  {journal} {\bibinfo
  {journal} {Phys. Rev. X}\ }\textbf {\bibinfo {volume} {14}},\ \bibinfo
  {pages} {031037} (\bibinfo {year} {2024})}\BibitemShut {NoStop}%
\bibitem [{\citenamefont {Sufyan}\ \emph {et~al.}(2026)\citenamefont {Sufyan},
  \citenamefont {Marfoua}, \citenamefont {Larsson}, \citenamefont {van Loon},\
  and\ \citenamefont {Armiento}}]{Sufyan2026}%
  \BibitemOpen
  \bibfield  {author} {\bibinfo {author} {\bibfnamefont {A.}~\bibnamefont
  {Sufyan}}, \bibinfo {author} {\bibfnamefont {B.}~\bibnamefont {Marfoua}},
  \bibinfo {author} {\bibfnamefont {J.~A.}\ \bibnamefont {Larsson}}, \bibinfo
  {author} {\bibfnamefont {E.}~\bibnamefont {van Loon}},\ and\ \bibinfo
  {author} {\bibfnamefont {R.}~\bibnamefont {Armiento}},\ }\bibfield  {title}
  {\bibinfo {title} {High-throughput quantification of altermagnetic band
  splitting},\ }\href {https://doi.org/10.1103/mmdm-hrj4} {\bibfield  {journal}
  {\bibinfo  {journal} {Phys. Rev. Mater.}\ }\textbf {\bibinfo {volume} {10}},\
  \bibinfo {pages} {044407} (\bibinfo {year} {2026})}\BibitemShut {NoStop}%
\bibitem [{\citenamefont {Krempask{\'y}}\ \emph {et~al.}(2024)\citenamefont
  {Krempask{\'y}}, \citenamefont {{\v S}mejkal}, \citenamefont {D'Souza},
  \citenamefont {Hajlaoui}, \citenamefont {Springholz}, \citenamefont
  {Uhl{\'i}{\v r}ov{\'a}}, \citenamefont {Alarab}, \citenamefont
  {Constantinou}, \citenamefont {Strocov}, \citenamefont {Usanov},
  \citenamefont {Pudelko}, \citenamefont {{Gonz{\'a}lez-Hern{\'a}ndez}},
  \citenamefont {Birk~Hellenes}, \citenamefont {Jansa}, \citenamefont
  {Reichlov{\'a}}, \citenamefont {{\v S}ob{\'a}{\v n}}, \citenamefont
  {Gonzalez~Betancourt}, \citenamefont {Wadley}, \citenamefont {Sinova},
  \citenamefont {Kriegner}, \citenamefont {Min{\'a}r}, \citenamefont {Dil},\
  and\ \citenamefont {Jungwirth}}]{Krempasky2024}%
  \BibitemOpen
  \bibfield  {author} {\bibinfo {author} {\bibfnamefont {J.}~\bibnamefont
  {Krempask{\'y}}}, \bibinfo {author} {\bibfnamefont {L.}~\bibnamefont {{\v
  S}mejkal}}, \bibinfo {author} {\bibfnamefont {S.~W.}\ \bibnamefont
  {D'Souza}}, \bibinfo {author} {\bibfnamefont {M.}~\bibnamefont {Hajlaoui}},
  \bibinfo {author} {\bibfnamefont {G.}~\bibnamefont {Springholz}}, \bibinfo
  {author} {\bibfnamefont {K.}~\bibnamefont {Uhl{\'i}{\v r}ov{\'a}}}, \bibinfo
  {author} {\bibfnamefont {F.}~\bibnamefont {Alarab}}, \bibinfo {author}
  {\bibfnamefont {P.~C.}\ \bibnamefont {Constantinou}}, \bibinfo {author}
  {\bibfnamefont {V.}~\bibnamefont {Strocov}}, \bibinfo {author} {\bibfnamefont
  {D.}~\bibnamefont {Usanov}}, \bibinfo {author} {\bibfnamefont {W.~R.}\
  \bibnamefont {Pudelko}}, \bibinfo {author} {\bibfnamefont {R.}~\bibnamefont
  {{Gonz{\'a}lez-Hern{\'a}ndez}}}, \bibinfo {author} {\bibfnamefont
  {A.}~\bibnamefont {Birk~Hellenes}}, \bibinfo {author} {\bibfnamefont
  {Z.}~\bibnamefont {Jansa}}, \bibinfo {author} {\bibfnamefont
  {H.}~\bibnamefont {Reichlov{\'a}}}, \bibinfo {author} {\bibfnamefont
  {Z.}~\bibnamefont {{\v S}ob{\'a}{\v n}}}, \bibinfo {author} {\bibfnamefont
  {R.~D.}\ \bibnamefont {Gonzalez~Betancourt}}, \bibinfo {author}
  {\bibfnamefont {P.}~\bibnamefont {Wadley}}, \bibinfo {author} {\bibfnamefont
  {J.}~\bibnamefont {Sinova}}, \bibinfo {author} {\bibfnamefont
  {D.}~\bibnamefont {Kriegner}}, \bibinfo {author} {\bibfnamefont
  {J.}~\bibnamefont {Min{\'a}r}}, \bibinfo {author} {\bibfnamefont {J.~H.}\
  \bibnamefont {Dil}},\ and\ \bibinfo {author} {\bibfnamefont {T.}~\bibnamefont
  {Jungwirth}},\ }\bibfield  {title} {\bibinfo {title} {Altermagnetic lifting
  of {{Kramers}} spin degeneracy},\ }\href
  {https://doi.org/10.1038/s41586-023-06907-7} {\bibfield  {journal} {\bibinfo
  {journal} {Nature}\ }\textbf {\bibinfo {volume} {626}},\ \bibinfo {pages}
  {517} (\bibinfo {year} {2024})}\BibitemShut {NoStop}%
\bibitem [{\citenamefont {Lee}\ \emph {et~al.}(2024)\citenamefont {Lee},
  \citenamefont {Lee}, \citenamefont {Jung}, \citenamefont {Jung},
  \citenamefont {Kim}, \citenamefont {Lee}, \citenamefont {Seok}, \citenamefont
  {Kim}, \citenamefont {Park}, \citenamefont {\ifmmode~\check{S}\else
  \v{S}\fi{}mejkal}, \citenamefont {Kang},\ and\ \citenamefont
  {Kim}}]{Lee2024}%
  \BibitemOpen
  \bibfield  {author} {\bibinfo {author} {\bibfnamefont {S.}~\bibnamefont
  {Lee}}, \bibinfo {author} {\bibfnamefont {S.}~\bibnamefont {Lee}}, \bibinfo
  {author} {\bibfnamefont {S.}~\bibnamefont {Jung}}, \bibinfo {author}
  {\bibfnamefont {J.}~\bibnamefont {Jung}}, \bibinfo {author} {\bibfnamefont
  {D.}~\bibnamefont {Kim}}, \bibinfo {author} {\bibfnamefont {Y.}~\bibnamefont
  {Lee}}, \bibinfo {author} {\bibfnamefont {B.}~\bibnamefont {Seok}}, \bibinfo
  {author} {\bibfnamefont {J.}~\bibnamefont {Kim}}, \bibinfo {author}
  {\bibfnamefont {B.~G.}\ \bibnamefont {Park}}, \bibinfo {author}
  {\bibfnamefont {L.}~\bibnamefont {\ifmmode~\check{S}\else \v{S}\fi{}mejkal}},
  \bibinfo {author} {\bibfnamefont {C.-J.}\ \bibnamefont {Kang}},\ and\
  \bibinfo {author} {\bibfnamefont {C.}~\bibnamefont {Kim}},\ }\bibfield
  {title} {\bibinfo {title} {Broken {K}ramers degeneracy in altermagnetic
  {M}n{T}e},\ }\href {https://doi.org/10.1103/PhysRevLett.132.036702}
  {\bibfield  {journal} {\bibinfo  {journal} {Phys. Rev. Lett.}\ }\textbf
  {\bibinfo {volume} {132}},\ \bibinfo {pages} {036702} (\bibinfo {year}
  {2024})}\BibitemShut {NoStop}%
\bibitem [{\citenamefont {Reimers}\ \emph {et~al.}(2024)\citenamefont
  {Reimers}, \citenamefont {Odenbreit}, \citenamefont {{\v S}mejkal},
  \citenamefont {Strocov}, \citenamefont {Constantinou}, \citenamefont
  {Hellenes}, \citenamefont {Jaeschke~Ubiergo}, \citenamefont {Campos},
  \citenamefont {Bharadwaj}, \citenamefont {Chakraborty}, \citenamefont
  {Denneulin}, \citenamefont {Shi}, \citenamefont {Dunin-Borkowski},
  \citenamefont {Das}, \citenamefont {Kl{\"a}ui}, \citenamefont {Sinova},\ and\
  \citenamefont {Jourdan}}]{Reimers2024}%
  \BibitemOpen
  \bibfield  {author} {\bibinfo {author} {\bibfnamefont {S.}~\bibnamefont
  {Reimers}}, \bibinfo {author} {\bibfnamefont {L.}~\bibnamefont {Odenbreit}},
  \bibinfo {author} {\bibfnamefont {L.}~\bibnamefont {{\v S}mejkal}}, \bibinfo
  {author} {\bibfnamefont {V.~N.}\ \bibnamefont {Strocov}}, \bibinfo {author}
  {\bibfnamefont {P.}~\bibnamefont {Constantinou}}, \bibinfo {author}
  {\bibfnamefont {A.~B.}\ \bibnamefont {Hellenes}}, \bibinfo {author}
  {\bibfnamefont {R.}~\bibnamefont {Jaeschke~Ubiergo}}, \bibinfo {author}
  {\bibfnamefont {W.~H.}\ \bibnamefont {Campos}}, \bibinfo {author}
  {\bibfnamefont {V.~K.}\ \bibnamefont {Bharadwaj}}, \bibinfo {author}
  {\bibfnamefont {A.}~\bibnamefont {Chakraborty}}, \bibinfo {author}
  {\bibfnamefont {T.}~\bibnamefont {Denneulin}}, \bibinfo {author}
  {\bibfnamefont {W.}~\bibnamefont {Shi}}, \bibinfo {author} {\bibfnamefont
  {R.~E.}\ \bibnamefont {Dunin-Borkowski}}, \bibinfo {author} {\bibfnamefont
  {S.}~\bibnamefont {Das}}, \bibinfo {author} {\bibfnamefont {M.}~\bibnamefont
  {Kl{\"a}ui}}, \bibinfo {author} {\bibfnamefont {J.}~\bibnamefont {Sinova}},\
  and\ \bibinfo {author} {\bibfnamefont {M.}~\bibnamefont {Jourdan}},\
  }\bibfield  {title} {\bibinfo {title} {Direct observation of altermagnetic
  band splitting in {C}r{S}b thin films},\ }\href
  {https://doi.org/10.1038/s41467-024-46476-5} {\bibfield  {journal} {\bibinfo
  {journal} {Nature Communications}\ }\textbf {\bibinfo {volume} {15}},\
  \bibinfo {pages} {2116} (\bibinfo {year} {2024})}\BibitemShut {NoStop}%
\bibitem [{\citenamefont {Osumi}\ \emph {et~al.}(2024)\citenamefont {Osumi},
  \citenamefont {Souma}, \citenamefont {Aoyama}, \citenamefont {Yamauchi},
  \citenamefont {Honma}, \citenamefont {Nakayama}, \citenamefont {Takahashi},
  \citenamefont {Ohgushi},\ and\ \citenamefont {Sato}}]{Osumi2024}%
  \BibitemOpen
  \bibfield  {author} {\bibinfo {author} {\bibfnamefont {T.}~\bibnamefont
  {Osumi}}, \bibinfo {author} {\bibfnamefont {S.}~\bibnamefont {Souma}},
  \bibinfo {author} {\bibfnamefont {T.}~\bibnamefont {Aoyama}}, \bibinfo
  {author} {\bibfnamefont {K.}~\bibnamefont {Yamauchi}}, \bibinfo {author}
  {\bibfnamefont {A.}~\bibnamefont {Honma}}, \bibinfo {author} {\bibfnamefont
  {K.}~\bibnamefont {Nakayama}}, \bibinfo {author} {\bibfnamefont
  {T.}~\bibnamefont {Takahashi}}, \bibinfo {author} {\bibfnamefont
  {K.}~\bibnamefont {Ohgushi}},\ and\ \bibinfo {author} {\bibfnamefont
  {T.}~\bibnamefont {Sato}},\ }\bibfield  {title} {\bibinfo {title}
  {Observation of a giant band splitting in altermagnetic {M}n{T}e},\ }\href
  {https://doi.org/10.1103/PhysRevB.109.115102} {\bibfield  {journal} {\bibinfo
   {journal} {Phys. Rev. B}\ }\textbf {\bibinfo {volume} {109}},\ \bibinfo
  {pages} {115102} (\bibinfo {year} {2024})}\BibitemShut {NoStop}%
\bibitem [{\citenamefont {Hajlaoui}\ \emph {et~al.}(2024)\citenamefont
  {Hajlaoui}, \citenamefont {Wilfred~D'Souza}, \citenamefont {Šmejkal},
  \citenamefont {Kriegner}, \citenamefont {Krizman}, \citenamefont {Zakusylo},
  \citenamefont {Olszowska}, \citenamefont {Caha}, \citenamefont {Michalička},
  \citenamefont {Sánchez-Barriga}, \citenamefont {Marmodoro}, \citenamefont
  {Výborný}, \citenamefont {Ernst}, \citenamefont {Cinchetti}, \citenamefont
  {Minar}, \citenamefont {Jungwirth},\ and\ \citenamefont
  {Springholz}}]{Hajlaoui2024}%
  \BibitemOpen
  \bibfield  {author} {\bibinfo {author} {\bibfnamefont {M.}~\bibnamefont
  {Hajlaoui}}, \bibinfo {author} {\bibfnamefont {S.}~\bibnamefont
  {Wilfred~D'Souza}}, \bibinfo {author} {\bibfnamefont {L.}~\bibnamefont
  {Šmejkal}}, \bibinfo {author} {\bibfnamefont {D.}~\bibnamefont {Kriegner}},
  \bibinfo {author} {\bibfnamefont {G.}~\bibnamefont {Krizman}}, \bibinfo
  {author} {\bibfnamefont {T.}~\bibnamefont {Zakusylo}}, \bibinfo {author}
  {\bibfnamefont {N.}~\bibnamefont {Olszowska}}, \bibinfo {author}
  {\bibfnamefont {O.}~\bibnamefont {Caha}}, \bibinfo {author} {\bibfnamefont
  {J.}~\bibnamefont {Michalička}}, \bibinfo {author} {\bibfnamefont
  {J.}~\bibnamefont {Sánchez-Barriga}}, \bibinfo {author} {\bibfnamefont
  {A.}~\bibnamefont {Marmodoro}}, \bibinfo {author} {\bibfnamefont
  {K.}~\bibnamefont {Výborný}}, \bibinfo {author} {\bibfnamefont
  {A.}~\bibnamefont {Ernst}}, \bibinfo {author} {\bibfnamefont
  {M.}~\bibnamefont {Cinchetti}}, \bibinfo {author} {\bibfnamefont
  {J.}~\bibnamefont {Minar}}, \bibinfo {author} {\bibfnamefont
  {T.}~\bibnamefont {Jungwirth}},\ and\ \bibinfo {author} {\bibfnamefont
  {G.}~\bibnamefont {Springholz}},\ }\bibfield  {title} {\bibinfo {title}
  {Temperature dependence of relativistic valence band splitting induced by an
  altermagnetic phase transition},\ }\href
  {https://doi.org/https://doi.org/10.1002/adma.202314076} {\bibfield
  {journal} {\bibinfo  {journal} {Advanced Materials}\ }\textbf {\bibinfo
  {volume} {36}},\ \bibinfo {pages} {2314076} (\bibinfo {year}
  {2024})}\BibitemShut {NoStop}%
\bibitem [{\citenamefont {Zeng}\ \emph {et~al.}(2024)\citenamefont {Zeng},
  \citenamefont {Zhu}, \citenamefont {Zhu}, \citenamefont {Liu}, \citenamefont
  {Ma}, \citenamefont {Hao}, \citenamefont {Liu}, \citenamefont {Qu},
  \citenamefont {Yang}, \citenamefont {Jiang}, \citenamefont {Yamagami},
  \citenamefont {Arita}, \citenamefont {Zhang}, \citenamefont {Shao},
  \citenamefont {Dai}, \citenamefont {Shimada}, \citenamefont {Liu},
  \citenamefont {Ye}, \citenamefont {Huang}, \citenamefont {Liu},\ and\
  \citenamefont {Liu}}]{Zeng2024}%
  \BibitemOpen
  \bibfield  {author} {\bibinfo {author} {\bibfnamefont {M.}~\bibnamefont
  {Zeng}}, \bibinfo {author} {\bibfnamefont {M.-Y.}\ \bibnamefont {Zhu}},
  \bibinfo {author} {\bibfnamefont {Y.-P.}\ \bibnamefont {Zhu}}, \bibinfo
  {author} {\bibfnamefont {X.-R.}\ \bibnamefont {Liu}}, \bibinfo {author}
  {\bibfnamefont {X.-M.}\ \bibnamefont {Ma}}, \bibinfo {author} {\bibfnamefont
  {Y.-J.}\ \bibnamefont {Hao}}, \bibinfo {author} {\bibfnamefont
  {P.}~\bibnamefont {Liu}}, \bibinfo {author} {\bibfnamefont {G.}~\bibnamefont
  {Qu}}, \bibinfo {author} {\bibfnamefont {Y.}~\bibnamefont {Yang}}, \bibinfo
  {author} {\bibfnamefont {Z.}~\bibnamefont {Jiang}}, \bibinfo {author}
  {\bibfnamefont {K.}~\bibnamefont {Yamagami}}, \bibinfo {author}
  {\bibfnamefont {M.}~\bibnamefont {Arita}}, \bibinfo {author} {\bibfnamefont
  {X.}~\bibnamefont {Zhang}}, \bibinfo {author} {\bibfnamefont {T.-H.}\
  \bibnamefont {Shao}}, \bibinfo {author} {\bibfnamefont {Y.}~\bibnamefont
  {Dai}}, \bibinfo {author} {\bibfnamefont {K.}~\bibnamefont {Shimada}},
  \bibinfo {author} {\bibfnamefont {Z.}~\bibnamefont {Liu}}, \bibinfo {author}
  {\bibfnamefont {M.}~\bibnamefont {Ye}}, \bibinfo {author} {\bibfnamefont
  {Y.}~\bibnamefont {Huang}}, \bibinfo {author} {\bibfnamefont
  {Q.}~\bibnamefont {Liu}},\ and\ \bibinfo {author} {\bibfnamefont
  {C.}~\bibnamefont {Liu}},\ }\bibfield  {title} {\bibinfo {title} {Observation
  of spin splitting in room-temperature metallic antiferromagnet {C}r{S}b},\
  }\href {https://doi.org/https://doi.org/10.1002/advs.202406529} {\bibfield
  {journal} {\bibinfo  {journal} {Advanced Science}\ }\textbf {\bibinfo
  {volume} {11}},\ \bibinfo {pages} {2406529} (\bibinfo {year}
  {2024})}\BibitemShut {NoStop}%
\bibitem [{\citenamefont {Ding}\ \emph {et~al.}(2024)\citenamefont {Ding},
  \citenamefont {Jiang}, \citenamefont {Chen}, \citenamefont {Tao},
  \citenamefont {Liu}, \citenamefont {Li}, \citenamefont {Liu}, \citenamefont
  {Sun}, \citenamefont {Cheng}, \citenamefont {Liu}, \citenamefont {Yang},
  \citenamefont {Zhang}, \citenamefont {Deng}, \citenamefont {Jing},
  \citenamefont {Huang}, \citenamefont {Shi}, \citenamefont {Ye}, \citenamefont
  {Qiao}, \citenamefont {Wang}, \citenamefont {Guo}, \citenamefont {Feng},\
  and\ \citenamefont {Shen}}]{Ding2024}%
  \BibitemOpen
  \bibfield  {author} {\bibinfo {author} {\bibfnamefont {J.}~\bibnamefont
  {Ding}}, \bibinfo {author} {\bibfnamefont {Z.}~\bibnamefont {Jiang}},
  \bibinfo {author} {\bibfnamefont {X.}~\bibnamefont {Chen}}, \bibinfo {author}
  {\bibfnamefont {Z.}~\bibnamefont {Tao}}, \bibinfo {author} {\bibfnamefont
  {Z.}~\bibnamefont {Liu}}, \bibinfo {author} {\bibfnamefont {T.}~\bibnamefont
  {Li}}, \bibinfo {author} {\bibfnamefont {J.}~\bibnamefont {Liu}}, \bibinfo
  {author} {\bibfnamefont {J.}~\bibnamefont {Sun}}, \bibinfo {author}
  {\bibfnamefont {J.}~\bibnamefont {Cheng}}, \bibinfo {author} {\bibfnamefont
  {J.}~\bibnamefont {Liu}}, \bibinfo {author} {\bibfnamefont {Y.}~\bibnamefont
  {Yang}}, \bibinfo {author} {\bibfnamefont {R.}~\bibnamefont {Zhang}},
  \bibinfo {author} {\bibfnamefont {L.}~\bibnamefont {Deng}}, \bibinfo {author}
  {\bibfnamefont {W.}~\bibnamefont {Jing}}, \bibinfo {author} {\bibfnamefont
  {Y.}~\bibnamefont {Huang}}, \bibinfo {author} {\bibfnamefont
  {Y.}~\bibnamefont {Shi}}, \bibinfo {author} {\bibfnamefont {M.}~\bibnamefont
  {Ye}}, \bibinfo {author} {\bibfnamefont {S.}~\bibnamefont {Qiao}}, \bibinfo
  {author} {\bibfnamefont {Y.}~\bibnamefont {Wang}}, \bibinfo {author}
  {\bibfnamefont {Y.}~\bibnamefont {Guo}}, \bibinfo {author} {\bibfnamefont
  {D.}~\bibnamefont {Feng}},\ and\ \bibinfo {author} {\bibfnamefont
  {D.}~\bibnamefont {Shen}},\ }\bibfield  {title} {\bibinfo {title} {Large band
  splitting in $g$-wave altermagnet {C}r{S}b},\ }\href
  {https://doi.org/10.1103/PhysRevLett.133.206401} {\bibfield  {journal}
  {\bibinfo  {journal} {Phys. Rev. Lett.}\ }\textbf {\bibinfo {volume} {133}},\
  \bibinfo {pages} {206401} (\bibinfo {year} {2024})}\BibitemShut {NoStop}%
\bibitem [{\citenamefont {Santhosh}\ \emph {et~al.}(2025)\citenamefont
  {Santhosh}, \citenamefont {Corbae}, \citenamefont {Yánez-Parreño},
  \citenamefont {Ghosh}, \citenamefont {Jensen}, \citenamefont {Fedorov},
  \citenamefont {Hashimoto}, \citenamefont {Lu}, \citenamefont {Borchers},
  \citenamefont {Grutter}, \citenamefont {Charlton}, \citenamefont {Islam},
  \citenamefont {Golovanova}, \citenamefont {Zhao}, \citenamefont {Tauraso},
  \citenamefont {Richardella}, \citenamefont {Yan}, \citenamefont {Mkhoyan},
  \citenamefont {Palmstrøm}, \citenamefont {Ou},\ and\ \citenamefont
  {Samarth}}]{Santhosh2025}%
  \BibitemOpen
  \bibfield  {author} {\bibinfo {author} {\bibfnamefont {S.}~\bibnamefont
  {Santhosh}}, \bibinfo {author} {\bibfnamefont {P.}~\bibnamefont {Corbae}},
  \bibinfo {author} {\bibfnamefont {W.~J.}\ \bibnamefont {Yánez-Parreño}},
  \bibinfo {author} {\bibfnamefont {S.}~\bibnamefont {Ghosh}}, \bibinfo
  {author} {\bibfnamefont {C.~J.}\ \bibnamefont {Jensen}}, \bibinfo {author}
  {\bibfnamefont {A.~V.}\ \bibnamefont {Fedorov}}, \bibinfo {author}
  {\bibfnamefont {M.}~\bibnamefont {Hashimoto}}, \bibinfo {author}
  {\bibfnamefont {D.}~\bibnamefont {Lu}}, \bibinfo {author} {\bibfnamefont
  {J.~A.}\ \bibnamefont {Borchers}}, \bibinfo {author} {\bibfnamefont {A.~J.}\
  \bibnamefont {Grutter}}, \bibinfo {author} {\bibfnamefont {T.~R.}\
  \bibnamefont {Charlton}}, \bibinfo {author} {\bibfnamefont {S.}~\bibnamefont
  {Islam}}, \bibinfo {author} {\bibfnamefont {D.}~\bibnamefont {Golovanova}},
  \bibinfo {author} {\bibfnamefont {Y.}~\bibnamefont {Zhao}}, \bibinfo {author}
  {\bibfnamefont {A.}~\bibnamefont {Tauraso}}, \bibinfo {author} {\bibfnamefont
  {A.}~\bibnamefont {Richardella}}, \bibinfo {author} {\bibfnamefont
  {B.}~\bibnamefont {Yan}}, \bibinfo {author} {\bibfnamefont {K.~A.}\
  \bibnamefont {Mkhoyan}}, \bibinfo {author} {\bibfnamefont {C.~J.}\
  \bibnamefont {Palmstrøm}}, \bibinfo {author} {\bibfnamefont
  {Y.}~\bibnamefont {Ou}},\ and\ \bibinfo {author} {\bibfnamefont
  {N.}~\bibnamefont {Samarth}},\ }\bibfield  {title} {\bibinfo {title}
  {Altermagnetic band splitting in 10 nm epitaxial {C}r{S}b thin films},\
  }\href {https://doi.org/https://doi.org/10.1002/adma.202508977} {\bibfield
  {journal} {\bibinfo  {journal} {Advanced Materials}\ }\textbf {\bibinfo
  {volume} {37}},\ \bibinfo {pages} {e08977} (\bibinfo {year}
  {2025})}\BibitemShut {NoStop}%
\bibitem [{\citenamefont {Yang}\ \emph {et~al.}(2025)\citenamefont {Yang},
  \citenamefont {Li}, \citenamefont {Yang}, \citenamefont {Li}, \citenamefont
  {Zheng}, \citenamefont {Zhu}, \citenamefont {Pan}, \citenamefont {Xu},
  \citenamefont {Cao}, \citenamefont {Zhao}, \citenamefont {Jana},
  \citenamefont {Zhang}, \citenamefont {Ye}, \citenamefont {Song},
  \citenamefont {Hu}, \citenamefont {Yang}, \citenamefont {Fujii},
  \citenamefont {Vobornik}, \citenamefont {Shi}, \citenamefont {Yuan},
  \citenamefont {Zhang}, \citenamefont {Xu},\ and\ \citenamefont
  {Liu}}]{Yang2025}%
  \BibitemOpen
  \bibfield  {author} {\bibinfo {author} {\bibfnamefont {G.}~\bibnamefont
  {Yang}}, \bibinfo {author} {\bibfnamefont {Z.}~\bibnamefont {Li}}, \bibinfo
  {author} {\bibfnamefont {S.}~\bibnamefont {Yang}}, \bibinfo {author}
  {\bibfnamefont {J.}~\bibnamefont {Li}}, \bibinfo {author} {\bibfnamefont
  {H.}~\bibnamefont {Zheng}}, \bibinfo {author} {\bibfnamefont
  {W.}~\bibnamefont {Zhu}}, \bibinfo {author} {\bibfnamefont {Z.}~\bibnamefont
  {Pan}}, \bibinfo {author} {\bibfnamefont {Y.}~\bibnamefont {Xu}}, \bibinfo
  {author} {\bibfnamefont {S.}~\bibnamefont {Cao}}, \bibinfo {author}
  {\bibfnamefont {W.}~\bibnamefont {Zhao}}, \bibinfo {author} {\bibfnamefont
  {A.}~\bibnamefont {Jana}}, \bibinfo {author} {\bibfnamefont {J.}~\bibnamefont
  {Zhang}}, \bibinfo {author} {\bibfnamefont {M.}~\bibnamefont {Ye}}, \bibinfo
  {author} {\bibfnamefont {Y.}~\bibnamefont {Song}}, \bibinfo {author}
  {\bibfnamefont {L.-H.}\ \bibnamefont {Hu}}, \bibinfo {author} {\bibfnamefont
  {L.}~\bibnamefont {Yang}}, \bibinfo {author} {\bibfnamefont {J.}~\bibnamefont
  {Fujii}}, \bibinfo {author} {\bibfnamefont {I.}~\bibnamefont {Vobornik}},
  \bibinfo {author} {\bibfnamefont {M.}~\bibnamefont {Shi}}, \bibinfo {author}
  {\bibfnamefont {H.}~\bibnamefont {Yuan}}, \bibinfo {author} {\bibfnamefont
  {Y.}~\bibnamefont {Zhang}}, \bibinfo {author} {\bibfnamefont
  {Y.}~\bibnamefont {Xu}},\ and\ \bibinfo {author} {\bibfnamefont
  {Y.}~\bibnamefont {Liu}},\ }\bibfield  {title} {\bibinfo {title}
  {Three-dimensional mapping of the altermagnetic spin splitting in {C}r{S}b},\
  }\href {https://doi.org/10.1038/s41467-025-56647-7} {\bibfield  {journal}
  {\bibinfo  {journal} {Nature Communications}\ }\textbf {\bibinfo {volume}
  {16}},\ \bibinfo {pages} {1442} (\bibinfo {year} {2025})}\BibitemShut
  {NoStop}%
\bibitem [{\citenamefont {Jiang}\ \emph {et~al.}(2025)\citenamefont {Jiang},
  \citenamefont {Hu}, \citenamefont {Bai}, \citenamefont {Song}, \citenamefont
  {Mu}, \citenamefont {Qu}, \citenamefont {Li}, \citenamefont {Zhu},
  \citenamefont {Pi}, \citenamefont {Wei}, \citenamefont {Sun}, \citenamefont
  {Huang}, \citenamefont {Zheng}, \citenamefont {Peng}, \citenamefont {He},
  \citenamefont {Li}, \citenamefont {Luo}, \citenamefont {Li}, \citenamefont
  {Chen}, \citenamefont {Li}, \citenamefont {Weng},\ and\ \citenamefont
  {Qian}}]{Jiang2025}%
  \BibitemOpen
  \bibfield  {author} {\bibinfo {author} {\bibfnamefont {B.}~\bibnamefont
  {Jiang}}, \bibinfo {author} {\bibfnamefont {M.}~\bibnamefont {Hu}}, \bibinfo
  {author} {\bibfnamefont {J.}~\bibnamefont {Bai}}, \bibinfo {author}
  {\bibfnamefont {Z.}~\bibnamefont {Song}}, \bibinfo {author} {\bibfnamefont
  {C.}~\bibnamefont {Mu}}, \bibinfo {author} {\bibfnamefont {G.}~\bibnamefont
  {Qu}}, \bibinfo {author} {\bibfnamefont {W.}~\bibnamefont {Li}}, \bibinfo
  {author} {\bibfnamefont {W.}~\bibnamefont {Zhu}}, \bibinfo {author}
  {\bibfnamefont {H.}~\bibnamefont {Pi}}, \bibinfo {author} {\bibfnamefont
  {Z.}~\bibnamefont {Wei}}, \bibinfo {author} {\bibfnamefont {Y.-J.}\
  \bibnamefont {Sun}}, \bibinfo {author} {\bibfnamefont {Y.}~\bibnamefont
  {Huang}}, \bibinfo {author} {\bibfnamefont {X.}~\bibnamefont {Zheng}},
  \bibinfo {author} {\bibfnamefont {Y.}~\bibnamefont {Peng}}, \bibinfo {author}
  {\bibfnamefont {L.}~\bibnamefont {He}}, \bibinfo {author} {\bibfnamefont
  {S.}~\bibnamefont {Li}}, \bibinfo {author} {\bibfnamefont {J.}~\bibnamefont
  {Luo}}, \bibinfo {author} {\bibfnamefont {Z.}~\bibnamefont {Li}}, \bibinfo
  {author} {\bibfnamefont {G.}~\bibnamefont {Chen}}, \bibinfo {author}
  {\bibfnamefont {H.}~\bibnamefont {Li}}, \bibinfo {author} {\bibfnamefont
  {H.}~\bibnamefont {Weng}},\ and\ \bibinfo {author} {\bibfnamefont
  {T.}~\bibnamefont {Qian}},\ }\bibfield  {title} {\bibinfo {title} {A metallic
  room-temperature $d$-wave altermagnet},\ }\href
  {https://doi.org/10.1038/s41567-025-02822-y} {\bibfield  {journal} {\bibinfo
  {journal} {Nature Physics}\ }\textbf {\bibinfo {volume} {21}},\ \bibinfo
  {pages} {754} (\bibinfo {year} {2025})}\BibitemShut {NoStop}%
\bibitem [{\citenamefont {Zhang}\ \emph {et~al.}(2025)\citenamefont {Zhang},
  \citenamefont {Cheng}, \citenamefont {Yin}, \citenamefont {Liu},
  \citenamefont {Deng}, \citenamefont {Qiao}, \citenamefont {Shi},
  \citenamefont {Zhang}, \citenamefont {Lin}, \citenamefont {Liu},
  \citenamefont {Ye}, \citenamefont {Huang}, \citenamefont {Meng},
  \citenamefont {Zhang}, \citenamefont {Okuda}, \citenamefont {Shimada},
  \citenamefont {Cui}, \citenamefont {Zhao}, \citenamefont {Cao}, \citenamefont
  {Qiao}, \citenamefont {Liu},\ and\ \citenamefont {Chen}}]{Zhang2025}%
  \BibitemOpen
  \bibfield  {author} {\bibinfo {author} {\bibfnamefont {F.}~\bibnamefont
  {Zhang}}, \bibinfo {author} {\bibfnamefont {X.}~\bibnamefont {Cheng}},
  \bibinfo {author} {\bibfnamefont {Z.}~\bibnamefont {Yin}}, \bibinfo {author}
  {\bibfnamefont {C.}~\bibnamefont {Liu}}, \bibinfo {author} {\bibfnamefont
  {L.}~\bibnamefont {Deng}}, \bibinfo {author} {\bibfnamefont {Y.}~\bibnamefont
  {Qiao}}, \bibinfo {author} {\bibfnamefont {Z.}~\bibnamefont {Shi}}, \bibinfo
  {author} {\bibfnamefont {S.}~\bibnamefont {Zhang}}, \bibinfo {author}
  {\bibfnamefont {J.}~\bibnamefont {Lin}}, \bibinfo {author} {\bibfnamefont
  {Z.}~\bibnamefont {Liu}}, \bibinfo {author} {\bibfnamefont {M.}~\bibnamefont
  {Ye}}, \bibinfo {author} {\bibfnamefont {Y.}~\bibnamefont {Huang}}, \bibinfo
  {author} {\bibfnamefont {X.}~\bibnamefont {Meng}}, \bibinfo {author}
  {\bibfnamefont {C.}~\bibnamefont {Zhang}}, \bibinfo {author} {\bibfnamefont
  {T.}~\bibnamefont {Okuda}}, \bibinfo {author} {\bibfnamefont
  {K.}~\bibnamefont {Shimada}}, \bibinfo {author} {\bibfnamefont
  {S.}~\bibnamefont {Cui}}, \bibinfo {author} {\bibfnamefont {Y.}~\bibnamefont
  {Zhao}}, \bibinfo {author} {\bibfnamefont {G.-H.}\ \bibnamefont {Cao}},
  \bibinfo {author} {\bibfnamefont {S.}~\bibnamefont {Qiao}}, \bibinfo {author}
  {\bibfnamefont {J.}~\bibnamefont {Liu}},\ and\ \bibinfo {author}
  {\bibfnamefont {C.}~\bibnamefont {Chen}},\ }\bibfield  {title} {\bibinfo
  {title} {Crystal-symmetry-paired spin--valley locking in a layered
  room-temperature metallic altermagnet candidate},\ }\href
  {https://doi.org/10.1038/s41567-025-02864-2} {\bibfield  {journal} {\bibinfo
  {journal} {Nature Physics}\ }\textbf {\bibinfo {volume} {21}},\ \bibinfo
  {pages} {760} (\bibinfo {year} {2025})}\BibitemShut {NoStop}%
\bibitem [{\citenamefont {Sun}\ \emph {et~al.}(2025)\citenamefont {Sun},
  \citenamefont {Guo}, \citenamefont {Wang}, \citenamefont {Abernathy},
  \citenamefont {Tian},\ and\ \citenamefont {Li}}]{Sun2025}%
  \BibitemOpen
  \bibfield  {author} {\bibinfo {author} {\bibfnamefont {Q.}~\bibnamefont
  {Sun}}, \bibinfo {author} {\bibfnamefont {J.}~\bibnamefont {Guo}}, \bibinfo
  {author} {\bibfnamefont {D.}~\bibnamefont {Wang}}, \bibinfo {author}
  {\bibfnamefont {D.~L.}\ \bibnamefont {Abernathy}}, \bibinfo {author}
  {\bibfnamefont {W.}~\bibnamefont {Tian}},\ and\ \bibinfo {author}
  {\bibfnamefont {C.}~\bibnamefont {Li}},\ }\bibfield  {title} {\bibinfo
  {title} {Observation of chiral magnon band splitting in altermagnetic
  hematite},\ }\href {https://doi.org/10.1103/7yhz-jptc} {\bibfield  {journal}
  {\bibinfo  {journal} {Phys. Rev. Lett.}\ }\textbf {\bibinfo {volume} {135}},\
  \bibinfo {pages} {186703} (\bibinfo {year} {2025})}\BibitemShut {NoStop}%
\bibitem [{\citenamefont {Faure}\ \emph {et~al.}(2025)\citenamefont {Faure},
  \citenamefont {Bounoua}, \citenamefont {Balédent}, \citenamefont {Gukasov},
  \citenamefont {Garlea}, \citenamefont {Ribeiro}, \citenamefont {Rau},
  \citenamefont {Petit},\ and\ \citenamefont {McClarty}}]{Faure2025}%
  \BibitemOpen
  \bibfield  {author} {\bibinfo {author} {\bibfnamefont {Q.}~\bibnamefont
  {Faure}}, \bibinfo {author} {\bibfnamefont {D.}~\bibnamefont {Bounoua}},
  \bibinfo {author} {\bibfnamefont {V.}~\bibnamefont {Balédent}}, \bibinfo
  {author} {\bibfnamefont {A.}~\bibnamefont {Gukasov}}, \bibinfo {author}
  {\bibfnamefont {V.~O.}\ \bibnamefont {Garlea}}, \bibinfo {author}
  {\bibfnamefont {A.}~\bibnamefont {Ribeiro}}, \bibinfo {author} {\bibfnamefont
  {J.~G.}\ \bibnamefont {Rau}}, \bibinfo {author} {\bibfnamefont
  {S.}~\bibnamefont {Petit}},\ and\ \bibinfo {author} {\bibfnamefont
  {P.}~\bibnamefont {McClarty}},\ }\href {https://arxiv.org/abs/2509.07087}
  {\bibinfo {title} {Altermagnetism revealed by polarized neutrons in
  {M}n{F}$_2$}} (\bibinfo {year} {2025}),\ \Eprint
  {https://arxiv.org/abs/2509.07087} {arXiv:2509.07087 [cond-mat.str-el]}
  \BibitemShut {NoStop}%
\bibitem [{\citenamefont {Singh}\ \emph {et~al.}(2026)\citenamefont {Singh},
  \citenamefont {Heinsdorf}, \citenamefont {Mancilla}, \citenamefont {Bannies},
  \citenamefont {Maity}, \citenamefont {Kolesnikov}, \citenamefont {Matsuda},
  \citenamefont {Stone}, \citenamefont {Franz}, \citenamefont {Gaudet},\ and\
  \citenamefont {Hallas}}]{Singh2026}%
  \BibitemOpen
  \bibfield  {author} {\bibinfo {author} {\bibfnamefont {A.~K.}\ \bibnamefont
  {Singh}}, \bibinfo {author} {\bibfnamefont {N.}~\bibnamefont {Heinsdorf}},
  \bibinfo {author} {\bibfnamefont {A.~A.}\ \bibnamefont {Mancilla}}, \bibinfo
  {author} {\bibfnamefont {J.}~\bibnamefont {Bannies}}, \bibinfo {author}
  {\bibfnamefont {A.}~\bibnamefont {Maity}}, \bibinfo {author} {\bibfnamefont
  {A.~I.}\ \bibnamefont {Kolesnikov}}, \bibinfo {author} {\bibfnamefont
  {M.}~\bibnamefont {Matsuda}}, \bibinfo {author} {\bibfnamefont {M.~B.}\
  \bibnamefont {Stone}}, \bibinfo {author} {\bibfnamefont {M.}~\bibnamefont
  {Franz}}, \bibinfo {author} {\bibfnamefont {J.}~\bibnamefont {Gaudet}},\ and\
  \bibinfo {author} {\bibfnamefont {A.~M.}\ \bibnamefont {Hallas}},\ }\href
  {https://arxiv.org/abs/2511.16086} {\bibinfo {title} {Coherent high-velocity
  chiral magnons in the metallic altermagnet {C}r{S}b}} (\bibinfo {year}
  {2026}),\ \Eprint {https://arxiv.org/abs/2511.16086} {arXiv:2511.16086
  [cond-mat.mtrl-sci]} \BibitemShut {NoStop}%
\bibitem [{\citenamefont {Sears}\ \emph {et~al.}(2026)\citenamefont {Sears},
  \citenamefont {Garlea}, \citenamefont {Lederman}, \citenamefont {Tranquada},\
  and\ \citenamefont {Zaliznyak}}]{Sears2026}%
  \BibitemOpen
  \bibfield  {author} {\bibinfo {author} {\bibfnamefont {J.}~\bibnamefont
  {Sears}}, \bibinfo {author} {\bibfnamefont {V.~O.}\ \bibnamefont {Garlea}},
  \bibinfo {author} {\bibfnamefont {D.}~\bibnamefont {Lederman}}, \bibinfo
  {author} {\bibfnamefont {J.~M.}\ \bibnamefont {Tranquada}},\ and\ \bibinfo
  {author} {\bibfnamefont {I.~A.}\ \bibnamefont {Zaliznyak}},\ }\bibfield
  {title} {\bibinfo {title} {Altermagnetic and dipolar splitting of magnons in
  {${\mathrm{FeF}}_{2}$}},\ }\href {https://doi.org/10.1103/g6dt-rf8c}
  {\bibfield  {journal} {\bibinfo  {journal} {Phys. Rev. Lett.}\ }\textbf
  {\bibinfo {volume} {136}},\ \bibinfo {pages} {226701} (\bibinfo {year}
  {2026})}\BibitemShut {NoStop}%
\bibitem [{\citenamefont {Asai}\ \emph {et~al.}(2026)\citenamefont {Asai},
  \citenamefont {Hu}, \citenamefont {Liu}, \citenamefont {Itoh}, \citenamefont
  {Ueta}, \citenamefont {Matsuda}, \citenamefont {Hatanaka}, \citenamefont
  {Watanabe}, \citenamefont {Arita},\ and\ \citenamefont {Masuda}}]{Asai2026}%
  \BibitemOpen
  \bibfield  {author} {\bibinfo {author} {\bibfnamefont {S.}~\bibnamefont
  {Asai}}, \bibinfo {author} {\bibfnamefont {J.}~\bibnamefont {Hu}}, \bibinfo
  {author} {\bibfnamefont {Z.}~\bibnamefont {Liu}}, \bibinfo {author}
  {\bibfnamefont {S.}~\bibnamefont {Itoh}}, \bibinfo {author} {\bibfnamefont
  {D.}~\bibnamefont {Ueta}}, \bibinfo {author} {\bibfnamefont {J.}~\bibnamefont
  {Matsuda}}, \bibinfo {author} {\bibfnamefont {T.}~\bibnamefont {Hatanaka}},
  \bibinfo {author} {\bibfnamefont {H.}~\bibnamefont {Watanabe}}, \bibinfo
  {author} {\bibfnamefont {R.}~\bibnamefont {Arita}},\ and\ \bibinfo {author}
  {\bibfnamefont {T.}~\bibnamefont {Masuda}},\ }\bibfield  {title} {\bibinfo
  {title} {Realization of a two-dimensional $d$-wave altermagnet in
  {La}$_{2}${O}$_{3}${Mn}$_{2}${Se}$_{2}$},\ }\href
  {https://doi.org/10.1103/q863-3sfx} {\bibfield  {journal} {\bibinfo
  {journal} {Phys. Rev. Mater.}\ }\textbf {\bibinfo {volume} {10}},\ \bibinfo
  {pages} {L011401} (\bibinfo {year} {2026})}\BibitemShut {NoStop}%
\bibitem [{\citenamefont {Asgharpour}\ \emph {et~al.}(2025)\citenamefont
  {Asgharpour}, \citenamefont {Koopmans},\ and\ \citenamefont
  {Duine}}]{Asgharpour2025}%
  \BibitemOpen
  \bibfield  {author} {\bibinfo {author} {\bibfnamefont {A.}~\bibnamefont
  {Asgharpour}}, \bibinfo {author} {\bibfnamefont {B.}~\bibnamefont
  {Koopmans}},\ and\ \bibinfo {author} {\bibfnamefont {R.~A.}\ \bibnamefont
  {Duine}},\ }\bibfield  {title} {\bibinfo {title} {Synthetic altermagnets},\
  }\href {https://doi.org/10.1103/PhysRevB.111.094412} {\bibfield  {journal}
  {\bibinfo  {journal} {Phys. Rev. B}\ }\textbf {\bibinfo {volume} {111}},\
  \bibinfo {pages} {094412} (\bibinfo {year} {2025})}\BibitemShut {NoStop}%
\bibitem [{\citenamefont {Gallardo}\ \emph {et~al.}(2026)\citenamefont
  {Gallardo}, \citenamefont {León}, \citenamefont {Lindner},\ and\
  \citenamefont {González}}]{Gallardo2026}%
  \BibitemOpen
  \bibfield  {author} {\bibinfo {author} {\bibfnamefont {R.~A.}\ \bibnamefont
  {Gallardo}}, \bibinfo {author} {\bibfnamefont {A.~M.}\ \bibnamefont {León}},
  \bibinfo {author} {\bibfnamefont {J.}~\bibnamefont {Lindner}},\ and\ \bibinfo
  {author} {\bibfnamefont {J.~W.}\ \bibnamefont {González}},\ }\href
  {https://arxiv.org/abs/2606.11481} {\bibinfo {title} {Synthetic
  altermagnetism beyond the crystal limit}} (\bibinfo {year} {2026}),\ \Eprint
  {https://arxiv.org/abs/2606.11481} {arXiv:2606.11481 [cond-mat.other]}
  \BibitemShut {NoStop}%
\bibitem [{\citenamefont {Duine}\ \emph {et~al.}(2018)\citenamefont {Duine},
  \citenamefont {Lee}, \citenamefont {Parkin},\ and\ \citenamefont
  {Stiles}}]{Duine:Synthetic:2018}%
  \BibitemOpen
  \bibfield  {author} {\bibinfo {author} {\bibfnamefont {R.~A.}\ \bibnamefont
  {Duine}}, \bibinfo {author} {\bibfnamefont {K.-J.}\ \bibnamefont {Lee}},
  \bibinfo {author} {\bibfnamefont {S.~S.~P.}\ \bibnamefont {Parkin}},\ and\
  \bibinfo {author} {\bibfnamefont {M.~D.}\ \bibnamefont {Stiles}},\ }\bibfield
   {title} {\bibinfo {title} {Synthetic antiferromagnetic spintronics},\ }\href
  {https://doi.org/10.1038/s41567-018-0050-y} {\bibfield  {journal} {\bibinfo
  {journal} {Nature Physics}\ }\textbf {\bibinfo {volume} {14}},\ \bibinfo
  {pages} {217} (\bibinfo {year} {2018})}\BibitemShut {NoStop}%
\bibitem [{\citenamefont {Ishibashi}\ \emph {et~al.}(2020)\citenamefont
  {Ishibashi}, \citenamefont {Shiota}, \citenamefont {Li}, \citenamefont
  {Funada}, \citenamefont {Moriyama},\ and\ \citenamefont
  {Ono}}]{Ishibashi:Switchable:2020}%
  \BibitemOpen
  \bibfield  {author} {\bibinfo {author} {\bibfnamefont {M.}~\bibnamefont
  {Ishibashi}}, \bibinfo {author} {\bibfnamefont {Y.}~\bibnamefont {Shiota}},
  \bibinfo {author} {\bibfnamefont {T.}~\bibnamefont {Li}}, \bibinfo {author}
  {\bibfnamefont {S.}~\bibnamefont {Funada}}, \bibinfo {author} {\bibfnamefont
  {T.}~\bibnamefont {Moriyama}},\ and\ \bibinfo {author} {\bibfnamefont
  {T.}~\bibnamefont {Ono}},\ }\bibfield  {title} {\bibinfo {title} {Switchable
  giant nonreciprocal frequency shift of propagating spin waves in synthetic
  antiferromagnets},\ }\href {https://doi.org/10.1126/sciadv.aaz6931}
  {\bibfield  {journal} {\bibinfo  {journal} {Science Advances}\ }\textbf
  {\bibinfo {volume} {6}},\ \bibinfo {pages} {eaaz6931} (\bibinfo {year}
  {2020})}\BibitemShut {NoStop}%
\bibitem [{\citenamefont {Litvin}\ and\ \citenamefont
  {Opechowski}(1974)}]{Litvin1974}%
  \BibitemOpen
  \bibfield  {author} {\bibinfo {author} {\bibfnamefont {D.}~\bibnamefont
  {Litvin}}\ and\ \bibinfo {author} {\bibfnamefont {W.}~\bibnamefont
  {Opechowski}},\ }\bibfield  {title} {\bibinfo {title} {Spin groups},\ }\href
  {https://doi.org/https://doi.org/10.1016/0031-8914(74)90157-8} {\bibfield
  {journal} {\bibinfo  {journal} {Physica}\ }\textbf {\bibinfo {volume} {76}},\
  \bibinfo {pages} {538} (\bibinfo {year} {1974})}\BibitemShut {NoStop}%
\bibitem [{\citenamefont {Campos}\ \emph {et~al.}(2026)\citenamefont {Campos},
  \citenamefont {Mbognou}, \citenamefont {Hellenes}, \citenamefont {Poata},
  \citenamefont {Chen}, \citenamefont {Priessnitz},\ and\ \citenamefont
  {Šmejkal}}]{Campos2026}%
  \BibitemOpen
  \bibfield  {author} {\bibinfo {author} {\bibfnamefont {W.~H.}\ \bibnamefont
  {Campos}}, \bibinfo {author} {\bibfnamefont {F.~C.~F.}\ \bibnamefont
  {Mbognou}}, \bibinfo {author} {\bibfnamefont {A.~B.}\ \bibnamefont
  {Hellenes}}, \bibinfo {author} {\bibfnamefont {J.}~\bibnamefont {Poata}},
  \bibinfo {author} {\bibfnamefont {T.}~\bibnamefont {Chen}}, \bibinfo {author}
  {\bibfnamefont {J.}~\bibnamefont {Priessnitz}},\ and\ \bibinfo {author}
  {\bibfnamefont {L.}~\bibnamefont {Šmejkal}},\ }\href
  {https://arxiv.org/abs/2603.12223} {\bibinfo {title} {Persistent
  altermagnetism}} (\bibinfo {year} {2026}),\ \Eprint
  {https://arxiv.org/abs/2603.12223} {arXiv:2603.12223 [cond-mat.mes-hall]}
  \BibitemShut {NoStop}%
\bibitem [{\citenamefont {Politi}\ and\ \citenamefont
  {Pini}(2002)}]{Politi2002}%
  \BibitemOpen
  \bibfield  {author} {\bibinfo {author} {\bibfnamefont {P.}~\bibnamefont
  {Politi}}\ and\ \bibinfo {author} {\bibfnamefont {M.~G.}\ \bibnamefont
  {Pini}},\ }\bibfield  {title} {\bibinfo {title} {Dipolar interaction between
  two-dimensional magnetic particles},\ }\href
  {https://doi.org/10.1103/PhysRevB.66.214414} {\bibfield  {journal} {\bibinfo
  {journal} {Phys. Rev. B}\ }\textbf {\bibinfo {volume} {66}},\ \bibinfo
  {pages} {214414} (\bibinfo {year} {2002})}\BibitemShut {NoStop}%
\bibitem [{\citenamefont {Puszkarski}\ and\ \citenamefont
  {Krawczyk}(2003)}]{Puszkarski2003}%
  \BibitemOpen
  \bibfield  {author} {\bibinfo {author} {\bibfnamefont {H.}~\bibnamefont
  {Puszkarski}}\ and\ \bibinfo {author} {\bibfnamefont {M.}~\bibnamefont
  {Krawczyk}},\ }\bibfield  {title} {\bibinfo {title} {Magnonic crystals—the
  magnetic counterpart of photonic crystals},\ }\href
  {https://doi.org/10.4028/www.scientific.net/SSP.94.125} {\bibfield  {journal}
  {\bibinfo  {journal} {Solid State Phenomena}\ }\textbf {\bibinfo {volume}
  {94}},\ \bibinfo {pages} {125} (\bibinfo {year} {2003})}\BibitemShut
  {NoStop}%
\bibitem [{\citenamefont {Krawczyk}\ and\ \citenamefont
  {Grundler}(2014)}]{Krawczyk2014}%
  \BibitemOpen
  \bibfield  {author} {\bibinfo {author} {\bibfnamefont {M.}~\bibnamefont
  {Krawczyk}}\ and\ \bibinfo {author} {\bibfnamefont {D.}~\bibnamefont
  {Grundler}},\ }\bibfield  {title} {\bibinfo {title} {Review and prospects of
  magnonic crystals and devices with reprogrammable band structure},\ }\href
  {https://doi.org/10.1088/0953-8984/26/12/123202} {\bibfield  {journal}
  {\bibinfo  {journal} {Journal of Physics: Condensed Matter}\ }\textbf
  {\bibinfo {volume} {26}},\ \bibinfo {pages} {123202} (\bibinfo {year}
  {2014})}\BibitemShut {NoStop}%
\bibitem [{\citenamefont {Chumak}\ \emph {et~al.}(2017)\citenamefont {Chumak},
  \citenamefont {Serga},\ and\ \citenamefont {Hillebrands}}]{Chumak2017}%
  \BibitemOpen
  \bibfield  {author} {\bibinfo {author} {\bibfnamefont {A.~V.}\ \bibnamefont
  {Chumak}}, \bibinfo {author} {\bibfnamefont {A.~A.}\ \bibnamefont {Serga}},\
  and\ \bibinfo {author} {\bibfnamefont {B.}~\bibnamefont {Hillebrands}},\
  }\bibfield  {title} {\bibinfo {title} {Magnonic crystals for data
  processing},\ }\href {https://doi.org/10.1088/1361-6463/aa6a65} {\bibfield
  {journal} {\bibinfo  {journal} {Journal of Physics D: Applied Physics}\
  }\textbf {\bibinfo {volume} {50}},\ \bibinfo {pages} {244001} (\bibinfo
  {year} {2017})}\BibitemShut {NoStop}%
\bibitem [{\citenamefont {Skj{\ae}rv{\o}}\ \emph {et~al.}(2020)\citenamefont
  {Skj{\ae}rv{\o}}, \citenamefont {Marrows}, \citenamefont {Stamps},\ and\
  \citenamefont {Heyderman}}]{Skjaervo2020}%
  \BibitemOpen
  \bibfield  {author} {\bibinfo {author} {\bibfnamefont {S.~H.}\ \bibnamefont
  {Skj{\ae}rv{\o}}}, \bibinfo {author} {\bibfnamefont {C.~H.}\ \bibnamefont
  {Marrows}}, \bibinfo {author} {\bibfnamefont {R.~L.}\ \bibnamefont
  {Stamps}},\ and\ \bibinfo {author} {\bibfnamefont {L.~J.}\ \bibnamefont
  {Heyderman}},\ }\bibfield  {title} {\bibinfo {title} {Advances in artificial
  spin ice},\ }\href {https://doi.org/10.1038/s42254-019-0118-3} {\bibfield
  {journal} {\bibinfo  {journal} {Nature Reviews Physics}\ }\textbf {\bibinfo
  {volume} {2}},\ \bibinfo {pages} {13} (\bibinfo {year} {2020})}\BibitemShut
  {NoStop}%
\bibitem [{\citenamefont {Lendinez}\ and\ \citenamefont
  {Jungfleisch}(2019)}]{Lendinez2020}%
  \BibitemOpen
  \bibfield  {author} {\bibinfo {author} {\bibfnamefont {S.}~\bibnamefont
  {Lendinez}}\ and\ \bibinfo {author} {\bibfnamefont {M.~B.}\ \bibnamefont
  {Jungfleisch}},\ }\bibfield  {title} {\bibinfo {title} {Magnetization
  dynamics in artificial spin ice},\ }\href
  {https://doi.org/10.1088/1361-648X/ab3e78} {\bibfield  {journal} {\bibinfo
  {journal} {Journal of Physics: Condensed Matter}\ }\textbf {\bibinfo {volume}
  {32}},\ \bibinfo {pages} {013001} (\bibinfo {year} {2019})}\BibitemShut
  {NoStop}%
\bibitem [{\citenamefont {Kaffash}\ \emph {et~al.}(2021)\citenamefont
  {Kaffash}, \citenamefont {Lendinez},\ and\ \citenamefont
  {Jungfleisch}}]{Kaffash2021}%
  \BibitemOpen
  \bibfield  {author} {\bibinfo {author} {\bibfnamefont {M.~T.}\ \bibnamefont
  {Kaffash}}, \bibinfo {author} {\bibfnamefont {S.}~\bibnamefont {Lendinez}},\
  and\ \bibinfo {author} {\bibfnamefont {M.~B.}\ \bibnamefont {Jungfleisch}},\
  }\bibfield  {title} {\bibinfo {title} {Nanomagnonics with artificial spin
  ice},\ }\href
  {https://doi.org/https://doi.org/10.1016/j.physleta.2021.127364} {\bibfield
  {journal} {\bibinfo  {journal} {Physics Letters A}\ }\textbf {\bibinfo
  {volume} {402}},\ \bibinfo {pages} {127364} (\bibinfo {year}
  {2021})}\BibitemShut {NoStop}%
\bibitem [{\citenamefont {Verba}\ \emph {et~al.}(2012)\citenamefont {Verba},
  \citenamefont {Melkov}, \citenamefont {Tiberkevich},\ and\ \citenamefont
  {Slavin}}]{Verba2012}%
  \BibitemOpen
  \bibfield  {author} {\bibinfo {author} {\bibfnamefont {R.}~\bibnamefont
  {Verba}}, \bibinfo {author} {\bibfnamefont {G.}~\bibnamefont {Melkov}},
  \bibinfo {author} {\bibfnamefont {V.}~\bibnamefont {Tiberkevich}},\ and\
  \bibinfo {author} {\bibfnamefont {A.}~\bibnamefont {Slavin}},\ }\bibfield
  {title} {\bibinfo {title} {Collective spin-wave excitations in a
  two-dimensional array of coupled magnetic nanodots},\ }\href
  {https://doi.org/10.1103/PhysRevB.85.014427} {\bibfield  {journal} {\bibinfo
  {journal} {Phys. Rev. B}\ }\textbf {\bibinfo {volume} {85}},\ \bibinfo
  {pages} {014427} (\bibinfo {year} {2012})}\BibitemShut {NoStop}%
\bibitem [{\citenamefont {Hansen}\ and\ \citenamefont
  {Krumme}(1984)}]{Hansen1984}%
  \BibitemOpen
  \bibfield  {author} {\bibinfo {author} {\bibfnamefont {P.}~\bibnamefont
  {Hansen}}\ and\ \bibinfo {author} {\bibfnamefont {J.-P.}\ \bibnamefont
  {Krumme}},\ }\bibfield  {title} {\bibinfo {title} {Magnetic and
  magneto-optical properties of garnet films},\ }\href
  {https://doi.org/https://doi.org/10.1016/0040-6090(84)90337-7} {\bibfield
  {journal} {\bibinfo  {journal} {Thin Solid Films}\ }\textbf {\bibinfo
  {volume} {114}},\ \bibinfo {pages} {69} (\bibinfo {year} {1984})},\ \bibinfo
  {note} {special Issue on Magnetic Garnet Films}\BibitemShut {NoStop}%
\bibitem [{\citenamefont {Soumah}\ \emph {et~al.}(2018)\citenamefont {Soumah},
  \citenamefont {Beaulieu}, \citenamefont {Qassym}, \citenamefont
  {Carr{\'e}t{\'e}ro}, \citenamefont {Jacquet}, \citenamefont {Lebourgeois},
  \citenamefont {Ben~Youssef}, \citenamefont {Bortolotti}, \citenamefont
  {Cros},\ and\ \citenamefont {Anane}}]{Soumah2018}%
  \BibitemOpen
  \bibfield  {author} {\bibinfo {author} {\bibfnamefont {L.}~\bibnamefont
  {Soumah}}, \bibinfo {author} {\bibfnamefont {N.}~\bibnamefont {Beaulieu}},
  \bibinfo {author} {\bibfnamefont {L.}~\bibnamefont {Qassym}}, \bibinfo
  {author} {\bibfnamefont {C.}~\bibnamefont {Carr{\'e}t{\'e}ro}}, \bibinfo
  {author} {\bibfnamefont {E.}~\bibnamefont {Jacquet}}, \bibinfo {author}
  {\bibfnamefont {R.}~\bibnamefont {Lebourgeois}}, \bibinfo {author}
  {\bibfnamefont {J.}~\bibnamefont {Ben~Youssef}}, \bibinfo {author}
  {\bibfnamefont {P.}~\bibnamefont {Bortolotti}}, \bibinfo {author}
  {\bibfnamefont {V.}~\bibnamefont {Cros}},\ and\ \bibinfo {author}
  {\bibfnamefont {A.}~\bibnamefont {Anane}},\ }\bibfield  {title} {\bibinfo
  {title} {Ultra-low damping insulating magnetic thin films get
  perpendicular},\ }\href {https://doi.org/10.1038/s41467-018-05732-1}
  {\bibfield  {journal} {\bibinfo  {journal} {Nature Communications}\ }\textbf
  {\bibinfo {volume} {9}},\ \bibinfo {pages} {3355} (\bibinfo {year}
  {2018})}\BibitemShut {NoStop}%
\bibitem [{\citenamefont {Deb}\ \emph {et~al.}(2019)\citenamefont {Deb},
  \citenamefont {Popova}, \citenamefont {Hehn}, \citenamefont {Keller},
  \citenamefont {Petit-Watelot}, \citenamefont {Bargheer}, \citenamefont
  {Mangin},\ and\ \citenamefont {Malinowski}}]{Deb2019}%
  \BibitemOpen
  \bibfield  {author} {\bibinfo {author} {\bibfnamefont {M.}~\bibnamefont
  {Deb}}, \bibinfo {author} {\bibfnamefont {E.}~\bibnamefont {Popova}},
  \bibinfo {author} {\bibfnamefont {M.}~\bibnamefont {Hehn}}, \bibinfo {author}
  {\bibfnamefont {N.}~\bibnamefont {Keller}}, \bibinfo {author} {\bibfnamefont
  {S.}~\bibnamefont {Petit-Watelot}}, \bibinfo {author} {\bibfnamefont
  {M.}~\bibnamefont {Bargheer}}, \bibinfo {author} {\bibfnamefont
  {S.}~\bibnamefont {Mangin}},\ and\ \bibinfo {author} {\bibfnamefont
  {G.}~\bibnamefont {Malinowski}},\ }\bibfield  {title} {\bibinfo {title}
  {Damping of standing spin waves in bismuth-substituted yttrium iron garnet as
  seen via the time-resolved magneto-optical {K}err effect},\ }\href
  {https://doi.org/10.1103/PhysRevApplied.12.044006} {\bibfield  {journal}
  {\bibinfo  {journal} {Phys. Rev. Appl.}\ }\textbf {\bibinfo {volume} {12}},\
  \bibinfo {pages} {044006} (\bibinfo {year} {2019})}\BibitemShut {NoStop}%
\bibitem [{\citenamefont {Bhatti}\ \emph {et~al.}(2026)\citenamefont {Bhatti},
  \citenamefont {Bhatti}, \citenamefont {Carr\'et\'ero}, \citenamefont {Cros},
  \citenamefont {Bortolotti}, \citenamefont {Lebrun}, \citenamefont {Mantion},
  \citenamefont {Reyren},\ and\ \citenamefont {Anane}}]{Bhatti2026}%
  \BibitemOpen
  \bibfield  {author} {\bibinfo {author} {\bibfnamefont {I.~N.}\ \bibnamefont
  {Bhatti}}, \bibinfo {author} {\bibfnamefont {I.~N.}\ \bibnamefont {Bhatti}},
  \bibinfo {author} {\bibfnamefont {C.}~\bibnamefont {Carr\'et\'ero}}, \bibinfo
  {author} {\bibfnamefont {V.}~\bibnamefont {Cros}}, \bibinfo {author}
  {\bibfnamefont {P.}~\bibnamefont {Bortolotti}}, \bibinfo {author}
  {\bibfnamefont {R.}~\bibnamefont {Lebrun}}, \bibinfo {author} {\bibfnamefont
  {S.}~\bibnamefont {Mantion}}, \bibinfo {author} {\bibfnamefont
  {N.}~\bibnamefont {Reyren}},\ and\ \bibinfo {author} {\bibfnamefont
  {A.}~\bibnamefont {Anane}},\ }\bibfield  {title} {\bibinfo {title} {Tailoring
  magnetic damping and anisotropy in bismuth-doped yttrium iron garnet
  ({Bi}$_{1}${Y}$_{2}${Fe}$_{5}${O}$_{12}$) thin films},\ }\href
  {https://doi.org/10.1103/k842-sxhh} {\bibfield  {journal} {\bibinfo
  {journal} {Phys. Rev. Mater.}\ }\textbf {\bibinfo {volume} {10}},\ \bibinfo
  {pages} {064410} (\bibinfo {year} {2026})}\BibitemShut {NoStop}%
\bibitem [{\citenamefont {Luttinger}\ and\ \citenamefont
  {Tisza}(1946)}]{Luttinger1946}%
  \BibitemOpen
  \bibfield  {author} {\bibinfo {author} {\bibfnamefont {J.~M.}\ \bibnamefont
  {Luttinger}}\ and\ \bibinfo {author} {\bibfnamefont {L.}~\bibnamefont
  {Tisza}},\ }\bibfield  {title} {\bibinfo {title} {Theory of dipole
  interaction in crystals},\ }\href {https://doi.org/10.1103/PhysRev.70.954}
  {\bibfield  {journal} {\bibinfo  {journal} {Phys. Rev.}\ }\textbf {\bibinfo
  {volume} {70}},\ \bibinfo {pages} {954} (\bibinfo {year} {1946})}\BibitemShut
  {NoStop}%
\bibitem [{\citenamefont {Luttinger}(1951)}]{Luttinger1951}%
  \BibitemOpen
  \bibfield  {author} {\bibinfo {author} {\bibfnamefont {J.~M.}\ \bibnamefont
  {Luttinger}},\ }\bibfield  {title} {\bibinfo {title} {A note on the ground
  state in antiferromagnetics},\ }\href
  {https://doi.org/10.1103/PhysRev.81.1015} {\bibfield  {journal} {\bibinfo
  {journal} {Phys. Rev.}\ }\textbf {\bibinfo {volume} {81}},\ \bibinfo {pages}
  {1015} (\bibinfo {year} {1951})}\BibitemShut {NoStop}%
\bibitem [{\citenamefont {Holstein}\ and\ \citenamefont
  {Primakoff}(1940)}]{Holstein1940}%
  \BibitemOpen
  \bibfield  {author} {\bibinfo {author} {\bibfnamefont {T.}~\bibnamefont
  {Holstein}}\ and\ \bibinfo {author} {\bibfnamefont {H.}~\bibnamefont
  {Primakoff}},\ }\bibfield  {title} {\bibinfo {title} {Field dependence of the
  intrinsic domain magnetization of a ferromagnet},\ }\href
  {https://doi.org/10.1103/PhysRev.58.1098} {\bibfield  {journal} {\bibinfo
  {journal} {Phys. Rev.}\ }\textbf {\bibinfo {volume} {58}},\ \bibinfo {pages}
  {1098} (\bibinfo {year} {1940})}\BibitemShut {NoStop}%
\bibitem [{\citenamefont {Ashcroft}\ and\ \citenamefont
  {Mermin}(1976)}]{Ashcroft1978}%
  \BibitemOpen
  \bibfield  {author} {\bibinfo {author} {\bibfnamefont {N.~W.}\ \bibnamefont
  {Ashcroft}}\ and\ \bibinfo {author} {\bibfnamefont {N.~D.}\ \bibnamefont
  {Mermin}},\ }\href@noop {} {\emph {\bibinfo {title} {{Solid state
  physics}}}}\ (\bibinfo  {publisher} {Holt, Rinehart and Winston},\ \bibinfo
  {address} {New York, NY},\ \bibinfo {year} {1976})\BibitemShut {NoStop}%
\bibitem [{\citenamefont {Colpa}(1978)}]{Colpa1978}%
  \BibitemOpen
  \bibfield  {author} {\bibinfo {author} {\bibfnamefont {J.}~\bibnamefont
  {Colpa}},\ }\bibfield  {title} {\bibinfo {title} {Diagonalization of the
  quadratic boson hamiltonian},\ }\href
  {https://doi.org/https://doi.org/10.1016/0378-4371(78)90160-7} {\bibfield
  {journal} {\bibinfo  {journal} {Physica A: Statistical Mechanics and its
  Applications}\ }\textbf {\bibinfo {volume} {93}},\ \bibinfo {pages} {327}
  (\bibinfo {year} {1978})}\BibitemShut {NoStop}%
\bibitem [{SM()}]{SM}%
  \BibitemOpen
  \href@noop {} {}\bibinfo {note} {The Supplemental Material at [URL inserted
  by publisher] contains Refs.~\cite{Hoyer2025a, Bradley2010}, and further
  details and additional supporting data.}\BibitemShut {Stop}%
\bibitem [{\citenamefont {Hoyer}\ \emph {et~al.}(2025)\citenamefont {Hoyer},
  \citenamefont {Jaeschke-Ubiergo}, \citenamefont {Ahn}, \citenamefont
  {\ifmmode~\check{S}\else \v{S}\fi{}mejkal},\ and\ \citenamefont
  {Mook}}]{Hoyer2025a}%
  \BibitemOpen
  \bibfield  {author} {\bibinfo {author} {\bibfnamefont {R.}~\bibnamefont
  {Hoyer}}, \bibinfo {author} {\bibfnamefont {R.}~\bibnamefont
  {Jaeschke-Ubiergo}}, \bibinfo {author} {\bibfnamefont {K.-H.}\ \bibnamefont
  {Ahn}}, \bibinfo {author} {\bibfnamefont {L.}~\bibnamefont
  {\ifmmode~\check{S}\else \v{S}\fi{}mejkal}},\ and\ \bibinfo {author}
  {\bibfnamefont {A.}~\bibnamefont {Mook}},\ }\bibfield  {title} {\bibinfo
  {title} {Spontaneous crystal thermal {H}all effect in insulating
  altermagnets},\ }\href {https://doi.org/10.1103/PhysRevB.111.L020412}
  {\bibfield  {journal} {\bibinfo  {journal} {Phys. Rev. B}\ }\textbf {\bibinfo
  {volume} {111}},\ \bibinfo {pages} {L020412} (\bibinfo {year}
  {2025})}\BibitemShut {NoStop}%
\bibitem [{\citenamefont {Kamra}\ \emph {et~al.}(2017)\citenamefont {Kamra},
  \citenamefont {Agrawal},\ and\ \citenamefont {Belzig}}]{Kamra2017}%
  \BibitemOpen
  \bibfield  {author} {\bibinfo {author} {\bibfnamefont {A.}~\bibnamefont
  {Kamra}}, \bibinfo {author} {\bibfnamefont {U.}~\bibnamefont {Agrawal}},\
  and\ \bibinfo {author} {\bibfnamefont {W.}~\bibnamefont {Belzig}},\
  }\bibfield  {title} {\bibinfo {title} {Noninteger-spin magnonic excitations
  in untextured magnets},\ }\href {https://doi.org/10.1103/PhysRevB.96.020411}
  {\bibfield  {journal} {\bibinfo  {journal} {Phys. Rev. B}\ }\textbf {\bibinfo
  {volume} {96}},\ \bibinfo {pages} {020411(R)} (\bibinfo {year}
  {2017})}\BibitemShut {NoStop}%
\bibitem [{\citenamefont {Lee}\ \emph {et~al.}(2017)\citenamefont {Lee},
  \citenamefont {Brangham}, \citenamefont {Cheng}, \citenamefont {White},
  \citenamefont {Ruane}, \citenamefont {Esser}, \citenamefont {McComb},
  \citenamefont {Hammel},\ and\ \citenamefont {Yang}}]{lee_metallic_2017}%
  \BibitemOpen
  \bibfield  {author} {\bibinfo {author} {\bibfnamefont {A.~J.}\ \bibnamefont
  {Lee}}, \bibinfo {author} {\bibfnamefont {J.~T.}\ \bibnamefont {Brangham}},
  \bibinfo {author} {\bibfnamefont {Y.}~\bibnamefont {Cheng}}, \bibinfo
  {author} {\bibfnamefont {S.~P.}\ \bibnamefont {White}}, \bibinfo {author}
  {\bibfnamefont {W.~T.}\ \bibnamefont {Ruane}}, \bibinfo {author}
  {\bibfnamefont {B.~D.}\ \bibnamefont {Esser}}, \bibinfo {author}
  {\bibfnamefont {D.~W.}\ \bibnamefont {McComb}}, \bibinfo {author}
  {\bibfnamefont {P.~C.}\ \bibnamefont {Hammel}},\ and\ \bibinfo {author}
  {\bibfnamefont {F.}~\bibnamefont {Yang}},\ }\bibfield  {title} {\bibinfo
  {title} {Metallic ferromagnetic films with magnetic damping under $1.4 \times
  10^{-3}$},\ }\href {https://doi.org/10.1038/s41467-017-00332-x} {\bibfield
  {journal} {\bibinfo  {journal} {Nature Communications}\ }\textbf {\bibinfo
  {volume} {8}},\ \bibinfo {pages} {234} (\bibinfo {year} {2017})}\BibitemShut
  {NoStop}%
\bibitem [{\citenamefont {Das}\ \emph {et~al.}(2024)\citenamefont {Das},
  \citenamefont {Mansell}, \citenamefont {Flaj\ifmmode~\check{s}\else
  \v{s}\fi{}man}, \citenamefont {Yao}, \citenamefont {van~der Jagt},
  \citenamefont {Chen}, \citenamefont {Ravelosona}, \citenamefont
  {Herrera~Diez},\ and\ \citenamefont {van
  Dijken}}]{PhysRevMaterials.8.114419}%
  \BibitemOpen
  \bibfield  {author} {\bibinfo {author} {\bibfnamefont {S.}~\bibnamefont
  {Das}}, \bibinfo {author} {\bibfnamefont {R.}~\bibnamefont {Mansell}},
  \bibinfo {author} {\bibfnamefont {L.~c.~v.}\ \bibnamefont
  {Flaj\ifmmode~\check{s}\else \v{s}\fi{}man}}, \bibinfo {author}
  {\bibfnamefont {L.}~\bibnamefont {Yao}}, \bibinfo {author} {\bibfnamefont
  {J.~W.}\ \bibnamefont {van~der Jagt}}, \bibinfo {author} {\bibfnamefont
  {S.}~\bibnamefont {Chen}}, \bibinfo {author} {\bibfnamefont {D.}~\bibnamefont
  {Ravelosona}}, \bibinfo {author} {\bibfnamefont {L.}~\bibnamefont
  {Herrera~Diez}},\ and\ \bibinfo {author} {\bibfnamefont {S.}~\bibnamefont
  {van Dijken}},\ }\bibfield  {title} {\bibinfo {title} {Tuning of
  perpendicular magnetic anisotropy in bi-substituted yttrium iron garnet films
  by {${\mathrm{He}}^{+}$} ion irradiation},\ }\href
  {https://doi.org/10.1103/PhysRevMaterials.8.114419} {\bibfield  {journal}
  {\bibinfo  {journal} {Phys. Rev. Mater.}\ }\textbf {\bibinfo {volume} {8}},\
  \bibinfo {pages} {114419} (\bibinfo {year} {2024})}\BibitemShut {NoStop}%
\bibitem [{\citenamefont {Florio}\ \emph {et~al.}(2026)\citenamefont {Florio},
  \citenamefont {Vitali}, \citenamefont {Levati}, \citenamefont {Ishola},
  \citenamefont {Mavilla}, \citenamefont {Lecis}, \citenamefont {Dubs},
  \citenamefont {Bertacco}, \citenamefont {Madami}, \citenamefont {Tacchi},
  \citenamefont {Petti},\ and\ \citenamefont
  {Albisetti}}]{florio_programmable_2026}%
  \BibitemOpen
  \bibfield  {author} {\bibinfo {author} {\bibfnamefont {P.}~\bibnamefont
  {Florio}}, \bibinfo {author} {\bibfnamefont {M.}~\bibnamefont {Vitali}},
  \bibinfo {author} {\bibfnamefont {V.}~\bibnamefont {Levati}}, \bibinfo
  {author} {\bibfnamefont {R.~M.}\ \bibnamefont {Ishola}}, \bibinfo {author}
  {\bibfnamefont {L.~C.}\ \bibnamefont {Mavilla}}, \bibinfo {author}
  {\bibfnamefont {N.}~\bibnamefont {Lecis}}, \bibinfo {author} {\bibfnamefont
  {C.}~\bibnamefont {Dubs}}, \bibinfo {author} {\bibfnamefont {R.}~\bibnamefont
  {Bertacco}}, \bibinfo {author} {\bibfnamefont {M.}~\bibnamefont {Madami}},
  \bibinfo {author} {\bibfnamefont {S.}~\bibnamefont {Tacchi}}, \bibinfo
  {author} {\bibfnamefont {D.}~\bibnamefont {Petti}},\ and\ \bibinfo {author}
  {\bibfnamefont {E.}~\bibnamefont {Albisetti}},\ }\href
  {https://arxiv.org/abs/2605.00290} {\bibinfo {title} {Programmable integrated
  magnonic meshes}} (\bibinfo {year} {2026}),\ \Eprint
  {https://arxiv.org/abs/2605.00290} {arXiv:2605.00290 [physics.app-ph]}
  \BibitemShut {NoStop}%
\bibitem [{\citenamefont {Levati}\ \emph {et~al.}(2025)\citenamefont {Levati},
  \citenamefont {Vitali}, \citenamefont {Del~Giacco}, \citenamefont {Pellizzi},
  \citenamefont {Silvani}, \citenamefont {Ciaccarini~Mavilla}, \citenamefont
  {Madami}, \citenamefont {Biancardi}, \citenamefont {Girardi}, \citenamefont
  {Panzeri}, \citenamefont {Florio}, \citenamefont {Cocconcelli}, \citenamefont
  {Breitbach}, \citenamefont {Pirro}, \citenamefont {Rovatti}, \citenamefont
  {Lecis}, \citenamefont {Maspero}, \citenamefont {Bertacco}, \citenamefont
  {Corrielli}, \citenamefont {Osellame}, \citenamefont {Russo}, \citenamefont
  {Li~Bassi}, \citenamefont {Tacchi}, \citenamefont {Petti},\ and\
  \citenamefont {Albisetti}}]{levati_three-dimensional_2025}%
  \BibitemOpen
  \bibfield  {author} {\bibinfo {author} {\bibfnamefont {V.}~\bibnamefont
  {Levati}}, \bibinfo {author} {\bibfnamefont {M.}~\bibnamefont {Vitali}},
  \bibinfo {author} {\bibfnamefont {A.}~\bibnamefont {Del~Giacco}}, \bibinfo
  {author} {\bibfnamefont {N.}~\bibnamefont {Pellizzi}}, \bibinfo {author}
  {\bibfnamefont {R.}~\bibnamefont {Silvani}}, \bibinfo {author} {\bibfnamefont
  {L.}~\bibnamefont {Ciaccarini~Mavilla}}, \bibinfo {author} {\bibfnamefont
  {M.}~\bibnamefont {Madami}}, \bibinfo {author} {\bibfnamefont
  {I.}~\bibnamefont {Biancardi}}, \bibinfo {author} {\bibfnamefont
  {D.}~\bibnamefont {Girardi}}, \bibinfo {author} {\bibfnamefont
  {M.}~\bibnamefont {Panzeri}}, \bibinfo {author} {\bibfnamefont
  {P.}~\bibnamefont {Florio}}, \bibinfo {author} {\bibfnamefont
  {M.}~\bibnamefont {Cocconcelli}}, \bibinfo {author} {\bibfnamefont
  {D.}~\bibnamefont {Breitbach}}, \bibinfo {author} {\bibfnamefont
  {P.}~\bibnamefont {Pirro}}, \bibinfo {author} {\bibfnamefont
  {L.}~\bibnamefont {Rovatti}}, \bibinfo {author} {\bibfnamefont
  {N.}~\bibnamefont {Lecis}}, \bibinfo {author} {\bibfnamefont
  {F.}~\bibnamefont {Maspero}}, \bibinfo {author} {\bibfnamefont
  {R.}~\bibnamefont {Bertacco}}, \bibinfo {author} {\bibfnamefont
  {G.}~\bibnamefont {Corrielli}}, \bibinfo {author} {\bibfnamefont
  {R.}~\bibnamefont {Osellame}}, \bibinfo {author} {\bibfnamefont
  {V.}~\bibnamefont {Russo}}, \bibinfo {author} {\bibfnamefont
  {A.}~\bibnamefont {Li~Bassi}}, \bibinfo {author} {\bibfnamefont
  {S.}~\bibnamefont {Tacchi}}, \bibinfo {author} {\bibfnamefont
  {D.}~\bibnamefont {Petti}},\ and\ \bibinfo {author} {\bibfnamefont
  {E.}~\bibnamefont {Albisetti}},\ }\bibfield  {title} {\bibinfo {title}
  {Three-dimensional nanoscale control of magnetism in crystalline yttrium iron
  garnet},\ }\href {https://doi.org/10.1038/s41467-025-64630-5} {\bibfield
  {journal} {\bibinfo  {journal} {Nature Communications}\ }\textbf {\bibinfo
  {volume} {16}},\ \bibinfo {pages} {9602} (\bibinfo {year}
  {2025})}\BibitemShut {NoStop}%
\bibitem [{\citenamefont {Bradley}\ and\ \citenamefont
  {Cracknell}(2010)}]{Bradley2010}%
  \BibitemOpen
  \bibfield  {author} {\bibinfo {author} {\bibfnamefont {C.}~\bibnamefont
  {Bradley}}\ and\ \bibinfo {author} {\bibfnamefont {A.}~\bibnamefont
  {Cracknell}},\ }\href@noop {} {\emph {\bibinfo {title} {The Mathematical
  Theory of Symmetry in Solids: Representation Theory for Point Groups and
  Space Groups}}}\ (\bibinfo  {publisher} {Oxford University Press},\ \bibinfo
  {year} {2010})\BibitemShut {NoStop}%
\bibitem [{\citenamefont {Mortada}\ \emph {et~al.}(2025)\citenamefont
  {Mortada}, \citenamefont {Verba}, \citenamefont {Wang}, \citenamefont
  {Pirro},\ and\ \citenamefont {Hamadeh}}]{mortada_nonreciprocal_2025}%
  \BibitemOpen
  \bibfield  {author} {\bibinfo {author} {\bibfnamefont {H.}~\bibnamefont
  {Mortada}}, \bibinfo {author} {\bibfnamefont {R.}~\bibnamefont {Verba}},
  \bibinfo {author} {\bibfnamefont {Q.}~\bibnamefont {Wang}}, \bibinfo {author}
  {\bibfnamefont {P.}~\bibnamefont {Pirro}},\ and\ \bibinfo {author}
  {\bibfnamefont {A.~A.}\ \bibnamefont {Hamadeh}},\ }\bibfield  {title}
  {\bibinfo {title} {Nonreciprocal spin waves in out-of-plane magnetized
  coupled waveguides reconfigured by domain wall displacements},\ }\href
  {https://doi.org/https://doi.org/10.1002/aelm.202500575} {\bibfield
  {journal} {\bibinfo  {journal} {Advanced Electronic Materials}\ }\textbf
  {\bibinfo {volume} {11}},\ \bibinfo {pages} {e00575} (\bibinfo {year}
  {2025})}\BibitemShut {NoStop}%
\bibitem [{\citenamefont {Vansteenkiste}\ \emph {et~al.}(2014)\citenamefont
  {Vansteenkiste}, \citenamefont {Leliaert}, \citenamefont {Dvornik},
  \citenamefont {Helsen}, \citenamefont {Garcia-Sanchez},\ and\ \citenamefont
  {Van~Waeyenberge}}]{vansteenkiste_design_2014}%
  \BibitemOpen
  \bibfield  {author} {\bibinfo {author} {\bibfnamefont {A.}~\bibnamefont
  {Vansteenkiste}}, \bibinfo {author} {\bibfnamefont {J.}~\bibnamefont
  {Leliaert}}, \bibinfo {author} {\bibfnamefont {M.}~\bibnamefont {Dvornik}},
  \bibinfo {author} {\bibfnamefont {M.}~\bibnamefont {Helsen}}, \bibinfo
  {author} {\bibfnamefont {F.}~\bibnamefont {Garcia-Sanchez}},\ and\ \bibinfo
  {author} {\bibfnamefont {B.}~\bibnamefont {Van~Waeyenberge}},\ }\bibfield
  {title} {\bibinfo {title} {The design and verification of {M}u{M}ax3},\
  }\href {https://doi.org/10.1063/1.4899186} {\bibfield  {journal} {\bibinfo
  {journal} {AIP Advances}\ }\textbf {\bibinfo {volume} {4}},\ \bibinfo {pages}
  {107133} (\bibinfo {year} {2014})}\BibitemShut {NoStop}%
\bibitem [{ait()}]{aithericon}%
  \BibitemOpen
  \href {www.aithericon.com} {\bibinfo {title} {See www.aithericon.com for
  further information and details.}}\BibitemShut {Stop}%
\end{thebibliography}%


\begin{thebibliography}{2}%
\makeatletter
\providecommand \@ifxundefined [1]{%
 \@ifx{#1\undefined}
}%
\providecommand \@ifnum [1]{%
 \ifnum #1\expandafter \@firstoftwo
 \else \expandafter \@secondoftwo
 \fi
}%
\providecommand \@ifx [1]{%
 \ifx #1\expandafter \@firstoftwo
 \else \expandafter \@secondoftwo
 \fi
}%
\providecommand \natexlab [1]{#1}%
\providecommand \enquote  [1]{``#1''}%
\providecommand \bibnamefont  [1]{#1}%
\providecommand \bibfnamefont [1]{#1}%
\providecommand \citenamefont [1]{#1}%
\providecommand \href@noop [0]{\@secondoftwo}%
\providecommand \href [0]{\begingroup \@sanitize@url \@href}%
\providecommand \@href[1]{\@@startlink{#1}\@@href}%
\providecommand \@@href[1]{\endgroup#1\@@endlink}%
\providecommand \@sanitize@url [0]{\catcode `\\12\catcode `\$12\catcode
  `\&12\catcode `\#12\catcode `\^12\catcode `\_12\catcode `\%12\relax}%
\providecommand \@@startlink[1]{}%
\providecommand \@@endlink[0]{}%
\providecommand \url  [0]{\begingroup\@sanitize@url \@url }%
\providecommand \@url [1]{\endgroup\@href {#1}{\urlprefix }}%
\providecommand \urlprefix  [0]{URL }%
\providecommand \Eprint [0]{\href }%
\providecommand \doibase [0]{https://doi.org/}%
\providecommand \selectlanguage [0]{\@gobble}%
\providecommand \bibinfo  [0]{\@secondoftwo}%
\providecommand \bibfield  [0]{\@secondoftwo}%
\providecommand \translation [1]{[#1]}%
\providecommand \BibitemOpen [0]{}%
\providecommand \bibitemStop [0]{}%
\providecommand \bibitemNoStop [0]{.\EOS\space}%
\providecommand \EOS [0]{\spacefactor3000\relax}%
\providecommand \BibitemShut  [1]{\csname bibitem#1\endcsname}%
\let\auto@bib@innerbib\@empty
\bibitem [{\citenamefont {Hoyer}\ \emph {et~al.}(2025)\citenamefont {Hoyer},
  \citenamefont {Jaeschke-Ubiergo}, \citenamefont {Ahn}, \citenamefont
  {\ifmmode~\check{S}\else \v{S}\fi{}mejkal},\ and\ \citenamefont
  {Mook}}]{Hoyer2025a}%
  \BibitemOpen
  \bibfield  {author} {\bibinfo {author} {\bibfnamefont {R.}~\bibnamefont
  {Hoyer}}, \bibinfo {author} {\bibfnamefont {R.}~\bibnamefont
  {Jaeschke-Ubiergo}}, \bibinfo {author} {\bibfnamefont {K.-H.}\ \bibnamefont
  {Ahn}}, \bibinfo {author} {\bibfnamefont {L.}~\bibnamefont
  {\ifmmode~\check{S}\else \v{S}\fi{}mejkal}},\ and\ \bibinfo {author}
  {\bibfnamefont {A.}~\bibnamefont {Mook}},\ }\bibfield  {title} {\bibinfo
  {title} {Spontaneous crystal thermal {H}all effect in insulating
  altermagnets},\ }\href {https://doi.org/10.1103/PhysRevB.111.L020412}
  {\bibfield  {journal} {\bibinfo  {journal} {Phys. Rev. B}\ }\textbf {\bibinfo
  {volume} {111}},\ \bibinfo {pages} {L020412} (\bibinfo {year}
  {2025})}\BibitemShut {NoStop}%
\bibitem [{\citenamefont {Bradley}\ and\ \citenamefont
  {Cracknell}(2010)}]{Bradley2010}%
  \BibitemOpen
  \bibfield  {author} {\bibinfo {author} {\bibfnamefont {C.}~\bibnamefont
  {Bradley}}\ and\ \bibinfo {author} {\bibfnamefont {A.}~\bibnamefont
  {Cracknell}},\ }\href@noop {} {\emph {\bibinfo {title} {The Mathematical
  Theory of Symmetry in Solids: Representation Theory for Point Groups and
  Space Groups}}}\ (\bibinfo  {publisher} {Oxford University Press},\ \bibinfo
  {year} {2010})\BibitemShut {NoStop}%
\end{thebibliography}%

\appendix
\section{End matter}

\label{sec:demag-tensor}
\setcounter{equation}{0}
\renewcommand{\theequation}{A\arabic{equation}}
\textit{Mutual demagnetizing tensor}---We assume uniform magnetization in the planar ellipses. Their height $h \ll 1/k$ with $k$ being the in-plane wavelength of the magnon. This leads to a constant magnetization along the  $z$ direction and to magnon modes that are independent of $z$.
To derive the mutual demagnetizing tensor between two islands $i$ and $j$, we follow the derivation in Ref.~\cite{Verba2012} starting with
\begin{align}
    N_{\mu \nu}^{\alpha \beta} (\bm{r}_i - \bm{r}_j) = \frac{1}{V_i} \int \frac{d^3 \bm{\kappa}}{(2 \uppi)^3} D_i^\alpha (\bm{\kappa}) \left(D_j^\beta (\bm{\kappa})\right)^* \frac{\kappa_\mu \kappa_\nu}{\kappa^2} \mathrm{e}^{\mathrm{i} \bm{\kappa} \cdot (\bm{r}_i - \bm{r}_j)},
    \label{eq:demag_2}
\end{align}
where $\mu$, $\nu$ are the Cartesian vector components $x$, $y$, $z$, and $V_i$ is the volume of the nanoisland at site $i$. 
We integrate over the wave vector $\bm{\kappa}$ in three-dimensional reciprocal space. The shape amplitude reads
\begin{align}
    D_i^\alpha (\bm{\kappa}) = \int_{V_i^\alpha} \mathrm{e}^{-\mathrm{i} \bm{\kappa} \cdot \bm{r}} d^3 \bm{r}.
\end{align}
The magnetic islands on both sublattices have identical shape, rotated by $90^\circ$. Having the same height, the $z$ component of the vector $\bm{r}_i - \bm{r}_j$ vanishes. We write $\bm{\kappa} = \bm{k} + \kappa_z \bm{e}_z = k_x \bm{e}_x + k_y \bm{e}_y + \kappa_z \bm{e}_z$ and integrate Eq.~\eqref{eq:demag_2} over the wave vector $ \kappa_z$ perpendicular to the $xy$ plane to find
\begin{align}
    \hat{\bm{N}}^{\alpha \beta}(\bm{r}_i - \bm{r}_j) = \int \hat{\bm{N}}_{\bm{k}}^{\alpha \beta} \mathrm{e}^{\mathrm{i} \bm{k} \cdot (\bm{r}_i - \bm{r}_j)} \frac{d^2\bm{k}}{\left(2 \uppi \right)^2}.
    \label{eq:demag_tensor}
\end{align}
The Fourier image of the demagnetizing tensor reads
\begin{align}
    \hat{\bm{N}}_{\bm{k}}^{\alpha \beta} = \frac{\sigma_{\bm{k}}^\alpha \left(\sigma_{\bm{k}}^\beta\right)^*}{S} \begin{pmatrix}
    \frac{k_x^2}{k^2} f(k h) & \frac{k_{x} k_y}{k^2} f(k h) &  0 \\
    \frac{k_{x}k_y}{k^2} f(k h) & \frac{k_y^2}{k^2} f(k h) & 0 \\
    0 & 0 & 1- f(k h)
    \end{pmatrix},
    \label{eq:demag_tensor_k}
\end{align}
with $f(k h) = 1- \frac{1 - \mathrm{e}^{-k h}}{k h}$ and $h$ as the height of an island and its area $S = \uppi \, h_1 h_2$. The Fourier image of the island's shape on sublattice A reads
\begin{align}
    \sigma_{\bm{k}}^{\text{A}} &= S \frac{2 \mathrm{J}_1\left(\sqrt{k_x^2 h_1^2 + k_y^2 h_2^2}\right)}{\sqrt{k_x^2 h_1^2 + k_y^2 h_2^2}},
    \label{eq:shape_amplitude}
\end{align}
with $\mathrm{J}_1(x)$ as the Bessel function of first kind and $h_1$ ($h_2$) the long (short) half axis of the ellipses. For sublattice B we exchange $h_1 \leftrightarrow h_2$.

\setcounter{equation}{0}
\renewcommand{\theequation}{B\arabic{equation}}
\label{sec:luttinger-tisza}
\textit{Ground state verification}---We verify the staggered out-of-plane magnetization configuration as the ground state with the Luttinger-Tisza \cite{Luttinger1946,Luttinger1951} method. We directly Fourier transform the unitary macrospins in Eq.~\eqref{eq:HAM} with 
\begin{align}
    \bm{m}_{i} = \frac{1}{\sqrt{N}} \sum_{\bm{q}} \mathrm{e}^{\mathrm{i} \bm{q} \cdot \left( \bm{r}_i + \bm{\delta}_{\mu} \right)} \bm{m}_{\bm{q}},
\end{align}
where $\mu$ denotes the sublattice, and find the $6\times 6$ interaction matrix
\begin{align}
    \hat{\bm{N}}_{\bm{q}} = \mu_0 M_{\text{s}}V  \begin{pmatrix}
        \hat{\bm{F}}_{\bm{q}}^{\text{AA}} & \hat{\bm{F}}_{\bm{q}}^{\text{AB}} \\
        \hat{\bm{F}}_{\bm{q}}^{\text{BA}} & \hat{\bm{F}}_{\bm{q}}^{\text{BB}} 
    \end{pmatrix} 
    -  K V \left[\mathbbm{1}_2 \otimes
        \diag(0, 0, 1) \right],
\end{align}
where $\hat{\bm{F}}_{\bm{q}}^{\alpha \beta} = \sum_{\Delta(\bm{r})} \hat{\bm{N}}^{\alpha \beta} \Delta(\bm{r}) \mathrm{e}^{-\mathrm{i} \bm{q} \cdot \bm{r}_{\alpha \beta}}$ is a $3\times3$ matrix. We then diagonalize the interaction matrix for every $\bm{q}$ point in the first BZ for each parameter set of the three cases (one antiferromagnetic and two altermagnetic configurations). For these parameter sets, the minimum eigenvalue lies at $\bm{q}=(0, 0)$. The corresponding eigenvector holds information about the spatial orientation of the macrospins, which we find to be $\hat{\bm{m}}_{\text{A}}=(0,0,1)$ and $\hat{\bm{m}}_{\text{B}}=(0,0,-1)$. This confirms the staggered out-of-plane magnetization of the islands.

\label{sec:HP-trafo}
\setcounter{equation}{0}
\renewcommand{\theequation}{C\arabic{equation}}
\textit{Macrospin-to-boson transformation}---The total magnetic moment of island $i$ is $ \bm{\mu}_i = V M_{\mathrm{s}} \bm{m}_i = V \bm{M}_i$, where $\bm{M}_i= M_{\text{s}} \bm{m}_i$ is the macrospin magnetization. The magnetic moment connects to the spin with $\bm{\mu}_i = -|\gamma| \bm{S}_i$, where $\gamma= \frac{g}{\hbar} \mu_\text{B}$ is the gyromagnetic ratio, $\mu_\text{B}$ the Bohr magneton and $g$ the Landé factor. We then find
\begin{align}
    \bm{m}_i = -\frac{|\gamma|}{M_\text{s} V} \bm{S}_i
    \label{eq:m-s}
\end{align}
and take the absolute value to find $S = \frac{M_\text{s} V}{|\gamma| \hbar}$. We insert the Holstein-Primakoff transformation~\cite{Holstein1940} assuming collinearity
\begin{subequations}
\begin{align}
    \hat{S}_{i,x}^\alpha & \approx \frac{\sqrt{2 S}\hbar}{2} \left( \hat{c}_i + \hat{c}_i^\dagger \right),\\
    \hat{S}_{i,y}^\alpha & \approx (- 1)^n \frac{\sqrt{2 S}\hbar}{2 \mathrm{i}} \left( \hat{c}_i - \hat{c}_i^\dagger \right),\\
    \hat{S}_{i,z}^\alpha & = (- 1)^n \hbar \left(S - \hat{c}_i^\dagger \hat{c}_i \right)
\end{align}
\end{subequations}
into Eq.~\eqref{eq:m-s}.  We set $n$ even (odd) and $\hat{c}_i=\hat{a}_i$ ($\hat{b}_i$) for sublattice A (B) and find the macrospin-to-boson transformation from the main text
\begin{subequations}
\begin{align}
    \hat{m}_{i,x}^\alpha &\approx -\frac{1}{\sqrt{2 S}}\left( \hat{c}_i +  \hat{c}_i^\dagger\right), \\
    \hat{m}_{i,y}^\alpha &\approx (- 1)^n  \frac{\mathrm{i}}{\sqrt{2 S}}\left( \hat{c}_i - \hat{c}_i^\dagger \right), \\
    \hat{m}_{i,z}^\alpha &= (- 1)^n \left(-1 + \frac{1}{S}  \hat{c}_i^\dagger \hat{c}_i \right).
\end{align}
\end{subequations}

\begin{table}
\centering
\caption{Parameters used in Fig.~\ref{fig:model-1}. The saturation magnetization $M_\text{s}$ and the easy-axis anisotropy $K$ are taken from Ref.~\cite{mortada_nonreciprocal_2025}.}
\label{tab:parameters-1}
\begin{tabular} {cccc} 
\toprule
Parameter & Symbol & Value &  Unit\\
\hline 
Vacuum permeability & $\mu_0$ & $4 \uppi \times 10^{-7}$ & NA$^{-2}$ \\[0.4em] 

Saturation magnetization & $M_{\text{s}}$ &  $1.0 \times 10^{5}$ & Am$^{-1}$  \\[0.4em] 
 
Easy-axis anisotropy & $K$ & $12.5 \times 10^{3}$ & Jm$^{-3}$  \\[0.4em]

Lattice constant & $a$ & $1.1 \times 10^{-7}$ & m  \\[0.4em]

Long half-axis of island & $h_1$ & $4.0 \times 10^{-8}$ & m  \\[0.4em]

Short half-axis of island & $h_2$ & $1.5 \times 10^{-8}$ & m  \\[0.4em]

Height of island & $h$ & $3.0 \times 10^{-8}$ & m \\
\botrule
\end{tabular}
\end{table}
\label{sec:parameters}
\textit{Parameters}---We list the parameters used in the altermagnetic geometry 1 in Table~\ref{tab:parameters-1}.

\label{sec:micromagnetics_methods}
\textit{Micromagnetic simulations and signal analysis}---Simulations are performed using the GPU-accelerated software package mumax$^3$~\cite{vansteenkiste_design_2014} within the Aithericon platform~\cite{aithericon}. The simulation volume encompasses a layout of 20 lattice constants in the $x$-direction and 2 lattice constants in the $y$-direction, discretized using a cell size of $d_x \times d_y \times d_z \approx \qty{1.72}{\nano\meter} \times \qty{1.72}{\nano\meter} \times \qty{30}{\nano\meter}$. A single cell layer along the $z$-axis ($N_z = 1$) ensures uniform magnetization profile ($k_z = 0$). 

Periodic boundary conditions are applied along both the $x$ and $y$ directions to mimic an infinitely extended pattern. To suppress spin-wave reflections at the $x$ boundaries, absorbing regions are implemented by exponentially increasing the Gilbert damping parameter to $\alpha = 1$ in 60 discrete steps, each with a width of \qty{15}{\nano\meter}, at both boundaries.

For both the Bi:YIG (see Table~\ref{tab:parameters-1}) and the optimized parameter sets discussed in the main text, the exchange stiffness is fixed to $A_\text{ex} = \qty{4.2}{\pico\joule\per\meter}$ and the Gilbert damping constant is set to $\alpha = \num{1.3e-4}$ within the central active region~\cite{mortada_nonreciprocal_2025}.

After initializing the system in its antiferromagnetic ground state and relaxing to an energy minimum, standing spin waves matching the lattice periodicity ($\lambda_\text{magnon} = 2a$, $k_\text{magnon} = \uppi/a$) are excited. The excitation is driven by a spatially periodic tickle field $\vec{b}_\text{ext}(x, t) = b_0 \sin(k_\text{magnon} x + \phi_k) \text{sinc}(2\uppi f_\text{exc} (t-t_0))\hat{\vec{e}}_x$ with an amplitude $b_0 = \qty{10}{\micro\tesla}$, cut-off frequency $f_\text{exc} = \qty{10}{\giga\hertz}$, and a temporal offset $t_0 = \qty{250}{\nano\second}$. The spatial phase shift is fixed to $\phi_k = \uppi/4$, ensuring a modal overlap that allows the efficient excitation of both altermagnetic modes. The dynamic magnetization component $m_x(x,y,t)$ is collected at intervals of $\Delta t = \qty{50}{\pico\second}$ over a total duration of $T = \qty{500}{\nano\second}$. Spatial amplitude maps $|m_x(x,y,f)|$ are extracted via a fast Fourier transform (FFT) utilizing a Hann window. The mode frequencies $f_{\uparrow}$ and $f_{\downarrow}$ are subsequently evaluated via peak detection from the spatially averaged spectrum $S(f) = \langle |m_x(x,y,f)| \rangle_{x,y}$.


\end{document}


\title{Supplemental Material to: Altermagnetic Magnons in Dipolar Nanomagnet Arrays}

\author{Rhea Hoyer\,\orcidlink{0000-0003-2285-435X}}
\affiliation{Institut für Festkörpertheorie, Universität Münster, Wilhelm-Klemm-Straße 10, 48149 Münster, Germany}

\author{Ephraim Spindler\,\orcidlink{0009-0001-5656-525X}}
\affiliation{Fachbereich Physik and Landesforschungszentrum OPTIMAS, Rheinland-Pfälzische Technische Universität Kaiserslautern-Landau, 67663 Kaiserslautern, Germany}

\author{Lukas Körber\,\orcidlink{0000-0001-8332-9669}}
\affiliation{Institute for Molecules and Materials, Radboud University, Heyendaalseweg 135, 6525 AJ Nijmegen, The Netherlands}

\author{Tobias Wagner\,\orcidlink{0009-0006-6764-0979}}
\affiliation{Department of Physics, Johannes Gutenberg University Mainz, Staudingerweg 7, 55128 Mainz, Germany}

\author{Mathias Weiler\,\orcidlink{0000-0003-0537-9251}}
\affiliation{Fachbereich Physik and Landesforschungszentrum OPTIMAS, Rheinland-Pfälzische Technische Universität Kaiserslautern-Landau, 67663 Kaiserslautern, Germany}

\author{Alexander Mook\,\orcidlink{0000-0002-8599-9209}}
\affiliation{Institut für Festkörpertheorie, Universität Münster, Wilhelm-Klemm-Straße 10, 48149 Münster, Germany}

\begin{abstract}
 In this Supplemental Material we provide additional information on (\ref{sec:X-Y}) the high-symmetry points X and Y, including (\ref{sec:decoupling}) the sublattice decoupling and (\ref{sec:spin-exp-vals}) the magnon spin expectation values at these points, (\ref{sec:symmetries}) the magnetic layer group symmetries of (\ref{sec:afm}) the antiferromagnetic geometry, (\ref{sec:am-1}) the altermagnetic geometry 1 and (\ref{sec:am-2}) the altermagnetic geometry 2. Furthermore, additional data addressing (\ref{sec:anisotropy}) the dependence of the altermagnetic splitting on the perpendicular magnetic anisotropy strength as well as (\ref{sec:finite_size}) the mitigation of finite-size aspect-ratio effects using square simulation domains is provided.
\end{abstract}

\date{\today}
\maketitle

\tableofcontents
\setcounter{equation}{0}
\renewcommand{\theequation}{S\arabic{equation}}
\renewcommand{\thesection}{S\Roman{section}} 
\setcounter{table}{0}
\renewcommand{\thetable}{S\Roman{table}}
\setcounter{figure}{0}
\renewcommand{\thefigure}{S\arabic{figure}}

\section{High-symmetry points X and Y}
\label{sec:X-Y}
\subsection{Sublattice decoupling}
\label{sec:decoupling}
As claimed in the main text, the components $W_{\bm{X}} = Y_{\bm{X}}= 0$ in Eqs.~\eqref{eq:matrix-elements} at the X and Y points of a square lattice vanish, which we prove by pairwise cancellation of the sum over reciprocal lattice vectors in Eq.~\eqref{eq:F-tensor} with $\bm{G}=p_1\bm{b}_1 + p_2 \bm{b}_2$ and $p_i \in \mathbbm{Z}$ . 
The requirements for the pairwise cancellation are as follows: (1) The momenta $\bm{k}^*$ at which the tensor is evaluated need to be time-reversal invariant momenta, meaning that $ \bm{k}^* = \bm{G}_0/2 = (m_1 \bm{b}_1 + m_2\bm{b}_2 )/2$, where $\bm{G}_0$ is a reciprocal lattice vector and $m_i \in \mathbbm{Z}$. (2) The vector $\bm{r}_{\alpha\beta}$ fulfills the condition $\bm{r}_{\alpha\beta} = (n_1 \bm{a}_1 + n_2 \bm{a}_2)/2$ with $n_i \in \mathbbm{Z}$, and (3) the exponential $\mathrm{exp}[\mathrm{i} 2 \bm{k}^* \cdot \bm{r}_{\alpha\beta}]$ must equal $-1$. Let us define the momentum $\bm{Q} = \bm{k}^* + \bm{G} = \bm{G}_0/2 + \bm{G}$ and re-index the reciprocal lattice vector $\bm{G} \to -\bm{G}-\bm{G}_0$. With this new lattice vector, we send the momentum $\bm{Q} \to \bm{G}_0/2 + (-\bm{G}-\bm{G}_0) = -(\bm{G}_0/2 +\bm{G}) = -\bm{Q}$. The mutual demagnetizing tensor $\hat{N}_{\mu\nu}^{\alpha\beta}(\bm{Q}) \propto \bm{Q}_\mu \bm{Q}_\nu/Q^2$ is even under momentum reversal $\hat{N}_{\mu\nu}^{\alpha\beta}(\bm{Q}) = \hat{N}_{\mu\nu}^{\alpha\beta}(-\bm{Q})$, whereas the phase $\mathrm{exp}[-\mathrm{i}\bm{Q}\cdot \bm{r}_{\alpha \beta}]$ in Eq.~\eqref{eq:F-tensor} changes sign. Summing the phase factors pairwise gives $\mathrm{exp}[-\mathrm{i}\bm{Q}\cdot \bm{r}_{\alpha \beta}] + \mathrm{exp}[\mathrm{i}\bm{Q}\cdot \bm{r}_{\alpha \beta}] = 2 \cos(\bm{Q}\cdot \bm{r}_{\alpha \beta})$, where $\bm{Q}\cdot \bm{r}_{\alpha \beta} = (\bm{k}^* + \bm{G}) \cdot \bm{r}_{\alpha \beta}$. We use requirements (1) and (2) and find 
\begin{align}
    \bm{k}^* + \bm{G} &= (m_1 \bm{b}_1 + m_2 \bm{b}_2)/2 \cdot (n_1 \bm{a}_1 + n_2 \bm{a}_2)/2 
    + (p_1 \bm{b}_1 + p_2 \bm{b}_2)/2 \cdot (n_1 \bm{a}_1 + n_2 \bm{a}_2)/2, \nonumber \\
    & = (m_1 n_1 + m_2 n_2) \uppi/2 + l \,\uppi ,
\end{align}
where we used $\bm{b}_i \cdot \bm{a}_i =2\uppi \delta_{ij}$, and set $l = (p_1 n_1 + p_2 n_2) \in \mathbbm{Z}$. The cosine now reads
\begin{align}
    \cos( \bm{k}^* + \bm{G}) = (-1)^l \cos \bm{(} (m_1 n_1 + m_2 n_2) \uppi/2\bm{)}.
\end{align}
With requirement (3) we find that $m_1 n_1 + m_2 n_2$ must be odd, and the cosine vanishes.
We therefore find  
\begin{align}
    \hat{\bm{F}}^{\alpha \beta}_{\mu \nu}(\bm{k}^*) = 0.
\end{align}
In all three cases (antiferromagnetic configuration, altermagnetic geometry 1 and 2), we have 
\begin{align}
    \bm{k}^* = \left\{ \begin{aligned} 
  \bm{X} &=  \bm{b}_1/2\\
  \bm{Y} &=  \bm{b}_2/2\\
\end{aligned} \right. , \quad \bm{r}_{AB} = (\bm{a}_1 + \bm{a}_2)/2,
\end{align}
meaning that $m_1 n_1 + m_1 n_2 = 1$ is odd, the cosine vanishes, and therefore $W_{\bm{X}} = Y_{\bm{X}}= 0$.

\subsection{Magnon spin expectation value: Spin nonconserving vs. spin conserving cases}
\label{sec:spin-exp-vals}
To find the spin expectation values at the X point for the dipolar-coupled array studied in the main text, we set $\bm{k} = \left(\uppi, 0\right) = \bm{X}$ (the Y point follows analogously). In a basis $\vec{\Psi}^\dagger_{\vec{X}} = (\hat{a}^\dagger_{\vec{X}},\hat{a}_{-\vec{X}}, \hat{b}^\dagger_{\vec{X}}, \hat{b}_{-\vec{X}})$, we can write
\begin{align}
    \widetilde{\mathcal{H}}_{\bm{X}} = \begin{pmatrix}
        \widetilde{\mathcal{H}}_{\bm{X}}^\text{A} & 0 \\ 0 & \widetilde{\mathcal{H}}_{\bm{X}}^\text{B}
    \end{pmatrix}.
\end{align}
Two individual (2D) Bogoliubov transformations with $\left(\widetilde{\vec{\Psi}}_{\vec{X}}^\alpha\right)^\dagger = (\hat{\gamma}^\dagger_{\vec{X}},\hat{\gamma}_{-\vec{X}})$ and $\delta_{\bm{X}} = \hat{\alpha}_{\bm{X}}, \hat{\beta}_{\bm{X}}$ can be performed:
\begin{align}
    \quad \begin{pmatrix}
        \delta_{\bm{X}} \\ \delta^\dagger_{-\bm{X}}
    \end{pmatrix} &= \left(T^\alpha_{\bm{X}}\right)^{-1} \widetilde{\vec{\Psi}}^\alpha_{\vec{X}}, \quad \text{ with }
    T^\alpha_{\bm{X}}  = \begin{pmatrix}
        u_\alpha & -v_\alpha \\
        - v_\alpha^* & u_\alpha
    \end{pmatrix},
\end{align}
where the respective Bogoliubov coefficients are 
\begin{subequations}
\begin{align}
    u_\alpha^2 &= \frac{1}{2}\left(\frac{V_{\bm{X}}^\alpha}{\sqrt{(V_{\bm{X}}^\alpha)^2 - |X_{\bm{X}}^\alpha|^2}} + 1\right), \\
    |v_\alpha|^2 &= \frac{1}{2}\left(\frac{V_{\bm{X}}^\alpha}{\sqrt{(V_{\bm{X}}^\alpha)^2 - |X_{\bm{X}}^\alpha|^2}} - 1\right), \quad \text{with } v = |v| \mathrm{e}^{\mathrm{i} \mathrm{arg}(X_{\bm{X}}^\alpha)},
\end{align}
\end{subequations}
and obey the normalization $u_\alpha^2 - |v_\alpha|^2 = 1$. With $\left(T^\alpha_{\bm{X}} \right)^\dagger \widetilde{\mathcal{H}}_{\bm{X}}^\alpha T^\alpha_{\bm{X}} = \diag (\varepsilon_{\bm{X}}^\alpha, \varepsilon_{-\bm{X}}^\alpha)$, we obtain the energies
\begin{align}
\varepsilon_{\bm{X}}^\alpha &= \sqrt{(V_{\bm{X}}^\alpha )^2- | X^\alpha_{\bm{X}}|^2}.
    \label{eq:Xepsilon}
\end{align}
The magnon spin expectation value $\langle m_z^\alpha \rangle = \left[\left(T_{\bm{X}}^\alpha \right)^\dagger \Sigma^\alpha T_{\bm{X}}^\alpha \right]_{11}$, where $\Sigma^\alpha = \mp \diag(1, 1)$, is found to read 
    \begin{align}
        \langle m_z^\alpha \rangle &=  \mp ( u_\alpha^2 + |v_\alpha|^2 ) =  \mp \left(1 + 2 |v_\alpha|^2 \right) = \mp \frac{V_{\bm{X}}^\alpha}{\varepsilon_{\bm{X}}^\alpha}
    \end{align}
at the X point, where ``$-$'' (``$+$'') belongs to sublattice $\alpha = \text{A}$ ($\alpha = \text{B}$), using the normalization condition. 

In the case of a spin conserving two-atom altermagnet as in Ref.~\cite{Hoyer2025a}, we find that the Hamilton kernel can also be block diagonalized, however with different blocks in the bases ${\bm{\Phi}}^\dagger_{\bm{k},+} = \left(\hat{a}^\dagger_{\bm{k}}, \hat{b}_{-\bm{k}}\right)$ and ${\bm{\Phi}}^\dagger_{\bm{k},-} = \left(\hat{b}^\dagger_{\bm{k}}, \hat{a}_{-\bm{k}}\right)$
\begin{align}
    \mathcal{H}_{\bm{k},+} = \begin{pmatrix}
        A_{\bm{k}}^\text{A} & B_{\bm{k}}^\text{B} \\
        \left(B_{\bm{k}}^\text{B}\right)^* & A_{-\bm{k}}^\text{A}
    \end{pmatrix}, \quad
    \mathcal{H}_{\bm{k},-} = \begin{pmatrix}
        A_{\bm{k}}^\text{B} & B_{\bm{k}}^\text{A} \\
        \left(B_{\bm{k}}^\text{A}\right)^* & A_{-\bm{k}}^\text{B}
    \end{pmatrix}.
\end{align}
The Bogoliubov transformation reads
\begin{align}
    \begin{pmatrix}
        \hat{\alpha}_{\bm{k}} \\ \hat{\beta}^\dagger_{-\bm{k}}
    \end{pmatrix} &= \left(T_{\bm{k}}\right)^{-1} {\vec{\Phi}}_{\vec{k},+},  \quad \begin{pmatrix}
        \hat{\beta}_{\bm{k}} \\ \hat{\alpha}^\dagger_{-\bm{k}}
    \end{pmatrix} = \left(T_{\bm{k}}\right)^{-1} {\vec{\Phi}}_{\vec{k},-},
\end{align}
with $T_{\bm{k}}  = \begin{pmatrix}
        u^* & -v \\
        - v^* & u
    \end{pmatrix}$
and the normalization $|u|^2 - |v|^2 = 1$. The blocks are diagonalized as   
\begin{subequations}
\begin{align}
    T_{\bm{k}}^\dagger \mathcal{H}_{\bm{k},+} T_{\bm{k}} &= \diag (\varepsilon_{\bm{k},\alpha}, \varepsilon_{-\bm{k},\beta}), \\
    T_{\bm{k}}^\dagger \mathcal{H}_{\bm{k},-} T_{\bm{k}} &= \diag (\varepsilon_{\bm{k},\beta}, \varepsilon_{-\bm{k},\alpha}).
\end{align}
\end{subequations}
We compute the spin expectation values with $\langle m_z^\alpha \rangle = \left[T_{\bm{k}}^\dagger \Sigma^\alpha T_{\bm{k}} \right]_{11}$, where $\Sigma^\alpha = \mp \diag(-1, 1)$, which gives
    \begin{align}
        \langle m_z^\alpha \rangle  &= \mp \left(|u|^2 -|v|^2 \right) = \mp 1,
    \end{align}
where ``$-$'' (``$+$'') refers to sublattice $\alpha=$ A ($\alpha=$ B).

\section{Symmetries}
\label{sec:symmetries}
\subsection{Antiferromagnetic geometry}
\label{sec:afm}
We tabulate a choice of generators of the magnetic layer group (MLG) $p_c4/mmm$ of the antiferromagnetic configuration in Table~\ref{tab:generators-0}. To prove the degeneracies at X and Y mentioned in the main text (see Fig.~\ref{fig:model-0}), we show that a symmetry of the little group of X and Y has two orthogonal eigenstates with different eigenvalues, which however share the same energy, thereby leading to degeneracy.
We start by showing that two symmetries of the little group anti-commute. The symmetry elements of the little group of a $\bm{k}$ point leave $\bm{k}$ invariant and a magnetic symmetry $g = \{R \mid \bm{\tau}\}$ acts on a magnon eigenmode state $|\psi_{\bm{k}}\rangle$ as $g |\psi_{\bm{k}}\rangle = \mathrm{e}^{-\mathrm{i} (R \bm{k})\cdot \bm{\tau}} D(g) |\psi_{R\bm{k}}\rangle$, where $D(g)$ is the representation of the symmetry acting on the magnon eigenmodes and $R$ the vector matrix representation of the symmetry. The $D(g)$ is related to the representation $U(g)$ acting on the Holstein-Primakoff operators in the basis $(a_{\bm k}, b_{\bm k}, a_{-\bm k}^\dagger, b_{-\bm k}^\dagger )$ by $D(g) = T_{\bm k}^{-1} U(g) T_{\bm k}$ where $T_{\bm k}$ is the eigenvector matrix containing the Bogoliubov factors. 

We consider the little group of X which contains inversion $g = \mathcal{P} = \{R_g \mid \bm{\tau}_g \}$ with $R_g = - \diag(1, 1, 1)$ and $\bm \tau_g = \bm 0$. Since inversion connects the same sublattice and spin is a pseudo-vector, we find $U(\mathcal P) = \mathbbm{1}_4$, and it commutes with all other $U(h)$ in the little group. In particular, we can take the glide-mirror $h = \mathcal{M}_{100,\bm{\tau}} = \{R_h \mid \bm{\tau}_h\}$ with $R_h = \diag(-1, 1, 1)$ and $\bm \tau_h = (1/2, 1/2)$, which is composed of ${E}'_{\bm{\tau}}$ acting on $\mathcal{M}_{100}'$ (see Table~\ref{tab:generators-0}). Since $[U(g), U(h)] = 0$ it follows $[D(g), D(h)] = 0$, and the full representations including the translational part satisfy $gh = \eta_{gh} hg$, where $\eta_{gh} = \mathrm{exp}[-\mathrm{i} \bm{k} (\bm{\tau}_g + R_g \bm{\tau}_h - \bm{\tau}_h - R_h \bm{\tau}_g)]$ is the projective phase~\cite{Bradley2010}. We find  $\eta_{gh} = -1$ and therefore, the symmetries anti-commute. We pick $ |\psi\rangle$ as an eigenstate of $h$ with $ h |\psi\rangle = \lambda |\psi\rangle$, then apply $g$ and use the anti-commutation relation to find 
\begin{subequations}
\begin{align}
g h |\psi\rangle &= \lambda g |\psi\rangle \\
    h ( g|\psi\rangle) &= - \lambda ( g|\psi\rangle).
\end{align}
\end{subequations}
It follows that $|\psi\rangle$ and $g|\psi\rangle$ are eigenstates of $h$ with different eigenvalues, and therefore orthogonal eigenstates. They share the same energy and are therefore degenerate. For Y, this finding holds analogously with the symmetry $ h = \mathcal{M}_{010,\bm{\tau}}$.

To show the degeneracy at $\Gamma$ in Fig.~\ref{fig:model-0} of the main text we also show that $|\psi_{\Gamma}\rangle$ and $g|\psi_{\Gamma}\rangle$ are orthogonal eigenstates sharing the same energy, but with opposite eigenvalues. Here, we use the elements of the little group $h = C_{4z} = \{C_{4z} \mid \bm{0}\}$ and $g = {C_{2x,\bm{\tau}}} = \{C_{2x} \mid \bm{\tau}\}$ with $\bm \tau = (1/2, 1/2)$ modulo lattice vectors, and instead of using anti-commutation we use the more general conjugation of symmetries. We can choose $|\psi_{\Gamma} \rangle$ such that $h |\psi_{\Gamma} \rangle = \lambda |\psi_{\Gamma} \rangle$ with possible eigenvalues $\lambda \in \{\pm 1, \pm \mathrm{i}\}$ because $h^4 |\psi_{\Gamma} \rangle = |\psi_{\Gamma} \rangle$ leaves the state invariant. The Seitz multiplication of the conjugation yields $g h g^{-1}= \{ C_{4z}^{-1} \mid \bm 0 \}$ modulo a lattice vector. Applied to the $\Gamma$ point we find
\begin{subequations}
\begin{align}
    g h g^{-1} |\psi_{\Gamma} \rangle &  = D(C_{4z}^{-1}) |\psi_{\Gamma} \rangle = h^{-1} |\psi_{\Gamma} \rangle = \lambda^{-1}  |\psi_{\Gamma} \rangle.
\end{align}
\end{subequations}
Since $g h g^{-1} = h^{-1} $, we have 
\begin{subequations}
\begin{align}
      hgh|\psi_{\Gamma} \rangle )&= g|\psi_{\Gamma} \rangle  \\
      h(g |\psi_{\Gamma} \rangle )&= \lambda^{-1} (g|\psi_{\Gamma} \rangle ).
\end{align}
\end{subequations}

For $\lambda = \pm \mathrm i$ we find that $\lambda \ne \lambda^{-1}$ are opposite eigenvalues of the eigenstates of $h$, which makes the eigenstates orthogonal, and double degeneracy follows at $\Gamma$. For eigenvalues $\lambda = \pm 1$, no degeneracy is enforced. We now show that indeed $\lambda = \pm \mathrm i$ by establishing how a $C_{nz}$ rotation acts on the magnon spin. The spin is a pseudovector and transforms as $\bm{S}' = \det (R) R \bm{S}$, where $R$ is the matrix representation of the unitary symmetry we want to apply, so $R = C_{nz}$. We use the usual rotation matrix about $z$ which has determinant $\det R = 1$ and find 
\begin{align}
    \begin{pmatrix}
        \tilde{S}_{i}^x \\ \tilde{S}_{i}^y \\ \tilde{S}_{i}^z
    \end{pmatrix} =  \begin{pmatrix}
        \cos \alpha & - \sin \alpha & 0 \\
        \sin \alpha & \cos \alpha & 0 \\
        0 & 0 & 1
    \end{pmatrix} \begin{pmatrix}
        S_{i}^{x} \\ S_{i}^{y} \\ S_{i}^{z}
    \end{pmatrix},
    \label{eq:Srot}
\end{align}
where $\alpha = 2\uppi/n$ is the rotation angle.
Using the spin ladder operators $S_{i}^{\pm} = S_{i}^{x} \pm \mathrm{i} S_{i}^{y}$ we can express the transformation as
\begin{align}
    \tilde{S}_{i}^{\pm} = \mathrm{e}^{\pm i \alpha} S_{i}^{\pm},
\end{align}
as it leaves $S_i^z$ invariant. We therefore write
\begin{align}
    h S_{i,\mu}^{\pm} h^{-1} =  \mathrm{e}^{\pm \mathrm{i} \alpha} S_{h(i),h(\mu)}^{\pm},
\end{align}
where $h$ also acts on the sublattice $\mu$. For the Holstein-Primakoff bosons $a_i$ and $b_i$, we substitute $ S_{i\in \text{A}}^{+} \approx \sqrt{2S} a_i$, $ S_{i\in \text{A}}^{-} \approx \sqrt{2S} a_i^\dagger$, $ S_{i\in \text{B}}^{+} \approx \sqrt{2S} b_i^\dagger$, and $S_{i\in \text{B}}^{-} \approx \sqrt{2S} b_i$ and find in Fourier space, when the symmetry maps the lattice site onto the same sublattice $h(\mu) = \mu$,
\begin{subequations}
\begin{align}
     h a_{\bm{k}} h^{-1} & =  \mathrm{e}^{ \mathrm{i} \alpha} a_{h (\bm{k})}, \\
      h a_{-\bm{k}}^\dagger h^{-1} & =  \mathrm{e}^{ -\mathrm{i} \alpha} a_{h (-\bm{k})}^\dagger, \\
      h b_{\bm{k}} h^{-1} & =  \mathrm{e}^{ -\mathrm{i} \alpha} b_{h (\bm{k})}, \\
     h b_{-\bm{k}}^\dagger h^{-1} & =  \mathrm{e}^{ \mathrm{i} \alpha} b_{h (-\bm{k})}^\dagger.
\end{align}
\end{subequations}
In the case of $h = \{ \mathcal{C}_{4z} \mid \bm{0} \}$, the representation in the Holstein-Primakoff basis is $U(h) = \diag(\mathrm i, -\mathrm i, -\mathrm i, \mathrm i)$, which evidently has eigenvalues $\lambda = \pm \mathrm i$. Transforming to the eigenmode basis representation $D(h) = T_{\Gamma}^{-1} U(h) T_{\Gamma}$ preserves the eigenvalues. When $|\psi_\Gamma \rangle$ is an eigenstate of $h$ with eigenvalue $\lambda = \mathrm i$, then $g|\psi_\Gamma \rangle$ is an eigenstate of $h$ with the same energy but with eigenvalue $\lambda' = -\mathrm i$, which enforces double degeneracy at $\Gamma$. 

\begin{table}[H]
\centering
\caption{Generators of the MLG $p_c4/mmm$ with the same origin as the origin of the lattice vectors $\bm{a}_1=a\left(1, 0\right)$ and $\bm{a}_2=a\left(0, 1\right)$. Subscripts $100$ and $110$ are given in lattice coordinates and $\bm{\tau} = (1/2, 1/2)$ in fractional coordinates.}
\label{tab:generators-0}
\renewcommand{\arraystretch}{1.2}
\begin{tabular} {ccc} 
\toprule
Magn. symm. & lattice coords. &  Cartesian coords.\\
\hline 

$\mathcal{C}_{4z}$ 
& $\begin{pNiceMatrix}[columns-width = 0.3cm] 0 & -1 & 0\\ 1 & 0 & 0\\ 0 & 0 & 1 \end{pNiceMatrix}$
& $\begin{pNiceMatrix}[columns-width = 0.3cm] 0 & -1 & 0\\ 1 & 0 & 0\\ 0 & 0 & 1 \end{pNiceMatrix}$  \\

$\mathcal{M}_z$ & $\diag(1, 1, -1)$ & $\diag(1, 1, -1)$ \\

$\mathcal{M}_{100}'$ & $\diag(-1, 1, 1)$ & $\diag(-1, 1, 1)$  \\

$E_{\bm{\tau}}'$ & $\diag(1, 1, 1)$
& $\diag(1, 1, 1)$  \\

\botrule
\end{tabular}
\end{table}

\subsection{Altermagnetic geometry 1}
\label{sec:am-1}
We list a choice of generators for the symmorphic MLG $p4'/mm'm$ in Table~\ref{tab:generators-1}. The parent layer group is $p4/mmm$ (No. 61). To make the symmorphic nature of the group apparent, the generators contain an origin shift $\bm{\tau}_0 = (0, 1/2)$ with respect to the origin of the lattice vectors. The origin shift needs an adjustment of the MLG elements $\{R \mid \bm{\tau}' \}$ as follows~\cite{Bradley2010}
\begin{align}
    \{R \mid \bm{\tau} \} = \{R \mid \bm{\tau}' + R \bm{\tau}_0 - \bm{\tau}_0 \}.
\end{align}
Figure~\ref{fig:model-1} in the main text shows a vanishing spin expectation value in the Brillouin zone (BZ) diagonal $-$M-$\Gamma$-M, which we want to prove by showing that $\langle m_z \rangle = 0$ with the unitary mirror $g = \{\mathcal{M}_{1\bar{1}0} \mid \bm{\tau} \}$ with $\bm{\tau} = (1/2, -1/2)$. The action on an eigenstate $ | \psi_{\bm{\kappa}} \rangle$ on the diagonal (which is non-degenerate) with $\bm{\kappa} = (k, k)$ is
\begin{align}
    \{ \mathcal{M}_{1\bar{1}0} | \bm{\tau} \} |\psi_{\bm{\kappa}}\rangle 
    = \lambda_{\bm \kappa} |\psi_{\bm{\kappa}}\rangle,
\end{align}
with $|\lambda_{\bm{\kappa}}|^2 = 1$ as $g$ is unitary. We compute the spin expectation value $\langle m_z(\bm \kappa) \rangle = \langle \psi_{\bm \kappa} | m_z | \psi_{\bm \kappa} \rangle$ for the operator $m_z = \sum_i S_i^z$ and $g = \mathcal{M}_{1\bar{1}0, \bm \tau}$, where $m_z$ transforms as a pseudovector component
\begin{align}
    g m_z g^{-1} = \det (R_g) (R_g \bm{m})_z.
\end{align}
For $R_g= \begin{pmatrix}
    0 & 1 & 0 \\ 1 & 0 & 0 \\ 0 & 0 & 1
\end{pmatrix}$ we find $R_g z = z$ and $\det (R_g) = -1$, so the spin expectation value per band is
\begin{align}
   \langle m_z (\bm \kappa)\rangle = \langle \psi_{\bm \kappa} | m_z | \psi_{\bm \kappa} \rangle &= \langle \psi_{\bm \kappa} | g^{-1} g \,m_z \,g^{-1} g| \psi_{\bm \kappa} \rangle \nonumber \\
    & = \langle g \psi_{\bm \kappa} |  g \,m_z \,g^{-1}| g \psi_{\bm \kappa} \rangle \nonumber \\
    &=  |\lambda_{\bm \kappa}|^2 \langle \psi_{\bm \kappa} | (- m_z)| \psi_{\bm \kappa} \rangle\nonumber \\
    & =  - \langle \psi_{\bm \kappa} | m_z | \psi_{\bm \kappa} \rangle = -\langle m_z (\bm \kappa) \rangle.
\end{align}
To solve this condition $\langle m_z(\bm \kappa) \rangle = - \langle m_z(\bm \kappa) \rangle$, we need to set $\langle m_z(\bm \kappa) \rangle = 0$, and therefore the spin expectation value has to vanish. 
Analogous considerations with the perpendicular mirror lead to the spin expectation value of the other diagonal $\bm \kappa' = (k, -k)$ to vanish.
\begin{table}[H]
\centering
\caption{Generators of the MLG $p4'/mm'm$ with an origin-shift of $\bm{\tau}_0 = (0, 1/2)$ in fractional coordinates with respect to the origin of the lattice vectors $\bm{a}_1=a\left(1, 0\right)$ and $\bm{a}_2=a\left(0, 1\right)$. Subscripts $100$ and $1\bar{1}0$ are given in lattice coordinates.}
\label{tab:generators-1}
\renewcommand{\arraystretch}{1.2}
\begin{tabular} {ccc} 
\toprule
Magn. symm. & lattice coords. &  Cartesian coords.\\
\hline 

$\mathcal{C}_{4z}'$ 
& $\begin{pNiceMatrix}[columns-width = 0.3cm] 0 & -1 & 0\\ 1 & 0 & 0\\ 0 & 0 & 1 \end{pNiceMatrix}$
& $\begin{pNiceMatrix}[columns-width = 0.3cm] 0 & -1 & 0\\ 1 & 0 & 0\\ 0 & 0 & 1 \end{pNiceMatrix}$  \\

$\mathcal{M}_z$ & $\diag(1, 1, -1)$ & $\diag(1, 1, -1)$ \\

$\mathcal{M}_{100}'$ & $\diag(-1, 1, 1)$ & $\diag(-1, 1, 1)$  \\

$\mathcal{M}_{1\bar{1}0}$ & $\begin{pNiceMatrix}[columns-width = 0.3cm] 0 & 1 & 0\\ 1 & 0 & 0\\ 0 & 0 & 1 \end{pNiceMatrix}$
& $\begin{pNiceMatrix}[columns-width = 0.3cm] 0 & 1 & 0\\ 1 & 0 & 0\\ 0 & 0 & 1 \end{pNiceMatrix}$  \\

\botrule
\end{tabular}
\end{table}

\subsection{Altermagnetic geometry 2}
\label{sec:am-2}
The chosen generators for the nonsymmorphic MLG  $p4'/mbm'$ are listed in Table~\ref{tab:generators-2}. Here, the parent layer group is $p4/mbm$ (No. 63). The altermagnetic geometry 2 is rotated by 45$^\circ$ with respect to the standard setting. The vanishing spin expectation value along the nondegenerate diagonal of the BZ ($-$Y-$\Gamma$-Y, without the $\pm$Y points) shown in Fig.~\ref{fig:model-2} of the main text is enforced by the unitary glide-mirror $g = \mathcal{M}_{100,\bm{\tau}}$ with $\bm{\tau}=(1/2,1/2)$. The glide mirror switches $x$ and $y$ and therefore leaves $\bm{\kappa} = (k, k)$ invariant, for which we find 
\begin{align}
    \{ \mathcal{M}_{100} | \bm{\tau} \} |\psi_{\bm{\kappa}}\rangle = \mathrm{e}^{- \mathrm{i} \bm{\kappa} \cdot \bm{\tau}} D(\mathcal{M}_{100})|\psi_{\bm{\kappa}}\rangle = \lambda_{\bm{\kappa}} |\psi_{\bm{\kappa}}\rangle,
\end{align}
with eigenvalue $\lambda_{\bm{\kappa}}$ satisfying $|\lambda_{\bm{\kappa}}|^2 = 1$ as $g$ is unitary, and $ g m_z g^{-1} = - m_z$ applies. Therefore, the spin expectation has to vanish, as in the previous section~\ref{sec:am-1}. The other diagonal's vanishing spin expectation value is found similarly with $g = \left( \mathcal{C}_{4z, \bm{\tau}}'\right)^2 \mathcal{M}_{100,\bm{\tau}}= \mathcal{M}_{010,\bm{\tau}}$ with $\bm \tau = (1/2, 1/2)$ modulo lattice vectors, which leaves $\bm\kappa=(k,-k)$ invariant.

The degeneracies at X and Y in Fig.~\ref{fig:model-2} of the main text for the MLG $p4'/mbm'$ follow from the same anti-commutation argument as in the antiferromagnetic case with MLG $p_c4/mmm$. Although the present $p4'/mbm'$ setting is rotated by 45$^\circ$, its little group at X (Y) also contains the elements inversion $\mathcal P=\mathcal M_z(\mathcal C'_{4z,\bm\tau})^2$ and the glide mirror $\mathcal M_{100,\bm\tau}$ ($\mathcal M_{010,\bm\tau}$), and the projective phase evaluates to $\eta_{gh}=-1$, enforcing the double degeneracy. 

For the degeneracy at $\Gamma$ we will make use of Kramer's theorem, which requires an anti-unitary symmetry that squares to $-1$. Let $a = \{C_{4z} \mid \bm \tau \}\mathcal T$ with $a^2 = \{C_{2z}\mid \bm0\}$ modulo lattice vectors. Using the general transformation in Eq.~\eqref{eq:Srot} we have $U(a^2) = -\mathbbm{1}_4$. Since this matrix is proportional to the identity it preserves its form in the eigenmode basis $D(a^2) = - \mathbbm{1}_4$. With $|\psi_{\Gamma} \rangle$ an eigenstate of the Hamiltonian at $\Gamma$, we define $|\psi_{\Gamma}' \rangle = a|\psi_{\Gamma} \rangle$. Since $a^2 = -1$, applying $a$ again leads to $-a|\psi_{\Gamma}' \rangle = |\psi_{\Gamma} \rangle$. We now prove that the states $|\psi_{\Gamma} \rangle$ and $|\psi_{\Gamma}' \rangle$ are orthogonal with
\begin{align}
    \langle \psi_{\Gamma}' \mid \psi_{\Gamma} \rangle = - \langle a\psi_{\Gamma}\mid a\psi_{\Gamma}' \rangle = - \left(\langle \psi_{\Gamma}\mid \psi_{\Gamma}' \rangle\right)^* = -\langle \psi_{\Gamma}'\mid \psi_{\Gamma} \rangle.
\end{align}
To satisfy this condition, we find that $\langle \psi_{\Gamma}' \mid \psi_{\Gamma} \rangle = -\langle \psi_{\Gamma}'\mid \psi_{\Gamma} \rangle = 0$ and the states are orthogonal and degenerate.

\begin{table}[h]
\centering
\caption{Generators of the MLG $p4'/mbm'$ with $\bm{\tau}=\left(1/2, 1/2\right)$ in fractional coordinates and lattice vectors $\bm{a}_1=a\left(1, -1\right)$ and $\bm{a}_2=a\left(1, 1\right)$. The symmetry-origin agrees with the origin of the lattice vectors. Subscripts $100$ and $110$ are given in lattice coordinates.}
\label{tab:generators-2}
\renewcommand{\arraystretch}{1.2}
\begin{tabular}{ccc}
\toprule
Magn. symm. & lattice coords. &  Cartesian coords.\\ 
\hline
$\mathcal{C}_{4z, \bm{\tau}}'$ & 
$\begin{pNiceMatrix}[columns-width = 0.3cm] 0 & -1 & 0\\ 1 & 0 & 0\\ 0 & 0 & 1 \end{pNiceMatrix}$
& $\begin{pNiceMatrix}[columns-width = 0.3cm] 0 & -1 & 0\\ 1 & 0 & 0\\ 0 & 0 & 1 \end{pNiceMatrix}$  \\

$\mathcal{M}_z$ & $\diag(1, 1, -1)$ & $\diag(1, 1, -1)$ \\

$\mathcal{M}_{100,\bm{\tau}}$ & $\diag(-1, 1, 1)$ & $\begin{pNiceMatrix}[columns-width = 0.3cm] 0 & 1 & 0\\ 1 & 0 & 0\\ 0 & 0 & 1 \end{pNiceMatrix}$ \\

$\mathcal{M}_{110}'$ 
& $\begin{pNiceMatrix}[columns-width = 0.3cm] 0 & -1 & 0\\ -1 & 0 & 0\\ 0 & 0 & 1 \end{pNiceMatrix}$
& $\diag(-1, 1, 1)$  \\

\botrule
\end{tabular}
\end{table}

\newpage
\section{Perpendicular magnetic anisotropy optimization}
\label{sec:anisotropy}
\begin{figure*}[h]
    \centering
    \includegraphics[width=1\textwidth]{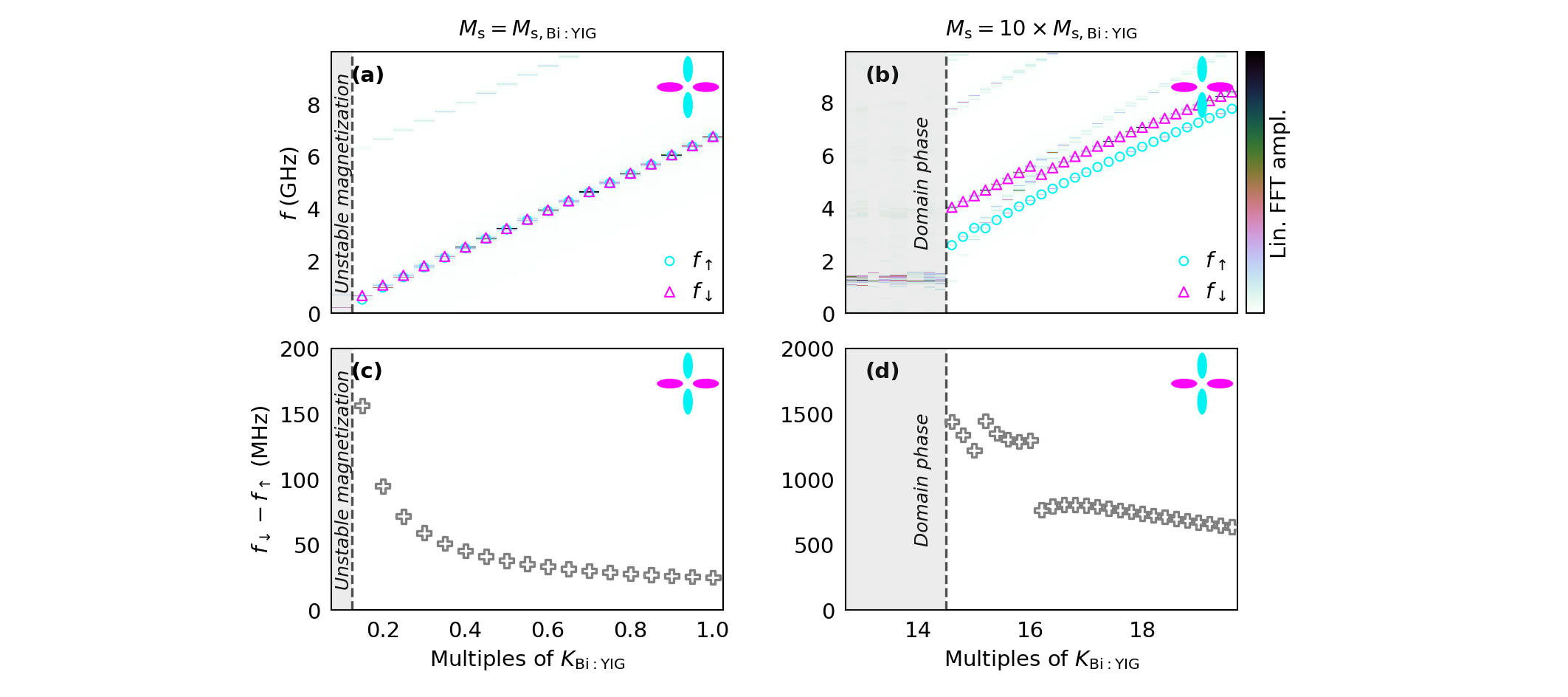}
    \caption{Dependence of mode frequencies and altermagnetic splitting on the perpendicular magnetic anisotropy $K$ for both $M_\mathrm{s}$ configurations. (a,~c) For the Bi:YIG saturation magnetization ($M_\mathrm{s} = M_\mathrm{s, Bi:YIG}$), reducing $K$ leads to a pronounced, non-linear increase in the zero-field altermagnetic splitting $f_\downarrow - f_\uparrow$. Below $K \approx 0.15 \times K_\mathrm{Bi:YIG}$, the magnetic texture becomes unstable and individual nanomagnet moments tilt in-plane. (b,~d) For the enhanced saturation magnetization ($M_\mathrm{s} = 10 \times M_\mathrm{s, Bi:YIG}$), the monotonic evolution of the splitting in (d) is disrupted by mode hybridization; avoided crossings appear between the fundamental altermagnetic modes and single-node intra-island modes at $K \approx 15 \times K_\mathrm{Bi:YIG}$ (for $f_\uparrow$) and $K \approx 16 \times K_\mathrm{Bi:YIG}$ (for $f_\downarrow$). Below $K \approx 14.5 \times K_\mathrm{Bi:YIG}$, the out-of-plane state destabilizes into a domain phase.}
    \label{fig:anisotropy}
\end{figure*}

\newpage
\section{Finite-size effects}
\label{sec:finite_size}

\begin{figure*}[h]
    \centering
    \includegraphics[width=1\textwidth]{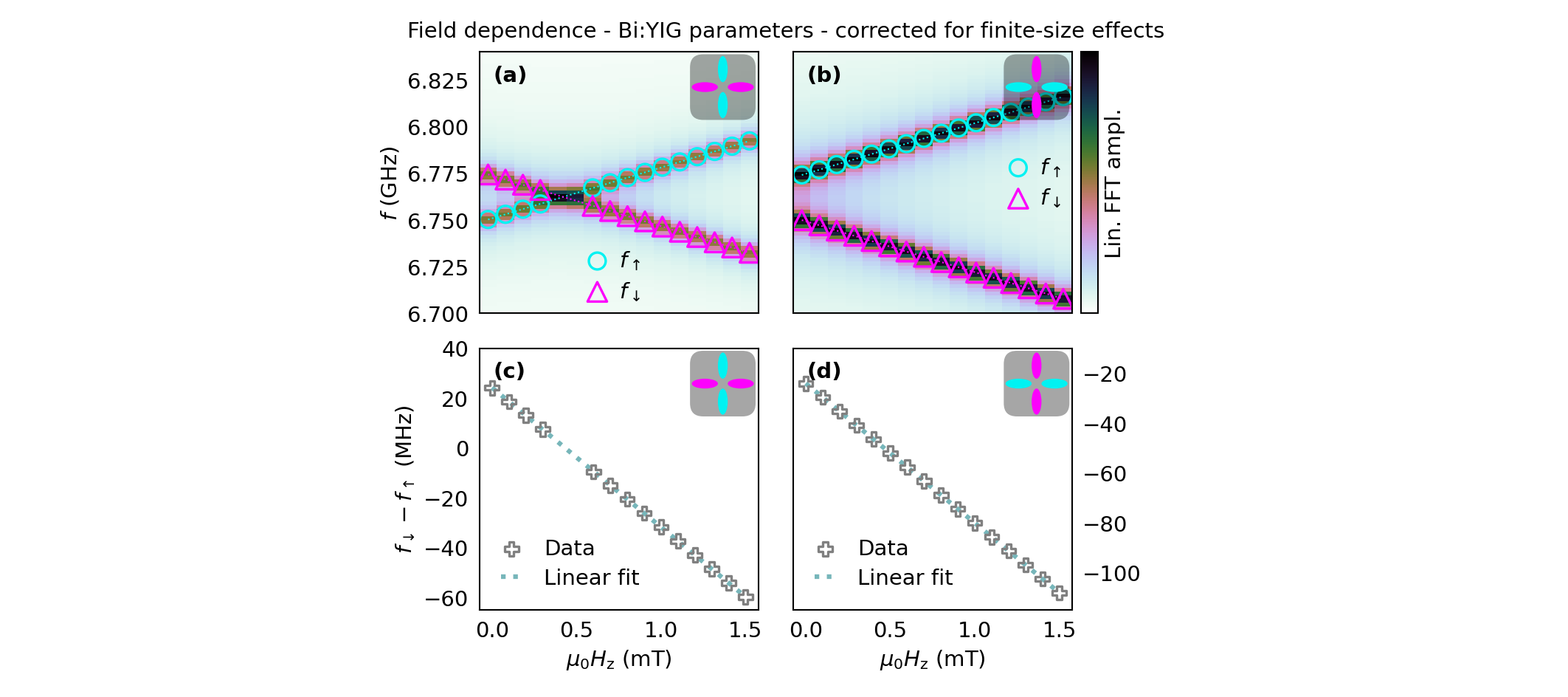}
    \caption{Mitigation of finite-size aspect-ratio effects using a square micromagnetic simulation domain ($4 \times 4$ unit cells, $128 \times 128$ grid resolution). (a,~b) Field dependence of the mode frequencies $f_\uparrow$ and $f_\downarrow$ for the regular configuration and its time-reversed counterpart employing the Bi:YIG parameter set (see Tab.~\ref{tab:parameters-1}, main text). Absorbing boundary regions are omitted since only non-propagating localized modes are excited. (c,~d) Frequency splitting $f_\downarrow - f_\uparrow$ as a function of the external perpendicular field $\mu_0 H_z$. By eliminating the geometric aspect-ratio asymmetry of the simulation window, the extracted magnon spin polarization values $|\langle m_z \rangle|$ become virtually symmetric under time inversion and approach unity, in excellent quantitative agreement with the analytical macrospin model.}
    \label{fig:finite_size_correction}
\end{figure*}

Correcting for finite-size effects (induced by the aspect ratio of the simulation window in the main text) by employing square simulation domains yields quantitative agreement with the theoretical model ($|\langle m_z \rangle| \approx 1.00$, see Fig.~\ref{fig:finite_size_correction} and Tab.~\ref{tab:spin_polarization_comparison}), confirming that minor polarization deviations under time inversion stem entirely from domain geometry rather than intrinsic physical symmetry breaking. Importantly, the weak anti-crossing between the altermagnetic modes vanishes as the spin polarization approaches unity, as is evident from the undisturbed linear evolution in Fig.~\ref{fig:finite_size_correction}(c).

\begin{table}[h!]
\centering
\caption{Comparison of the extracted magnon spin polarization components $\langle m_z \rangle$ between the rectangular simulation window (see main text, Fig.~\ref{fig:micromagnetics}) and the square, finite-size-corrected simulation window (Fig.~\ref{fig:finite_size_correction}).}
\label{tab:spin_polarization_comparison}
\begin{tabular}{l cc c cc}
\hline\hline
& \multicolumn{2}{c}{Rectangular Window ($20 \times 2$ unit cells)} & & \multicolumn{2}{c}{Square Window ($4 \times 4$ unit cells)} \\
\cline{2-3} \cline{5-6}
Polarization  & Regular & Time-Reversed & & Regular & Time-Reversed \\
\hline
$\langle m_z \rangle_{f_\uparrow}$ & $+1.107$ & $+0.979$ & & $+1.002$ & $+0.999$ \\
$\langle m_z \rangle_{f_\downarrow}$ & $-1.107$ & $-0.983$ & & $-0.998$ & $-1.003$ \\
\hline
$\mathbf{|\langle m_z \rangle|_\text{mean}}$ & $\mathbf{1.107}$ & $\mathbf{0.981}$ & & $\mathbf{1.000}$ & $\mathbf{1.001}$ \\
\hline\hline
\end{tabular}
\end{table}

\bibliography{bib}